\documentclass[11pt,a4paper]{article}
\usepackage[T1]{fontenc}
\usepackage[utf8]{inputenc}
\usepackage[ngerman,english]{babel}
\usepackage{cite}
\usepackage{amsmath}
\usepackage{amsfonts}
\usepackage{amssymb}
\usepackage{amsthm}
\usepackage{enumitem}
\usepackage{dcolumn}
\usepackage{bm}
\usepackage{bbm}
\usepackage{cancel}
\usepackage[pdftex]{graphicx}
\usepackage{tikz}
\usetikzlibrary{calc,arrows,positioning}
\usepackage{tabularx}
\usepackage{booktabs}
\usepackage{tensor}
\usepackage{scalerel}
\usepackage[small]{caption}
\usepackage{xcolor}
\usepackage{mathtools}
\usepackage{float}
\usepackage{xfrac}
\usepackage[pdfborder={0 0 0}]{hyperref}
\definecolor{darkblue}{rgb}{0,0,.5}
\definecolor{fhl}{rgb}{1,0,0}
\hypersetup{colorlinks=true, breaklinks=true, linkcolor=darkblue, menucolor=darkblue, urlcolor=darkblue, linktocpage=true}
\usepackage{times}
\usepackage{algpseudocode}
\usepackage{algorithm}
\usepackage{textpos}
\usepackage{geometry}
\DeclareMathSizes{11}{10}{8}{6} 
\usepackage[titletoc,title]{appendix}

\usepackage{authblk}
\makeatletter 
\newsavebox\myboxA 
\newsavebox\myboxB 
\newlength\mylenA 
 
\newcommand*\xoverline[2][0.75]{% 
    \sbox{\myboxA}{$\m@th#2$}%
    \setbox\myboxB\null% Phantom box 
    \ht\myboxB=\ht\myboxA% 
    \dp\myboxB=\dp\myboxA% 
    \wd\myboxB=#1\wd\myboxA% Scale phantom 
    \sbox\myboxB{$\m@th\overline{\copy\myboxB}$}%  Overlined phantom 
    \setlength\mylenA{\the\wd\myboxA}%   calc width diff 
    \addtolength\mylenA{-\the\wd\myboxB}% 
    \ifdim\wd\myboxB<\wd\myboxA% 
       \rlap{\hskip 0.5\mylenA\usebox\myboxB}{\usebox\myboxA}% 
    \else 
        \hskip -0.5\mylenA\rlap{\usebox\myboxA}{\hskip 0.5\mylenA\usebox\myboxB}% 
    \fi}
\makeatother

\numberwithin{equation}{section}

\renewcommand*\bar[1]{\ThisStyle{\SavedStyle\xoverline{\SavedStyle #1}}}

\let\oldhat\hat
\renewcommand*\hat[1]{\ThisStyle{\smash{\SavedStyle\oldhat{\SavedStyle#1}}\vphantom{\SavedStyle#1}}}
\newcommand\ohat[1]{\widehat{#1}}
\newcommand\obar[1]{\xoverline{#1}}
\let\originalleft\left
\let\originalright\right
\renewcommand{\left}{\mathopen{}\mathclose\bgroup\originalleft}
\renewcommand{\right}{\aftergroup\egroup\originalright}
\newcommand{\e}{\operatorname{e}}
\newcommand{\SU}[1]{\operatorname{SU}\left(#1\right)}
\newcommand{\On}[1]{\operatorname{O}\left(#1\right)}
\newcommand{\U}[1]{\operatorname{U}\left(#1\right)}
\newcommand{\CP}[1]{\operatorname{CP}\left(#1\right)}

\newcommand{\of}[1]{\left(\right.#1\left.\right)}
\newcommand{\bof}[1]{\biggl(\bigg.#1\bigg.\biggr)}
\newcommand{\sof}[1]{\bigl(\big.#1\big.\bigr)}
\newcommand{\ssof}[1]{(#1)}
\newcommand{\fof}[1]{\left[\right.#1\left.\right]}

\newcommand{\cof}[1]{\left\{\right.#1\left.\right\}}
\newcommand{\bcof}[1]{\biggl\{\bigg.#1\bigg.\biggr\}}
\newcommand{\scof}[1]{\bigl\{\big.#1\big.\bigr\}}

\newcommand{\avof}[1]{\left\langle #1\right\rangle}
\newcommand{\savof}[1]{\big\langle #1\big\rangle}

\newcommand{\ii}{\mathrm{i}}
\newcommand{\idd}[2]{\mathrm{d}^{#2}\,#1}

\newcommand{\DD}[1]{\mathcal{D}\left[#1\right]}

\newcommand{\partd}[2]{\frac{\partial #1}{\partial #2}}
\newcommand{\partdm}[3]{\frac{\partial^{#3} #1}{\partial #2^{#3}}}
\newcommand{\spartd}[3]{\frac{\partial^{2} #1}{\partial #2 \partial #3}}

\newcommand{\order}[1]{\mathcal{O}\big(#1\big)}

\newcommand{\abs}[1]{\left| #1\right|}
\newcommand{\sabs}[1]{\big| #1\big|}

\renewcommand*\[{\begin{equation}}
\renewcommand*\]{\end{equation}}
\let\oldstackrel\stackrel
\renewcommand*\stackrel[2]{{\scriptstyle\oldstackrel{#1}{#2}}}

\definecolor{emphcol}{RGB}{0,0,0}
\let\oldemph\emph
\renewcommand*\emph[1]{\oldemph{\textcolor{emphcol}{#1}}}
\newlength{\hatchspread}
\newlength{\hatchthickness}
\newlength{\hatchshift}
\newcommand{\hatchcolor}{}
\tikzset{hatchspread/.code={\setlength{\hatchspread}{#1}},
         hatchthickness/.code={\setlength{\hatchthickness}{#1}},
         hatchshift/.code={\setlength{\hatchshift}{#1}},% must be >= 0
         hatchcolor/.code={\renewcommand{\hatchcolor}{#1}}}
\tikzset{hatchspread=3pt,
         hatchthickness=0.4pt,
         hatchshift=0pt,% must be >= 0
         hatchcolor=black}
\pgfdeclarepatternformonly[\hatchspread,\hatchthickness,\hatchshift,\hatchcolor]% variables
   {custom north west lines}% name
   {\pgfqpoint{\dimexpr-2\hatchthickness}{\dimexpr-2\hatchthickness}}% lower left corner
   {\pgfqpoint{\dimexpr\hatchspread+2\hatchthickness}{\dimexpr\hatchspread+2\hatchthickness}}% upper right corner
   {\pgfqpoint{\dimexpr\hatchspread}{\dimexpr\hatchspread}}% tile size
   {% shape description
    \pgfsetlinewidth{\hatchthickness}
    \pgfpathmoveto{\pgfqpoint{\dimexpr0pt}{\dimexpr\hatchspread+\hatchshift}}
    \pgfpathlineto{\pgfqpoint{\dimexpr\hatchspread+0.15pt}{\dimexpr\hatchshift-0.15pt}}
    \ifdim \hatchshift > 0pt
      \pgfpathmoveto{\pgfqpoint{\dimexpr-0.15pt}{\dimexpr\hatchshift+0.15pt}}
      \pgfpathlineto{\pgfqpoint{\dimexpr\hatchshift+0.15pt}{\dimexpr-0.15pt}}
    \fi
    \pgfsetstrokecolor{\hatchcolor}
    \pgfusepath{stroke}
   }

\pgfdeclarepatternformonly[\hatchspread,\hatchthickness,\hatchshift,\hatchcolor]% variables
   {custom north east lines}% name
   {\pgfqpoint{\dimexpr-2\hatchthickness}{\dimexpr-2\hatchthickness}}% lower left corner
   {\pgfqpoint{\dimexpr\hatchspread+2\hatchthickness}{\dimexpr\hatchspread+2\hatchthickness}}% upper right corner
   {\pgfqpoint{\dimexpr\hatchspread}{\dimexpr\hatchspread}}% tile size
   {% shape description
    \pgfsetlinewidth{\hatchthickness}
    \pgfpathmoveto{\pgfqpoint{0pt}{\dimexpr\hatchshift}}
    \pgfpathlineto{\pgfqpoint{\dimexpr\hatchspread+0.15pt}{\dimexpr\hatchspread+0.15pt+\hatchshift}}
    \ifdim \hatchshift > 0pt
      \pgfpathmoveto{\pgfqpoint{\dimexpr\hatchspread-\hatchshift-0.15pt}{\dimexpr-0.15pt}}
      \pgfpathlineto{\pgfqpoint{\dimexpr\hatchspread+0.15pt}{\dimexpr\hatchshift+0.15pt}}
    \fi
    \pgfsetstrokecolor{\hatchcolor}
    \pgfusepath{stroke}
   }

\pgfdeclarepatternformonly[\hatchspread,\hatchthickness,\hatchshift,\hatchcolor]% variables
   {custom vertical lines}% name
   {\pgfqpoint{\dimexpr-2\hatchthickness}{\dimexpr-2\hatchthickness}}% lower left corner
   {\pgfqpoint{\dimexpr\hatchspread+2\hatchthickness}{\dimexpr\hatchspread+2\hatchthickness}}% upper right corner
   {\pgfqpoint{\dimexpr\hatchspread}{\dimexpr\hatchspread}}% tile size
   {% shape description
    \pgfsetlinewidth{\hatchthickness}
    \pgfpathmoveto{\pgfqpoint{\dimexpr\hatchshift}{0pt}}
    \pgfpathlineto{\pgfqpoint{\dimexpr\hatchshift}{\dimexpr\hatchspread+0.15pt}}
    \pgfsetstrokecolor{\hatchcolor}
    \pgfusepath{stroke}
   }

\pgfdeclarepatternformonly[\hatchspread,\hatchthickness,\hatchshift,\hatchcolor]% variables
   {custom horizontal lines}% name
   {\pgfqpoint{\dimexpr-2\hatchthickness}{\dimexpr-2\hatchthickness}}% lower left corner
   {\pgfqpoint{\dimexpr\hatchspread+2\hatchthickness}{\dimexpr\hatchspread+2\hatchthickness}}% upper right corner
   {\pgfqpoint{\dimexpr\hatchspread}{\dimexpr\hatchspread}}% tile size
   {% shape description
    \pgfsetlinewidth{\hatchthickness}
    \pgfpathmoveto{\pgfqpoint{0pt}{\dimexpr\hatchshift}}
    \pgfpathlineto{\pgfqpoint{\dimexpr\hatchspread+0.15pt}{\dimexpr\hatchshift}}
    \pgfsetstrokecolor{\hatchcolor}
    \pgfusepath{stroke}
   }

\begin{document}\selectlanguage{english}
\newcounter{romanPagenumber}
\newcounter{arabicPagenumber}
\pagenumbering{Roman}
\title{
\begin{textblock*}{100pt}(399pt,-96pt)
\textnormal{\small \texttt{CERN-TH-2016-056}}
\end{textblock*}Sampling of General Correlators in Worm-Algorithm Based Simulations}
\author[1]{Tobias Rindlisbacher \thanks{E-mail: rindlisbacher@itp.phys.ethz.ch}}
\author[1]{Oscar {\AA}kerlund \thanks{E-mail: oscara@itp.phys.ethz.ch}}
\author[1,2]{Philippe de Forcrand \thanks{E-mail: forcrand@itp.phys.ethz.ch}}
\affil[1]{\small ETH Z\"urich, Institute for Theoretical Physics, Wolfgang-Pauli-Str. 27, CH - 8093 Z\"urich, Switzerland}
\affil[2]{\small CERN, Physics Department, TH Unit, CH-1211 Gen\`eve 23, Switzerland}
\renewcommand\Authands{ and }
\maketitle
\begin{abstract}
%% Text of abstract
Using the complex $\phi^4$-model as a prototype for a system which is simulated by a worm algorithm, we show that not only the charged correlator $\avof{\phi^{*}\of{x}\phi\of{y}}$, but also more general correlators such as $\avof{\abs{\phi\of{x}}\abs{\phi\of{y}}}$ or $\avof{\arg\of{\phi\of{x}}\arg\of{\phi\of{y}}}$, as well as condensates like $\avof{\abs{\phi}}$, can be measured at every step of the Monte Carlo evolution of the worm instead of on closed-worm configurations only. The method generalizes straightforwardly to other systems simulated by worms, such as spin or sigma models.  
\end{abstract}

        
\newpage
\setcounter{romanPagenumber}{\value{page}} 
\pagenumbering{arabic}

\section{Introduction}\label{sec:intro}

In a paper from 2001 \cite{Prokofev}, Prokof'ev and Svistunov proposed the worm algorithm as an alternative to cluster algorithms \cite{Swendsen} to overcome critical slowing-down in the simulation of classical spin models like the Ising or the 3 state Potts model, but also of the complex $\phi^4$ model.\\
Critical slowing-down usually occurs at a second order phase transition where the typical length-scale over which the degrees of freedom (e.g. the spins of an Ising system) are correlated, diverges and the system develops long-range order. Local update algorithms then usually become very inefficient due to the high energy barrier that has to be overcome to flip individual spins against the field of its nearest neighbors and to the corresponding low acceptance rate of such updates. Cluster algorithms overcome this problem by being non-local: typical structures of correlated sites, the so called clusters, are grown and then flipped as a whole. Because the typical size of these clusters grows like the correlation length, critical slowing-down is averted. Unfortunately, the typical cluster size usually grows even further when leaving the critical region and going deeper into the ordered phase, resulting in a loss of efficiency of the cluster updates which becomes particularly dramatic when the typical cluster size approaches the size of the whole system\footnote{When the typical clusters consist of almost the whole system, the cluster updates essentially just flip the whole system back and forth.}. The more relevant problem with cluster algorithms is however that they are in general not useful if the system couples to an external field, as the large energy barrier that has to be overcome to flip a whole cluster \emph{against} the external field quickly leads to very low acceptance rates even for moderate cluster sizes.\\
Worm algorithms on the other hand are based on local updates and remain relatively efficient even in the ordered phase and in the presence of external fields. For bosonic systems, worm algorithms emerge quite naturally when expressing the partition function in terms of a particular kind of integer-valued "dual" variables, the so-called \emph{flux-variables} (see \cite{Prokofev,Endres,Endres2,Wolff,Wolff1,Gattringer1} and Sec.~\ref{ssec:phi4partf}). The working principle behind this dualization process is as follows: consider the simple system described by the following partition function:
\[
Z=\int\limits_{0}^{1}\idd{\phi}{}\e^{\phi+\lambda\,\phi}\ .\label{eq:toymodel}
\]
If we wanted to use Monte Carlo to compute for example expectation value and variance of the field $\phi$, i.e.
\[
\savof{\phi}\,=\,\partd{\log\of{Z}}{\lambda}\qquad\text{and}\qquad \savof{\phi^2}-\savof{\phi}^2\,=\,\partdm{\log\of{Z}}{\lambda}{2}\ ,
\]
the standard way to do this would be to interpret $\e^{\phi+\lambda\,\phi}$ as probability weight for doing importance sampling with respect to $\phi$. To obtain the dual formulation, we instead write the integrand of \eqref{eq:toymodel} in a power series in $\phi$, carry out the integral for each monomial (which can usually be done analytically),
\[
w_{n}\of{\lambda}\,=\,\int\limits_{0}^{1}\idd{\phi}{}\frac{\of{\of{1+\lambda}\phi}^{n}}{n!}\,=\,\frac{\of{1+\lambda}^{n}}{\of{n+1}!},
\]
such that
\[
Z=\sum\limits_{n=0}^{\infty}w_{n}\of{\lambda},
\]
and then use the $w_{n}\of{\lambda}$ as probability weights for doing importance sampling with respect to the monomial order $n$, which is our new, integer-valued configuration variable, in terms of which the above observables read
\[
\partd{\log\of{Z}}{\lambda}\,=\,\frac{\savof{n}}{1+\lambda}\qquad\text{and}\qquad \partdm{\log\of{Z}}{\lambda}{2}\,=\,\frac{\savof{n^2}-\savof{n}^2-\savof{n}}{\of{1+\lambda}^2}\ .
\]
This change of variables from $\phi\in\mathbb{R}$ to $n\in\mathbb{N}_{0}$ is of course only meaningful if the $w_{n}\of{\lambda}$ are real and non-negative for all $n$. In the above mentioned dual representation of bosonic systems, the \emph{flux variables}, which can be interpreted as representing the hopping of particles/excitations between neighboring sites, play a similar role as $n$. However if the excitations carry a conserved charge, these \emph{flux variables} are in general subject to a so-called \emph{closed loop constraint}, which makes it impossible to update them with ordinary local Metropolis. The worm algorithm deals with this problem by temporarily violating the constraint on two sites, referred to as head and tail of the worm, and updates the flux variables while the head is moved around from site to neighboring site using importance sampling. The constraint is restored as soon as head and tail meet each other again. This reformulation of the partition function in terms of flux variables also avoids the sign problem in many significant cases, notably in those where the sign problem is related to the introduction of a non-zero chemical potential (see \cite{Endres,Endres2,Gattringer,Gattringer1,Rindlisbacher1} and Sec.~\ref{ssec:phi4partf} below).\\
As pointed out for example in \cite{Wolff,Wolff2,Gattringer0}, the sampling of the location of the head of the worm (relative to its tail) during the worm update corresponds to the sampling of the correlator for the particle to whose hopping the updated flux variables correspond, and provides therefore an easy and very efficient way to estimate this correlator by recording a histogram for the head-tail distances realized during the worm updates.\\
In the present paper we show, using the complex $\phi^4$-model as a prototype for systems which can be simulated by worm algorithms, that not only the correlator $\avof{\phi^{*}\of{x}\phi\of{y}}$ of the charged fields $\phi$ and $\phi^{*}$ (which are the excitations to whose hopping the flux variables correspond), but also more general correlators such as $\avof{\abs{\phi\of{x}}\abs{\phi\of{y}}}$ and $\avof{\arg\of{\phi\of{x}}\arg\of{\phi\of{y}}}$ as well as condensates like $\avof{\abs{\phi}}$ can be measured at every step of the Monte Carlo evolution of a worm. However, as the $\phi$ field is integrated out exactly (and with it the $\U{1}$ symmetry), the direct measurement of such one and two-point functions requires in general the introduction of non-vanishing source terms\footnote{As an alternative to source terms, one could also introduce a background field: for the complex $\phi^4$ model discussed in this paper, one would change variables $\phi\rightarrow\phi+\phi_{0}$ where $\phi_{0}$ is the background field which would be set to the vacuum expectation value.} to the action, to which, as suggested in \cite{Gattringer0}, correspond additional dual variables representing so-called \emph{monomers} (see Sec. \ref{ssec:phi4partf}). Note that the necessity of a source term in order to obtain a non-vanishing result for the condensate is not a flaw of the dual formulation but a consequence of the definition of the partition function in terms of an Euclidean path-integral, which explicitly sums over all possible field configurations. This means that even in the ordered phase where the minimum of the free energy develops a $\U{1}$ degeneracy and one usually assumes (motivated by the fact that in an infinite system, vacua that correspond to different minima of the free energy do not interact) that the system undergoes spontaneous symmetry breaking and picks just one of the corresponding degenerate vacua, the partition function explicitly always sums homogeneously over all of these vacua, unless one specifies a source term in the action that gives dominant weight to just one of them. This will be discussed in more detail in Sec. \ref{ssec:measurecond} and \ref{ssec:effectofsource}.\\

The paper is organized as follows: in the remainder of this section, we first define in part~\ref{ssec:phi4partf} our lattice $\phi^4$ action and the corresponding partition function, then we derive, following \cite{Gattringer}, the flux-variable representation of the latter. In part~\ref{ssec:chargedcorr} we then discuss the relation between the charged correlator and the worm algorithm. Sec.~\ref{sec:generalcorr} is intended to demonstrate how some simple modifications of the worm algorithm allow one to measure also more general correlators as well as condensates during the worm evolution, provided a non-vanishing source term is added to the action. The section also contains a short discussion on why such a source term is necessary. A short summary follows in Sec.~\ref{sec:summary}. A detailed description of our simulation algorithm is given in appendix~\ref{sec:wormupdate}. 

\subsection{Complex $\phi^4$ with Source Terms}\label{ssec:phi4partf}

The Euclidean lattice action for a complex $\phi^4$ field, coupled to a chemical potential $\mu$ and a (complex) source term $s$ (which could be thought of as an external magnetic field), reads in $d$ dimensions:
\[
S\fof{\phi}\,=\,\sum\limits_{x}\bcof{-\kappa\,\sum\limits_{\nu=1}^{d}\of{\e^{2\,\mu\delta_{\nu,d}}\,\phi^{*}_{x}\,\phi_{x+\hat{\nu}}+\e^{-2\,\mu\delta_{\nu,d}}\,\phi^{*}_{x+\hat{\nu}}\,\phi_{x}}+2\,\abs{\phi_{x}}^2+\lambda\of{2\,\abs{\phi_{x}}^{2}-1}^{2}-\of{s^{*}\,\phi_{x}+s\,\phi^{*}_{x}}},\label{eq:phifouraction}
\]
where $\kappa$ and $\lambda$ are dimensionless lattice parameters\footnote{We use the convention $\phi=\frac{1}{\sqrt{2}}\of{\phi_{1}+\ii\,\phi_{2}}$, $\phi^{*}=\frac{1}{\sqrt{2}}\of{\phi_{1}-\ii\,\phi_{2}}$, which is convenient when generalizing the action to the $\On{N}$ case. Note also that our $\kappa$ is four times the "standard $\kappa$" used in other lattice formulations of $\phi^4$ theory.}. As \eqref{eq:phifouraction} is in general complex for non-zero values of the chemical potential $\mu$, the corresponding partition function,
\[
Z\,=\,\int\DD{\phi}\,\e^{-S\fof{\phi}},\label{eq:phifourpartf}
\]
has a so-called \emph{sign problem}, as a complex $\DD{\phi}\e^{-S\fof{\phi}}$ has no probabilistic meaning and can therefore not be used for importance sampling of configurations within a Monte Carlo simulation.\\

To overcome this problem we follow \cite{Gattringer} and first write $\e^{-S\fof{\phi}}$ in \eqref{eq:phifourpartf} as a product of individual exponentials,
\begin{multline}
Z\,=\,\int\DD{\phi}\prod\limits_{x}\bcof{\bof{\prod\limits_{\nu=1}^{d}\exp\sof{\kappa\,\e^{2\,\mu\delta_{\nu,d}}\,\phi^{*}_{x}\,\phi_{x+\hat{\nu}}}\exp\sof{\kappa\,\e^{-2\,\mu\delta_{\nu,d}}\,\phi^{*}_{x+\hat{\nu}}\,\phi_{x}}}\\
\exp\sof{-2\,\abs{\phi_{x}}^2-\lambda\of{2\,\abs{\phi_{x}}^{2}-1}^{2}}\exp\of{s^{*}\,\phi_{x}}\exp\of{s\,\phi^{*}_{x}}},\label{eq:phifourfluxrepd1}
\end{multline}
and then write these exponentials, except the ones that correspond to the potential part, in terms of their power-series,
\[
\exp\sof{\kappa\,\e^{2\,\mu\delta_{\nu,d}}\,\phi^{*}_{x}\,\phi_{x+\hat{\nu}}}\,=\,\sum\limits_{\xi_{x,\nu}}\frac{\of{\phi^{*}_{x}\,\phi_{x+\hat{\nu}}\,\kappa\,\e^{\mu\delta_{\nu,d}}}^{\xi_{x,\nu}}}{\xi_{x,\nu}!},
\]
\[
\exp\sof{\kappa\,\e^{-2\,\mu\delta_{\nu,d}}\,\phi^{*}_{x+\hat{\nu}}\,\phi_{x}}\,=\,\sum\limits_{\bar{\xi}_{x,\nu}}\frac{\of{\phi^{*}_{x+\hat{\nu}}\,\phi_{x}\,\kappa\,\e^{-\mu\delta_{\nu,d}}}^{\bar{\xi}_{x,\nu}}}{\bar{\xi}_{x,\nu}!},
\]
and
\[
\exp\of{s\,\phi^{*}_{x}}\,=\,\sum\limits_{m_{x}}\frac{\of{s\,\phi^{*}_{x}}^{m_{x}}}{m_{x}!}
\]
\[
\exp\of{s^{*}\,\phi_{x}}\,=\,\sum\limits_{\bar{m}_{x}}\frac{\of{s^{*}\,\phi_{x}}^{\bar{m}_{x}}}{\bar{m}_{x}!}.
\]
After writing $\phi_{x}$ in polar form $\phi_{x}=\frac{r_{x}\e^{\ii\,\theta_{x}}}{\sqrt{2}}$, the partition function \eqref{eq:phifourpartf} can then be written as
\begin{multline}
Z\,=\,\sum\limits_{\cof{\xi,\bar{\xi},m,\bar{m}}}\bcof{\bof{\prod\limits_{x,\nu}\frac{\of{\frac{\kappa}{2}}^{\xi_{x,\nu}+\bar{\xi}_{x,\nu}}}{\xi_{x,\nu}!\,\bar{\xi}_{x,\nu}!}}\bof{\prod\limits_{x}\frac{\sof{\frac{s}{\sqrt{2}}}^{m_{x}}\sof{\frac{s^{*}}{\sqrt{2}}}^{\bar{m}_{x}}}{m_{x}!\,\bar{m}_{x}!}}\\
\bof{\prod\limits_{x}\,\e^{2\,\mu\of{\xi_{x,d}-\bar{\xi}_{x,d}}}\,\int\limits_{-\pi}^{\pi}\idd{\theta_{x}}{}\,\e^{-\ii\,\theta_{x}\,\sof{m_{x}-\bar{m}_{x}+\sum\limits_{\nu}\of{\xi_{x,\nu}-\bar{\xi}_{x,\nu}-\of{\xi_{x-\hat{\nu},\nu}-\bar{\xi}_{x-\hat{\nu},\nu}}}}}}\\
\bof{\prod\limits_{x}\,\int\limits_{0}^{\infty}\idd{r_{x}}{}\,r_{x}^{1+m_{x}+\bar{m}_{x}+\sum\limits_{\nu}\sof{\xi_{x,\nu}+\bar{\xi}_{x,\nu}+\xi_{x-\hat{\nu},\nu}+\bar{\xi}_{x-\hat{\nu},\nu}}}\,\e^{-r_{x}^{2}-\lambda\of{r_{x}^{2}-1}^{2}}}}.\label{eq:phifourfluxrepd2}
\end{multline}
The non-negative intergers $\xi_{x,\nu}$, $\bar{\xi}_{x,\nu}$ and $m_{x}$, $\bar{m}_{x}$ are called \emph{flux} and \emph{monomer variables}, respectively: $\xi_{x,\nu}$ counts the number of hoppings of particles from site $x$ to site $\ssof{x+\hat{\nu}}$ and of antiparticles from site $\ssof{x+\hat{\nu}}$ to $x$, while $\bar{\xi}_{x,\nu}$ counts the corresponding inverse moves. It is therefore convenient \cite{Gattringer} to introduce new variables $k_{x,\nu}$ and $l_{x,\nu}$, such that
\[
\xi_{x,\nu}-\bar{\xi}_{x,\nu}\,=\,k_{x,\nu}\,\in\,\mathbb{Z}\qquad\text{and}\qquad \xi_{x,\nu}+\bar{\xi}_{x,\nu}\,=\,\abs{k_{x,\nu}}\,+\,2\,l_{x,\nu}\quad,\quad l_{x,\nu}\in\,\mathbb{N}_{0},\label{eq:subst1}
\]
where $k_{x,\nu}$ counts the net charge flowing from site $x$ to site $\ssof{x+\hat{\nu}}$ and $\ssof{\abs{k_{x,\nu}}+2\,l_{x,\nu}}$ counts the total number of particles and antiparticles hopping around between $x$ and $\of{x+\hat{\nu}}$. Similarly $m_{x}$ and $\bar{m}_{x}$ count the particle and anti-particle content of site $x$, and we define $p_{x}$ and $q_{x}$ as follows,
\[
m_{x}-\bar{m}_{x}\,=\,p_{x}\,\in\,\mathbb{Z}\qquad\text{and}\qquad m_{x}+\bar{m}_{x}\,=\,\abs{p_{x}}\,+\,2\,q_{x}\quad,\quad q_{x}\in \mathbb{N}_{0},\label{eq:subst2}
\]
so that $p_{x}$ counts the total charge content of site $x$ and $\ssof{\abs{p_{x}}+2\,q_{x}}$ its total particle content. Integrating now over the $\theta_{x}$ in \eqref{eq:phifourfluxrepd2} and using \eqref{eq:subst1} and \eqref{eq:subst2} yields the final form of the $\phi^{4}$ partition function (up to an irrelevant constant pre-factor): 
\begin{multline}
Z\,=\,\sum\limits_{\cof{k,l,p,q}}\prod\limits_{x}\bcof{\bof{\prod\limits_{\nu}\frac{\of{\frac{\kappa}{2}}^{\abs{k_{x,\nu}}+2\,l_{x,\nu}}}{\of{\abs{k_{x,\nu}}+l_{x,\nu}}!\,l_{x,\nu}!}}\,\frac{\sof{\frac{s}{\sqrt{2}}}^{\frac{1}{2}\of{\abs{p_{x}}+p_{x}}+q_{x}}\sof{\frac{s^{*}}{\sqrt{2}}}^{\frac{1}{2}\of{\abs{p_{x}}-p_{x}}+q_{x}}\,\e^{2\,\mu\,k_{x,d}}}{\of{\abs{p_{x}}+q_{x}}!\,q_{x}!}\\
\delta\sof{p_{x}+\sum\limits_{\nu}\of{k_{x,\nu}-k_{x-\hat{\nu},\nu}}}\,W_{\lambda}\sof{\abs{p_{x}}+2\,q_{x}+\sum\limits_{\nu}\sof{\abs{k_{x,\nu}}+\abs{k_{x-\hat{\nu},\nu}}+2\,\ssof{l_{x,\nu}+l_{x-\hat{\nu},\nu}}}}},\label{eq:phifourfluxreppartf}
\end{multline}
where
\[
W_{\lambda}\of{x}\,=\,\int\limits_{0}^{\infty}\idd{r}{}\,r^{1+x}\,\e^{-r^{2}-\lambda\of{r^{2}-1}^{2}}\,=\,\int\limits_{0}^{\infty}\idd{u}{}\,\frac{u^{x/2}\,\e^{-u-\lambda\of{u-1}^{2}}}{2}.\label{eq:weightf}
\]
At first sight it seems that the source terms in \eqref{eq:phifourfluxreppartf} give rise to a complex phase. By writing them in the form
\[
\sof{\frac{s}{\sqrt{2}}}^{\frac{1}{2}\of{\abs{p_{x}}+p_{x}}+q_{x}}\sof{\frac{s^{*}}{\sqrt{2}}}^{\frac{1}{2}\of{\abs{p_{x}}-p_{x}}+q_{x}}\,=\,\sof{\frac{\abs{s}}{\sqrt{2}}}^{\abs{p_{x}}+2\,q_{x}}\,\sof{\frac{s}{\abs{s}}}^{p_{x}}\,=\,\sof{\frac{r_{s}}{2}}^{\abs{p_{x}}+2\,q_{x}}\,\e^{\ii\,\theta_{s}\,p_{x}},\label{eq:polarsource}
\]
with the radial, $r_{s}=\sqrt{2}\,\abs{s}$, and the polar source, $\theta_{s}$, we see that locally this is indeed the case. But due to the delta-function constraint in \eqref{eq:phifourfluxreppartf}, we have $\sum_{x}\,p_{x}\,=\,0$\footnote{Due to the delta function constraints in \eqref{eq:phifourfluxreppartf} we have $p_{x}=-\sum_{\nu}\of{k_{x,\nu}-k_{x-\hat{\nu},\nu}}$. Furthermore we have $\sum_{x,\nu}\of{k_{x,\nu}-k_{x-\hat{\nu},\nu}}=0$, as each $k$-variable appears twice in this sum with opposite sign, and therefore $\sum_{x}\,p_{x}=0$.}, which implies $\prod_{x}\e^{\ii\,\theta_{s}\,p_{x}}=1$, and there is therefore no net complex phase\footnote{Note that if $s$ and $s^{*}$ were treated as independent variables, \eqref{eq:polarsource} would not be true and the overall complex phase would not vanish in general.}. All other terms in \eqref{eq:phifourfluxreppartf} are manifestly real and non-negative, and thus the \emph{flux representation} \eqref{eq:phifourfluxreppartf} of the partition function \eqref{eq:phifourpartf} is sign-problem free and can be sampled by Monte Carlo. 

\subsection{Charged Correlator and Worm Algorithm}\label{ssec:chargedcorr}
Due to the delta function constraints, configurations contributing to \eqref{eq:phifourfluxreppartf} cannot be sampled directly by a local Metropolis algorithm: the transition probabilities for local updates of the $p$ and $k$ variables would always vanish. For a non-zero source $r_{s}=\sqrt{2}\abs{s}$, one could sample the $k$ and $p$ variables simultaneously, such that the delta function constraint is always satisfied, but if $\abs{s}$ is small, this would be highly inefficient due to low acceptance rates. For $\abs{s}=0$, in order to satisfy the delta function constraint, one would have to update randomly chosen closed chains of $k$ variables, which is in general also very inefficient. The alternative is to use a worm algorithm, which, roughly speaking, is based on the idea \cite{Prokofev,Wolff2,Gattringer0,Gattringer} to update the $k$ variables while sampling configurations that contribute to the charged correlator $\avof{\phi^{*}\of{x}\,\phi\of{y}}$ instead of configurations that contribute to \eqref{eq:phifourfluxreppartf} itself, as will be explained in what follows.\\

\subsubsection{Charged Correlator}\label{sssec:chargedcorr}
In order to define correlation functions, we consider from now on the source $s$ in \eqref{eq:phifourfluxreppartf} as a local quantity $s_{x}$, keeping in mind, that the partition function (or any observable) will actually only be evaluated when $s_{x}=s\,\forall x$. The charged two-point function is then given by the following expression:
\[
\avof{\phi\of{x}\,\phi^{*}\of{y}}-\avof{\phi\of{x}}\avof{\phi^{*}\of{y}}\,=\,\spartd{\log\of{Z}}{s^{*}_{x}}{s_{y}}\,=\,\frac{1}{Z}\spartd{Z}{s^{*}_{x}}{s_{y}}-\frac{1}{Z}\partd{Z}{s^{*}_{x}}\frac{1}{Z}\partd{Z}{s_{y}}.\label{eq:chargedcorr}
\]
Carrying out the formal derivative of $Z$ with respect to $s^{*}_{x}$ in \eqref{eq:phifourfluxreppartf}, leads to (highlighting in red the changes from \eqref{eq:phifourfluxreppartf}):
\begin{multline}
\partd{Z}{s^{*}_{x}}\,=\,\sum\limits_{\cof{k,l,p,q}}\bcof{
{\color{fhl}\frac{\frac{1}{2}\sof{\abs{p_{x}}-p_{x}}+q_{x}}{\sqrt{2}}}
\bof{\prod\limits_{z}\frac{\sof{\frac{s}{\sqrt{2}}}^{\frac{1}{2}\of{\abs{p_{z}}+p_{z}}+q_{z}}\sof{\frac{s^{*}}{\sqrt{2}}}^{\frac{1}{2}\of{\abs{p_{z}}-p_{z}}+q_{z}{\color{fhl}-\delta_{x,z}}}\,\e^{2\,\mu\,k_{z,d}}}{\of{\abs{p_{z}}+q_{z}}!\,q_{z}!}\\
\bof{\prod\limits_{\nu}\frac{\of{\frac{\kappa}{2}}^{\abs{k_{x,\nu}}+2\,l_{x,\nu}}}{\of{\abs{k_{z,\nu}}+l_{z,\nu}}!\,l_{z,\nu}!}}\delta\sof{p_{z}+\sum\limits_{\nu}\of{k_{z,\nu}-k_{z-\hat{\nu},\nu}}}\\
W_{\lambda}\sof{\abs{p_{z}}+2\,q_{z}+\sum\limits_{\nu}\sof{\abs{k_{z,\nu}}+\abs{k_{z-\hat{\nu},\nu}}+2\,\ssof{l_{z,\nu}+l_{z-\hat{\nu},\nu}}}}}},\label{eq:deriv1}
\end{multline}
which, after dividing by $Z$, can be interpreted as,
\[
\frac{1}{Z}\partd{Z}{s^{*}_{x}}\,=\,\frac{1}{s^{*}}\avof{\frac{1}{2}\of{\abs{p_{x}}-p_{x}}+q_{x}},\label{eq:naivederiv1}
\]
Analogously the formal derivative of $Z$ with respect to $s_{y}$ yields
\[
\frac{1}{Z}\partd{Z}{s_{y}}\,=\,\frac{1}{s}\avof{\frac{1}{2}\of{\abs{p_{y}}+p_{y}}+q_{y}},\label{eq:naivederiv2}
\]
and for the mixed second derivative we find:
\[
\frac{1}{Z}\spartd{Z}{s^{*}_{x}}{s_{y}}\,=\,\frac{1}{\abs{s}^{2}}\avof{\sof{\frac{1}{2}\of{\abs{p_{x}}-p_{x}}+q_{x}}\sof{\frac{1}{2}\of{\abs{p_{y}}+p_{y}}+q_{y}}},\label{eq:naivesderiv1}
\]
where, we have set $s_{x}=s_{y}=s$ as stated above. Although equations \eqref{eq:naivederiv1}, \eqref{eq:naivederiv2} and \eqref{eq:naivesderiv1} are correct, they are only well defined if $\abs{s}\neq 0$. Also these expressions so far do not affect in any way the delta function constraint in \eqref{eq:phifourfluxreppartf} and will therefore not help in finding an efficient updating algorithm for the $k$-variables.\\

To bring \eqref{eq:deriv1} into a more useful form, we can absorb the factor of $\sof{\frac{1}{2}\sof{\abs{p_{x}}-p_{x}}+q_{x}}$ into a shift of the $p_{x}$ and $q_{x}$ summation variables\footnote{To improve readability, we continue to highlight in red the important changes from one equation to the next.}:
\begin{itemize}
\item Terms in \eqref{eq:deriv1} for which $p_{x}\geq 0$, such that $\sof{\frac{1}{2}\sof{\abs{p_{x}}-p_{x}}+q_{x}}=q_{x}$ can, for $q_{x}>0$, be written as
\begin{multline}
{\color{fhl}\frac{1}{\sqrt{2}}}\bof{\prod\limits_{z}\frac{\sof{\frac{s}{\sqrt{2}}}^{\frac{1}{2}\of{\abs{p_{z}}+p_{z}}+q_{z}}\sof{\frac{s^{*}}{\sqrt{2}}}^{\frac{1}{2}\of{\abs{p_{z}}-p_{z}}+q_{z}{\color{fhl}-\delta_{x,z}}}\,\e^{2\,\mu\,k_{z,d}}}{\of{\abs{p_{z}}+q_{z}}!\,\of{q_{z}{\color{fhl}-\delta_{x,z}}}!}\\
\bof{\prod\limits_{\nu}\frac{\kappa^{\abs{k_{x,\nu}}+2\,l_{x,\nu}}}{\of{\abs{k_{z,\nu}}+l_{z,\nu}}!\,l_{z,\nu}!}}\delta\sof{p_{z}+\sum\limits_{\nu}\of{k_{z,\nu}-k_{z-\hat{\nu},\nu}}}\\
W_{\lambda}\sof{\abs{p_{z}}+2\,q_{z}+\sum\limits_{\nu}\sof{\sabs{k_{z,\nu}}+\sabs{k_{z-\hat{\nu},\nu}}+2\,\ssof{l_{z,\nu}+l_{z-\hat{\nu},\nu}}}}}.\label{eq:derivs1}
\end{multline}
By introducing new variables $p'_{z}$ and $q'_{z}$, such that
\[
p_{z}=p'_{z}-\delta_{x,z}\qquad,\qquad q_{z}=q'_{z}+\delta_{x,z},\label{eq:varshift1}
\]
we have $\sabs{p_{x}}=\sabs{p'_{x}}-1$ and obtain from \eqref{eq:derivs1}:
\begin{multline}
{\color{fhl}\frac{1}{\sqrt{2}}}\bof{\prod\limits_{z}\frac{\sof{\frac{s}{\sqrt{2}}}^{\frac{1}{2}\of{\sabs{{\color{fhl}p'_{z}}}+{\color{fhl}p'_{z}}}+{\color{fhl}q'_{z}}}\sof{\frac{s^{*}}{\sqrt{2}}}^{\frac{1}{2}\of{\sabs{{\color{fhl}p'_{z}}}-{\color{fhl}p'_{z}}}+{\color{fhl}q'_{z}}}\,\e^{2\,\mu\,k_{z,d}}}{\of{\sabs{{\color{fhl}p'_{z}}}+{\color{fhl}q'_{z}}}!\,{\color{fhl}q'_{z}}!}\\
\bof{\prod\limits_{\nu}\frac{\of{\frac{\kappa}{2}}^{\abs{k_{x,\nu}}+2\,l_{x,\nu}}}{\of{\abs{k_{z,\nu}}+l_{z,\nu}}!\,l_{z,\nu}!}}\delta\sof{{\color{fhl}p'_{z}-\delta_{x,z}}+\sum\limits_{\nu}\of{k_{z,\nu}-k_{z-\hat{\nu},\nu}}}\\
W_{\lambda}\sof{\sabs{{\color{fhl}p'_{z}}}+2\,{\color{fhl}q'_{z}}{\color{fhl}+\delta_{x,z}}+\sum\limits_{\nu}\sof{\sabs{k_{z,\nu}}+\sabs{k_{z-\hat{\nu},\nu}}+2\,\ssof{l_{z,\nu}+l_{z-\hat{\nu},\nu}}}}}.\label{eq:derivs12}
\end{multline}
Note that for terms with $q_{x}=0$, the right hand side of \eqref{eq:varshift1} would require $q'_{x}=-1$, which is not valid as, according to \eqref{eq:subst2}, all $q$-variables have to be non-negative. But these terms are anyway zero in \eqref{eq:deriv1} and it is therefore completely fine that they can not be expressed in terms of valid primed variables $p'_{x}$, $q'_{x}$.
\item Terms in \eqref{eq:deriv1} with $p_{x}< 0$, for which $\frac{1}{2}\of{\abs{p_{x}}-p_{x}}=\abs{p_{x}}$, can be written as
\begin{multline}
{\color{fhl}\frac{1}{\sqrt{2}}}\bof{\prod\limits_{z}\frac{\sof{\frac{s}{\sqrt{2}}}^{\frac{1}{2}\of{\abs{p_{z}}+p_{z}}+q_{z}}\sof{\frac{s^{*}}{\sqrt{2}}}^{\frac{1}{2}\of{\abs{p_{z}}-p_{z}}+q_{z}{\color{fhl}-\delta_{x,z}}}\,\e^{2\,\mu\,k_{z,d}}}{\of{\abs{p_{z}}+q_{z}{\color{fhl}-\delta_{x,z}}}!\,q_{z}!}\\
\bof{\prod\limits_{\nu}\frac{\of{\frac{\kappa}{2}}^{\abs{k_{x,\nu}}+2\,l_{x,\nu}}}{\of{\abs{k_{z,\nu}}+l_{z,\nu}}!\,l_{z,\nu}!}}\delta\sof{p_{z}+\sum\limits_{\nu}\of{k_{z,\nu}-k_{z-\hat{\nu},\nu}}}\\
W_{\lambda}\sof{\abs{p_{z}}+2\,q_{z}+\sum\limits_{\nu}\sof{\sabs{k_{z,\nu}}+\sabs{k_{z-\hat{\nu},\nu}}+2\,\ssof{l_{z,\nu}+l_{z-\hat{\nu},\nu}}}}},\label{eq:derivs2}
\end{multline}
and by choosing $p'_{z}$ and $q'_{z}$ in order to satisfy
\[
p_{z}=p'_{z}-\delta_{x,z}\qquad,\qquad q_{z}=q'_{z},\label{eq:varshift2}
\]
such that this time $\abs{p_{x}}=\abs{p'_{x}}+1$, we obtain from \eqref{eq:derivs2} again
\begin{multline}
{\color{fhl}\frac{1}{\sqrt{2}}}\bof{\prod\limits_{z}\frac{\sof{\frac{s}{\sqrt{2}}}^{\frac{1}{2}\of{\sabs{{\color{fhl}p'_{z}}}+{\color{fhl}p'_{z}}}+{\color{fhl}q'_{z}}}\sof{\frac{s^{*}}{\sqrt{2}}}^{\frac{1}{2}\of{\sabs{{\color{fhl}p'_{z}}}-{\color{fhl}p'_{z}}}+{\color{fhl}q'_{z}}}\,\e^{2\,\mu\,k_{z,d}}}{\of{\sabs{{\color{fhl}p'_{z}}}+{\color{fhl}q'_{z}}}!\,{\color{fhl}q'_{z}}!}\\
\bof{\prod\limits_{\nu}\frac{\of{\frac{\kappa}{2}}^{\abs{k_{x,\nu}}+2\,l_{x,\nu}}}{\of{\abs{k_{z,\nu}}+l_{z,\nu}}!\,l_{z,\nu}!}}\delta\sof{{\color{fhl}p'_{z}-\delta_{x,z}}+\sum\limits_{\nu}\of{k_{z,\nu}-k_{z-\hat{\nu},\nu}}}\\
W_{\lambda}\sof{\sabs{{\color{fhl}p'_{z}}}+2\,{\color{fhl}q'_{z}+\delta_{x,z}}+\sum\limits_{\nu}\sof{\sabs{k_{z,\nu}}+\sabs{k_{z-\hat{\nu},\nu}}+2\,\ssof{l_{z,\nu}+l_{z-\hat{\nu},\nu}}}}}.\label{eq:derivs22}
\end{multline}
\end{itemize}
Dropping again the primes from the $p$ and $q$ variables, eq. \eqref{eq:deriv1} can therefore be written as
\begin{multline}
\partd{Z}{s^{*}_{x}}\,=\,{\color{fhl}\frac{1}{\sqrt{2}}}\sum\limits_{\cof{k,l,p,q}}\prod\limits_{z}\bcof{
\frac{\sof{\frac{s}{\sqrt{2}}}^{\frac{1}{2}\of{\abs{p_{z}}+p_{z}}+q_{z}}\sof{\frac{s^{*}}{\sqrt{2}}}^{\frac{1}{2}\of{\abs{p_{z}}-p_{z}}+q_{z}}\,\e^{2\,\mu\,k_{z,d}}}{\of{\abs{p_{z}}+q_{z}}!\,q_{z}!}\\
\bof{\prod\limits_{\nu}\frac{\of{\frac{\kappa}{2}}^{\abs{k_{x,\nu}}+2\,l_{x,\nu}}}{\of{\abs{k_{z,\nu}}+l_{z,\nu}}!\,l_{z,\nu}!}}\delta\sof{p_{z}{\color{fhl}-\delta_{x,z}}+\sum\limits_{\nu}\of{k_{z,\nu}-k_{z-\hat{\nu},\nu}}}\\
W_{\lambda}\sof{\abs{p_{z}}+2\,q_{z}{\color{fhl}+\delta_{x,z}}+\sum\limits_{\nu}\sof{\abs{k_{z,\nu}}+\abs{k_{z-\hat{\nu},\nu}}+2\,\ssof{l_{z,\nu}+l_{z-\hat{\nu},\nu}}}}},\label{eq:deriv1n}
\end{multline}
and in a similar way we find
\begin{multline}
\partd{Z}{s_{x}}\,=\,{\color{fhl}\frac{1}{\sqrt{2}}}\sum\limits_{\cof{k,l,p,q}}\prod\limits_{z}\bcof{
\frac{\sof{\frac{s}{\sqrt{2}}}^{\frac{1}{2}\of{\abs{p_{z}}+p_{z}}+q_{z}}\sof{\frac{s^{*}}{\sqrt{2}}}^{\frac{1}{2}\of{\abs{p_{z}}-p_{z}}+q_{z}}\,\e^{2\,\mu\,k_{z,d}}}{\of{\abs{p_{z}}+q_{z}}!\,q_{z}!}\\
\bof{\prod\limits_{\nu}\frac{\of{\frac{\kappa}{2}}^{\abs{k_{x,\nu}}+2\,l_{x,\nu}}}{\of{\abs{k_{z,\nu}}+l_{z,\nu}}!\,l_{z,\nu}!}}\delta\sof{p_{z}{\color{fhl}+\delta_{x,z}}+\sum\limits_{\nu}\of{k_{z,\nu}-k_{z-\hat{\nu},\nu}}}\\
W_{\lambda}\sof{\abs{p_{z}}+2\,q_{z}{\color{fhl}+\delta_{x,z}}+\sum\limits_{\nu}\sof{\abs{k_{z,\nu}}+\abs{k_{z-\hat{\nu},\nu}}+2\,\ssof{l_{z,\nu}+l_{z-\hat{\nu},\nu}}}}},\label{eq:deriv2n}
\end{multline}
and finally:
\begin{multline}
\spartd{Z}{s^{*}_{x}}{s_{y}}\,=\,{\color{fhl}\frac{1}{2}}\sum\limits_{\cof{k,l,p,q}}\prod\limits_{z}\bcof{\frac{\sof{\frac{s}{\sqrt{2}}}^{\frac{1}{2}\of{\abs{p_{z}}+p_{z}}+q_{z}}\sof{\frac{s^{*}}{\sqrt{2}}}^{\frac{1}{2}\of{\abs{p_{z}}-p_{z}}+q_{z}}\,\e^{2\,\mu\,k_{z,d}}}{\of{\abs{p_{z}}+q_{z}}!\,q_{z}!}\\
\bof{\prod\limits_{\nu}\frac{\of{\frac{\kappa}{2}}^{\abs{k_{x,\nu}}+2\,l_{x,\nu}}}{\of{\abs{k_{z,\nu}}+l_{z,\nu}}!\,l_{z,\nu}!}}\delta\sof{p_{z}{\color{fhl}-\delta_{x,z}+\delta_{y,z}}+\sum\limits_{\nu}\of{k_{z,\nu}-k_{z-\hat{\nu},\nu}}}\\
W_{\lambda}\sof{\abs{p_{z}}+2\,q_{z}{\color{fhl}+\delta_{x,z}+\delta_{y,z}}+\sum\limits_{\nu}\sof{\abs{k_{z,\nu}}+\abs{k_{z-\hat{\nu},\nu}}+2\,\ssof{l_{z,\nu}+l_{z-\hat{\nu},\nu}}}}}.\label{eq:sderivn}
\end{multline}

Equations \eqref{eq:deriv1n}, \eqref{eq:deriv2n} and \eqref{eq:sderivn} are now well defined even if $\abs{s}=0$ and we can deduce that in the flux representation formulation of the $\phi^4$ partition function, the right way to implement observables like
\[
\avof{\phi\of{x}}\,=\,\frac{1}{Z}\,\partd{Z}{s^{*}_{x}},
\]
\[
\avof{\phi^{*}\of{x}}\,=\,\frac{1}{Z}\,\partd{Z}{s_{x}}
\]
and
\[
\avof{\phi\of{x}\,\phi^{*}\of{y}}\,=\,\frac{1}{Z}\,\spartd{Z}{s^{*}_{x}}{s_{y}},\label{eq:chargedcorr2}
\]
is\cite{Wolff,Wolff1,Wolff2,Gattringer0}, to think of $\phi\of{x}$ and $\phi^{*}\of{x}$ as external source and sink, respectively, where the effect of adding $\phi\of{x}$ to the system is a change in the local constraint at $x$,
\[
\delta\sof{p_{x}+\sum\limits_{\nu}\of{k_{x,\nu}-k_{x-\hat{\nu},\nu}}}\,\longrightarrow\,\delta\sof{p_{x}{\color{fhl}-1}+\sum\limits_{\nu}\of{k_{x,\nu}-k_{x-\hat{\nu},\nu}}},
\] 
and in the local weight at $x$,
\begin{multline}
W_{\lambda}\sof{\abs{p_{x}}+2\,q_{x}+\sum\limits_{\nu}\sof{\abs{k_{x,\nu}}+\abs{k_{x-\hat{\nu},\nu}}+2\,\ssof{l_{x,\nu}+l_{x-\hat{\nu},\nu}}}}\\
\longrightarrow\,W_{\lambda}\sof{\abs{p_{x}}+2\,q_{x}{\color{fhl}+1}+\sum\limits_{\nu}\sof{\abs{k_{x,\nu}}+\abs{k_{x-\hat{\nu},\nu}}+2\,\ssof{l_{x,\nu}+l_{x-\hat{\nu},\nu}}}}.
\end{multline}
Similarly we have for the insertion of $\phi^{*}\of{x}$:
\[
\delta\sof{p_{x}+\sum\limits_{\nu}\of{k_{x,\nu}-k_{x-\hat{\nu},\nu}}}\,\longrightarrow\,\delta\sof{p_{x}{\color{fhl}+1}+\sum\limits_{\nu}\of{k_{x,\nu}-k_{x-\hat{\nu},\nu}}},
\] 
and,
\begin{multline}
W_{\lambda}\sof{\abs{p_{x}}+2\,q_{x}+\sum\limits_{\nu}\sof{\abs{k_{x,\nu}}+\abs{k_{x-\hat{\nu},\nu}}+2\,\ssof{l_{x,\nu}+l_{x-\hat{\nu},\nu}}}}\\
\longrightarrow\,W_{\lambda}\sof{\abs{p_{x}}+2\,q_{x}{\color{fhl}+1}+\sum\limits_{\nu}\sof{\abs{k_{x,\nu}}+\abs{k_{x-\hat{\nu},\nu}}+2\,\ssof{l_{x,\nu}+l_{x-\hat{\nu},\nu}}}}.
\end{multline}

\subsubsection{Worm Algorithm}\label{sssec:wormalgorithm}
\begin{figure}[H]
\centering
\begin{minipage}[t]{\linewidth}
\centering
\begin{tikzpicture}[scale=0.9,nodes={inner sep=0}]
  \pgfpointtransformed{\pgfpointxy{1}{1}};
  \pgfgetlastxy{\vx}{\vy}
  \begin{scope}[node distance=\vx and \vy]
    \foreach \i in {0,...,3} {
        \draw [very thin,gray] (\i,0) -- (\i,4)  node[below] at (\i,-0.3) {$\i$};
    }
    \foreach \i in {0,...,3} {
        \draw [very thin,gray] (0,\i) -- (4,\i) node[left] at (-0.2,\i) {$\i$};
    }
    \draw [ultra thick,red!100] (1.05,1) -- (1.95,1);
    \node[draw,circle,inner sep=1.25,fill,color=red] at (0.9,1) {};
    \node[draw,circle,inner sep=1.25,fill,color=red] at (2.1,1) {};
    \node[draw,circle,inner sep=0.5,fill,color=white] at (2.1,1) {};
    \node at (0.9,0.6) {$\phi\ssof{x_{0}}$};
    \node at (2.2,0.6) {$\phi^{*}\ssof{x}$};
  \end{scope}
  \begin{scope}[node distance=\vx and \vy]
    
    \foreach \i in {0,...,3} {
        \draw [very thin,gray] (\i+7,0) -- (\i+7,4)  node[below] at (\i+7,-0.3) {$\i$};
    }
    \foreach \i in {0,...,3} {
        \draw [very thin,gray] (0+7,\i) -- (4+7,\i) node[left] at (7-0.2,\i) {$\i$};
    }
    \draw [ultra thick,red!100] (7+1.05,1) -- (7+2,1) -- (7+2,1.95);
    \node[draw,circle,inner sep=1.25,fill,color=red] at (7+0.9,1) {};
    \node[draw,circle,inner sep=1.25,fill,color=red] at (7+2,2.1) {};
    \node[draw,circle,inner sep=0.5,fill,color=white] at (7+2,2.1) {};
    \node at (7+0.9,0.6) {$\phi\ssof{x_{0}}$};
    \node[right] at (7+2.2,2.1) {$\phi^{*}\ssof{x+\hat{\nu}}$};
  \end{scope}
\end{tikzpicture}
\end{minipage}
\caption{Starting from the configuration on the left-hand side, a standard worm-update consists of proposing to shift $\phi^{*}\of{x}$ from $x$ to some random neighboring site, $x+\ohat{\nu}$, say, and to compensate for the charge-displacement by increasing the flux-variable $k_{x,\nu}$, as depicted in the right-hand figure.}
  \label{fig:worm1}
\end{figure}
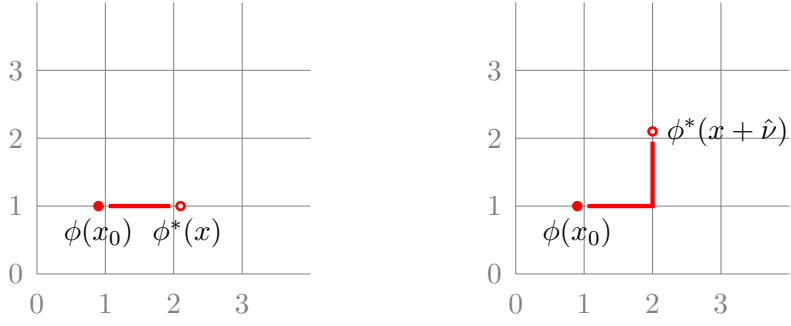

The idea behind the worm algorithm is now as follows (cf. \cite{Prokofev,Wolff2,Gattringer0,Gattringer}): first propose to insert at some site $x_{0}$ an external source/sink pair $\phi\of{x_{0}}$, $\phi^{*}\of{x_{0}}$. The $-1$ in the delta function constraint at $x_{0}$, coming from the $\phi\of{x_{0}}$ insertion and the $+1$ from the insertion of $\phi^{*}\of{x_{0}}$ cancel each other and the insertion of the external source/sink pair changes therefore only the argument of $W_{\lambda}$ at $x_{0}$:
\begin{multline}
W_{\lambda}\sof{\abs{p_{x_{0}}}+2\,q_{x_{0}}+\sum\limits_{\nu}\sof{\sabs{k_{x_{0},\nu}}+\sabs{k_{x_{0}-\hat{\nu},\nu}}+2\,\ssof{l_{x_{0},\nu}+l_{x_{0}-\hat{\nu},\nu}}}}\\
\longrightarrow\,W_{\lambda}\sof{\abs{p_{x_{0}}}+2\,q_{x_{0}}{\color{fhl}+2}+\sum\limits_{\nu}\sof{\sabs{k_{x_{0},\nu}}+\sabs{k_{x_{0}-\hat{\nu},\nu}}+2\,\ssof{l_{x_{0},\nu}+l_{x_{0}-\hat{\nu},\nu}}}}.
\end{multline}
If this insertion is rejected, the proposal counts as an attempted worm update. If the insertion is accepted, the next step could either consist of proposing to remove the source/sink pair again from the system, or to move $\phi^{*}$ from $x_{0}$ to a randomly chosen neighboring site, e.g. $x=x_{0}+\hat{\nu}$. The latter update would change the product of local weights for the sites $x_{0}$ and $x$ as follows: 
\begin{multline}
\delta\sof{p_{x_{0}}+\sum\limits_{\mu}\of{k_{x_{0},\mu}-k_{x_{0}-\hat{\mu},\mu}}}\delta\sof{p_{x}+\sum\limits_{\mu}\of{k_{x,\mu}-k_{x-\hat{\mu},\mu}}}\\
\cdot W_{\lambda}\sof{\abs{p_{x_{0}}}+2\,q_{x_{0}}{\color{fhl}+2}+\sum\limits_{\mu}\sof{\sabs{k_{x_{0},\mu}}+\sabs{k_{x_{0}-\hat{\mu},\mu}}+2\,\ssof{l_{x_{0},\mu}+l_{x_{0}-\hat{\mu},\mu}}}}\\
\cdot W_{\lambda}\sof{\abs{p_{x}}+2\,q_{x}+\sum\limits_{\mu}\sof{\abs{k_{x,\mu}}+\abs{k_{x-\hat{\mu},\mu}}+2\,\ssof{l_{x,\mu}+l_{x-\hat{\mu},\mu}}}}\\
\longrightarrow \delta\sof{p_{x_{0}}{\color{fhl}-1}+\sum\limits_{\mu}\of{k_{x_{0},\mu}-k_{x_{0}-\hat{\mu},\mu}}}\delta\sof{p_{x}{\color{fhl}+1}+\sum\limits_{\mu}\of{k_{x,\mu}-k_{x-\hat{\mu},\mu}}}\\
\cdot W_{\lambda}\sof{\abs{p_{x_{0}}}+2\,q_{x_{0}}{\color{fhl}+1}+\sum\limits_{\mu}\sof{\sabs{k_{x_{0},\mu}}+\sabs{k_{x_{0}-\hat{\mu},\mu}}+2\,\ssof{l_{x_{0},\mu}+l_{x_{0}-\hat{\mu},\mu}}}}\\
\cdot W_{\lambda}\sof{\abs{p_{x}}+2\,q_{x}{\color{fhl}+1}+\sum\limits_{\mu}\sof{\abs{k_{x,\mu}}+\abs{k_{x-\hat{\mu},\mu}}+2\,\ssof{l_{x,\mu}+l_{x-\hat{\mu},\mu}}}},\label{eq:weightchange1}
\end{multline}
and in order to get a non-zero result on the right hand side of the arrow in \eqref{eq:weightchange1}, this update has to be combined with a simultaneous shift of the corresponding flux variable, $k_{x,\nu}\rightarrow k_{x,\nu}+1$ (assuming for the moment that the $p$ variables are kept fixed) as depicted in Fig. \ref{fig:worm1}. If this move of $\phi^{*}$ from $x_{0}$ to $x$ and the simultaneous shift $k_{x,\nu}\rightarrow k_{x,\nu}+1$ are accepted, one can propose to move $\phi^{*}$ further to a random nearest-neighboring site of $x$, and so on. In this way, one can update the $k$-variables while sampling the charged correlator $\avof{\phi\of{x_{0}}\phi^{*}\of{x}}$. Whenever, during this procedure, the head of the worm hits the site $x_{0}$ where also the worm's tail is located, one can in addition to the head-shift also propose to remove the external source/sink pair again. If this latter proposal is accepted, the worm terminates: we have completed a whole \emph{worm-update} and are left with a new closed-worm configuration which contributes to the partition function \eqref{eq:phifourfluxreppartf}\footnote{As an alternative to the insertion and removal of the external source/sink pair, one can do importance sampling with respect to the location of the pair and just reweight it away when evaluating observables on closed-worm configurations.}.\\

\begin{figure}[H]
\centering
\begin{minipage}[t]{\linewidth}
\centering
\begin{tikzpicture}[scale=0.9,nodes={inner sep=0}]
  \pgfpointtransformed{\pgfpointxy{1}{1}};
  \pgfgetlastxy{\vx}{\vy}
  \begin{scope}[node distance=\vx and \vy]
    \foreach \i in {0,...,3} {
        \draw [very thin,gray] (\i,0) -- (\i,4)  node[below] at (\i,-0.3) {$\i$};
    }
    \foreach \i in {0,...,3} {
        \draw [very thin,gray] (0,\i) -- (4,\i) node[left] at (-0.2,\i) {$\i$};
    }
    \draw [ultra thick,red!100] (1.05,1) -- (1.95,1);
    \node[draw,circle,inner sep=1.25,fill,color=red] at (0.9,1) {};
    \node[draw,circle,inner sep=1.25,fill,color=red] at (2.1,1) {};
    \node[draw,circle,inner sep=0.5,fill,color=white] at (2.1,1) {};
    \node at (0.9,0.6) {$\phi\ssof{x_{0}}$};
    \node at (2.2,0.6) {$\phi^{*}\ssof{x}$};
  \end{scope}
  \begin{scope}[node distance=\vx and \vy]
    
    \foreach \i in {0,...,3} {
        \draw [very thin,gray] (\i+7,0) -- (\i+7,4)  node[below] at (\i+7,-0.3) {$\i$};
    }
    \foreach \i in {0,...,3} {
        \draw [very thin,gray] (0+7,\i) -- (4+7,\i) node[left] at (7-0.2,\i) {$\i$};
    }
    \draw [ultra thick,red!100] (7+1.05,1) -- (7+1.95,1);
    \node[draw,circle,inner sep=1.25,fill,color=red] at (7+0.9,1) {};
    \node[draw,circle,inner sep=1.25,fill,color=blue] at (7+2.1,1) {};
    \node[draw,circle,inner sep=0.5,fill,color=white] at (7+2.1,1) {};

    \node[draw,circle,inner sep=1.25,fill,color=blue] at (7+2.9,3) {};
    \node[draw,circle,inner sep=1.25,fill,color=red] at (7+3.1,3) {};
    \node[draw,circle,inner sep=0.5,fill,color=white] at (7+3.1,3) {};
    \node at (7+0.9,0.6) {$\phi\ssof{x_{0}}$};
    \node at (7+3.5,2.6) {$\phi^{*}\ssof{y}$};
  \end{scope}
\end{tikzpicture}
\end{minipage}
\caption{For $\abs{s}>0$, starting from the configuration on the left-hand side, the disconnected worm-update consists of proposing to shift $\phi^{*}\of{x}$ from $x$ to some random site $y$ and to compensate for the charge-displacement by increasing $p_{x}$ (open blue circle) and decreasing $p_{y}$ (filled blue circle), as depicted on the right-hand side of the figure.}
  \label{fig:worm2}
\end{figure}
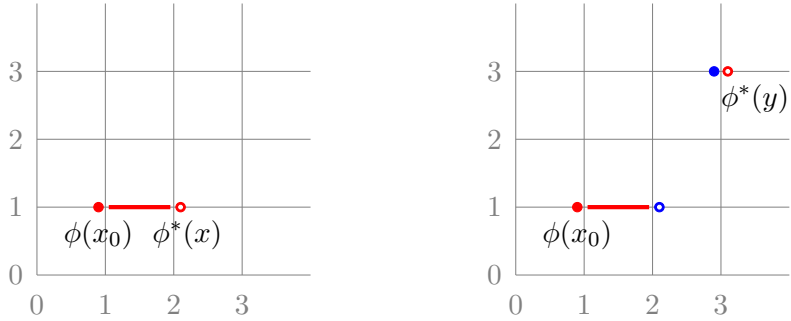

If $\abs{s}$ is non-zero, one also has to include the update of the $p$-variables into the worm algorithm in order to correctly sample the charged correlator. This can in principle be done by adding a new move to the worm-update algorithm, in which one replaces the shift in the $k$-variable, $k_{x,\nu}\rightarrow k_{x,\nu}+1$, required to compensate for the $\pm 1$ in the delta functions in \eqref{eq:weightchange1}, by the two shifts $p_{x}\rightarrow p_{x}+1$ and $p_{x+\hat{\nu}}\rightarrow p_{x+\hat{\nu}}-1$, which could be interpreted as compensating for the displacement from $x$ to $x+\hat{\nu}$ of the external charge $\phi^{*}$, by inserting a pair of oppositely charged (dynamical) monomers at $x$ and $x+\hat{\nu}$. However, in this case, there is no need to restrict the set of possible locations to which the head, $\phi^{*}$, of the worm can be shifted to the nearest-neighboring sites of $x$, and one can instead choose a random site $y$ as the target location (see Fig. \ref{fig:worm2}). If such a move is accepted, the worm just continues from the new location as before proposing either to move the head to a nearest-neighboring site by changing a $k$-variable, or to move the head to a random new site by changing two $p$-variables. But still, the worm can only be terminated if the head hits again the site where its tail is located and the removal of the external source/sink pair is accepted.\\
The measurement of the two-point functions, e.g. of the component $\avof{\phi\of{x}\phi^{*}\of{y}}$, happens by measuring the fractional Monte Carlo time during which the external source/sink pair is present in the system and located at sites $x$ and $y$ respectively.\\

So far, the algorithm is analogous to the \emph{closed-worm} algorithm described in \cite{Gattringer2} for the case of a Potts model in the presence of a chemical potential and an external magnetic field. In the next section we supplement this algorithm with additional moves which allow for the sampling of more two-point functions as well as of one-point functions. For the latter the moves will be analogous to the \emph{open-worm} moves of \cite{Gattringer2}, but unlike \cite{Gattringer2}, where simulations are performed using closed-worm or open-worm a priori, here the choice between closed-worm and open-worm will be performed stochastically.\\

\section{Sampling General Correlators and Condensates During the Worm}\label{sec:generalcorr}

In the previous section we have seen that the worm algorithm can be interpreted as sampling the two-point function $\frac{1}{Z}\spartd{Z}{s^{*}_{x}}{s_{y}}$. The latter can therefore be measured in a very efficient way during the worm-update by keeping track of the fractional Monte Carlo time the external source/sink pair is present in the system and source and sink are located at $x$ and $y$ respectively.\\
Other observables on the other hand, can only be measured on "closed-worm" configurations, obtained in between successive \emph{worm updates}\footnote{A \emph{worm update} consists of either a failed attempt to insert an external source/sink pair into the system, or, if the insertion succeeds, of all local updates between insertion and removal of the source/sink pair.}. The auto-correlation times of such observables do not depend on the number of \emph{worm updates} but on the number of "micro-steps"\cite{Wolff2} done during these \emph{worm updates} (i.e. the number of \emph{local updates} of the flux variables). It therefore makes sense to choose the number of \emph{worm updates} between successive measurements (on closed-worm configurations) so that the average number of \emph{local updates} is of the order of the number of degrees of freedom in the system, which corresponds to the usual definition of a \emph{sweep} used in ordinary non-worm based Monte Carlo simulations. As the average \emph{worm length}\footnote{The average \emph{worm length} is the average number of attempted \emph{local updates} between start and end of a worm, i.e. between insertion and removal of the external source/sink pair.} varies as a function of the simulation parameters ($\kappa$, $\lambda$, $s$), so does the optimal number of \emph{worm updates} which are necessary to obtain a certain average number of \emph{local updates} between successive measurements.\\

\begin{figure}[h!]
\centering
\begin{minipage}[t]{0.49\linewidth}
\centering
\includegraphics[width=\linewidth]{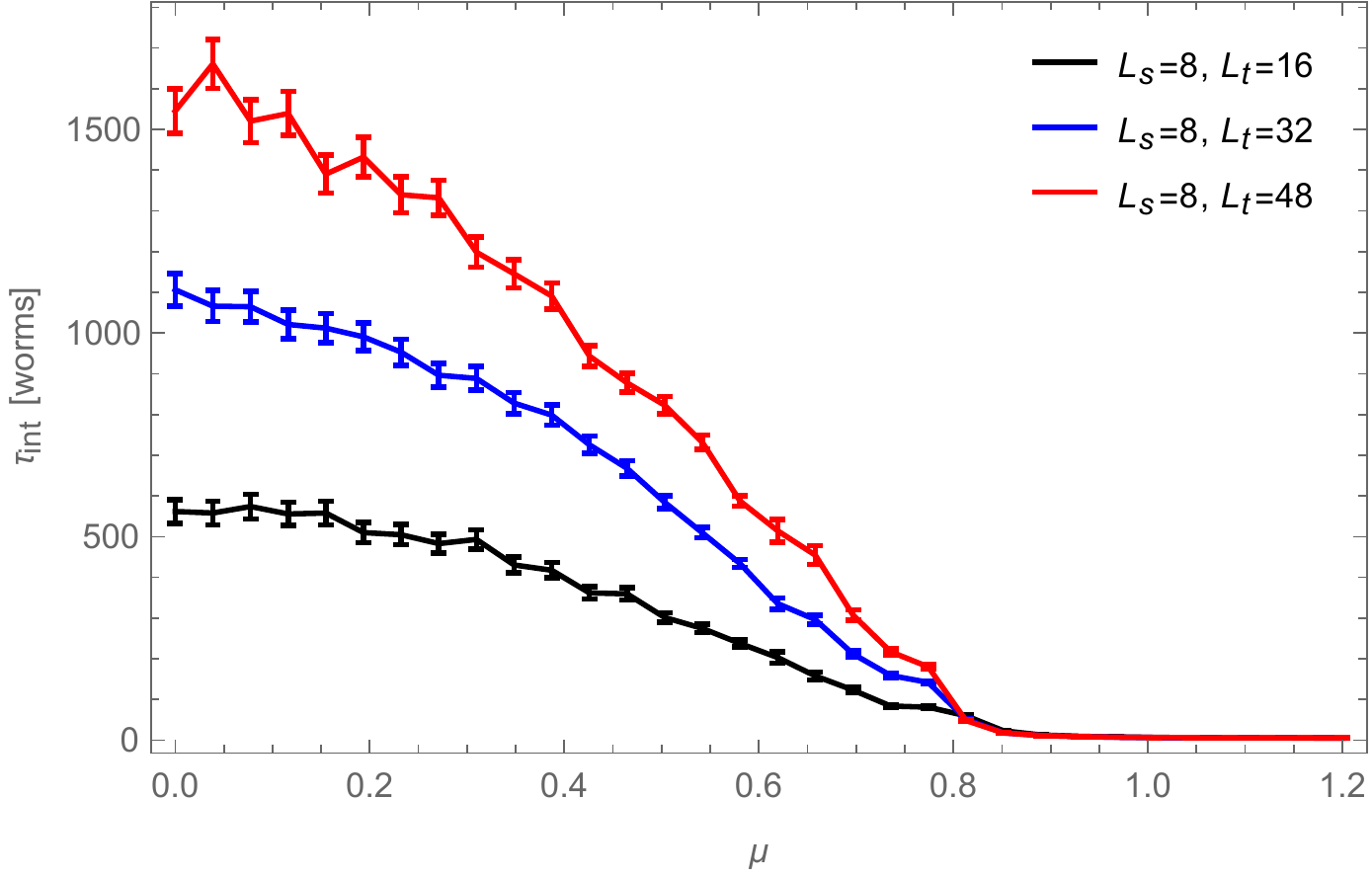}
\end{minipage}\hfill
\begin{minipage}[t]{0.49\linewidth}
\centering
\includegraphics[width=\linewidth]{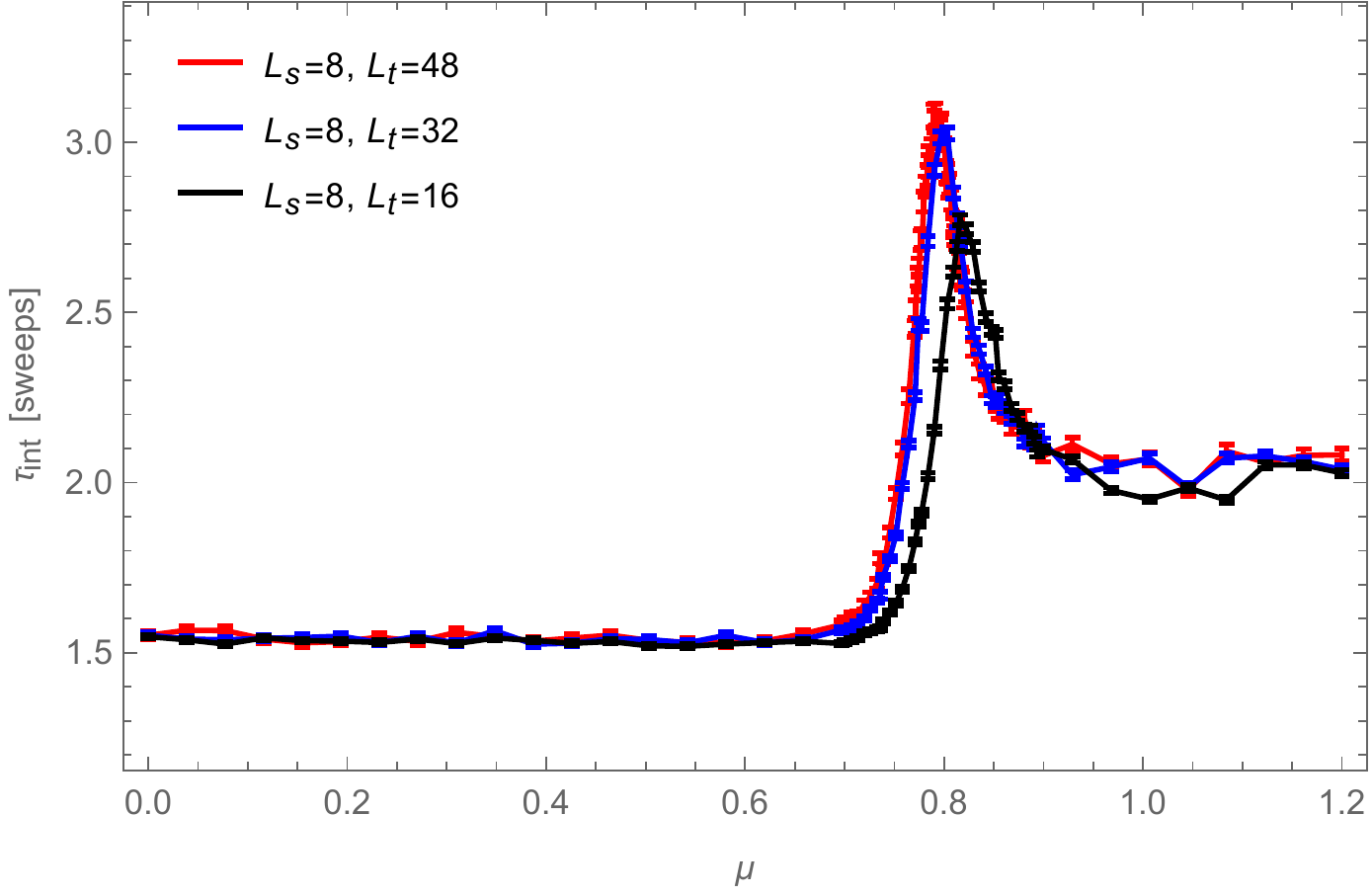}
\end{minipage}
\caption{The figure shows for a (2+1)-dimensional system of spatial size $N_{s}=8$, with couplings $\kappa=0.25$, $\lambda=7.5$ and source $r_{s}=0.01$, the integrated auto-correlation time $\tau_{int}$ for the average energy as a function of $\mu$ for three different values of $N_{t}\in\cof{16,\,32,\,48}$. In the left-hand figure, $\tau_{int}$ is measured in units of $2\,d\,N$ \emph{worms}\textsuperscript{\ref{fn:prefact}}. With increasing $\mu$, the average worm-length increases, such that the average number of worms required to de-correlate two configurations decreases. In the right-hand figure, $\tau_{int}$ is measured in units of sweeps, where a sweep is defined as the number of \emph{worm updates}, such that the total number of \emph{local updates} during the \emph{worm updates} is of the order of the number of degrees of freedom in the system. In these units, the integrated auto-correlation time seems to depend only weakly on the volume, even at the pseudo critical points around $\mu\sim 0.8$.}
  \label{fig:tauintsweeps}
\end{figure}

\begin{figure}[h!]
\centering
\includegraphics[width=0.6\linewidth]{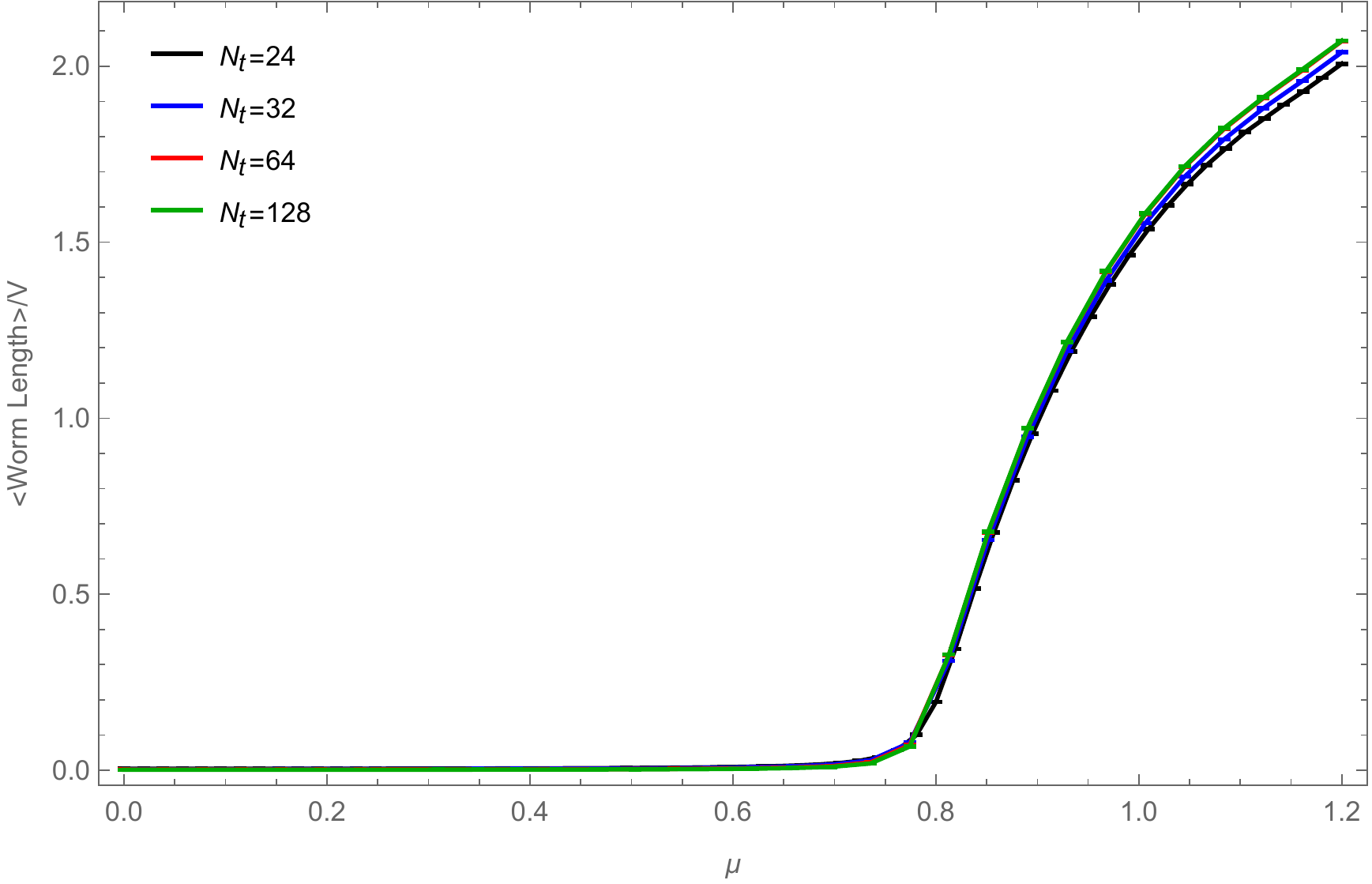}
\caption{The figure shows for a (2+1)-dimensional system of spatial size $N_{s}=8$, with couplings $\kappa=0.25$, $\lambda=7.5$ and source $r_{s}=0.01$, the average worm length divided by the system volume as a function of $\mu$ for four different values of $N_{t}\in\cof{24,\,32,\,64,\,128}$. As can be seen, for $\mu>0.8$, when the system is in the ordered phase (broken symmetry), the different curves almost coincide, which means that the worm length is proportional to the volume. The latter is not true in the symmetric phase (unbroken symmetry), but as the worms are much shorter in this phase, the different curves cannot be distinguished on the given scale.}
  \label{fig:wormlength}
\end{figure}

In Fig.~\ref{fig:tauintsweeps} we show the integrated auto-correlation time for the average "energy"\footnote{Note that we call $E$ the "energy" just for simplicity: it is only one part of the full energy of the system.},
\[
\avof{E}\,=\,\partd{\log\of{Z}}{\kappa}\ ,\label{eq:avenergy}
\]
as a function of $\mu$, once in units of $2\,d\,N$ worms\footnote{\label{fn:prefact}The pre-factor $2\,d\,N$ is just a normalization: $d=2+1$ is the number of space-time dimensions, while $N$ is the number of independent components of $\phi$, i.e. in the present case, we have $N=2$, as a complex scalar field can be understood as an $\On{2}$ field. The factor $N$ reflects the fact that the worm can start with $N$ different choices for the head and $2\,d$ reflects the number of different directions in which the worm can start on each site.} (left) and once in units of \emph{sweeps} (right), where we now define a \emph{sweep}, as suggested above, as the number $L_{sweep}$ of worms which are necessary on average to proceed as many local updates as there are degrees of freedom in the system. In the first case, where the auto-correlation time is given in units of \emph{worm-updates}, it can be seen that the number of such worms required to de-correlate two configurations decreases with increasing $\mu$,  and that in the disordered phase, $\tau_{int}$ is proportional to the system size, while in the ordered phase no volume-dependency is visible. This has of course to do with the fact that the average number of \emph{local updates} required to complete a worm, increases with $\mu$ and becomes proportional to the system volume in the ordered phase (see Fig.~\ref{fig:wormlength}). In the second case, where $\tau_{int}$ is measured in terms of sweeps, no volume dependency is visible, except at the pseudo-critical point. And even there, the volume dependency seems to be rather mild.\\

It is highly convenient to use the above definition of a sweep as measure for the time between successive measurements, because the computational cost for such sweeps is almost independent of $\mu$ (at least for a reasonably broad interval of $\mu$ values) and it can therefore also efficiently be used in combination with parallel tempering \cite{Geyer}, where it avoids long waiting times among the replicas before swap moves can take place.\\

Note that although we tune the number of worms per sweep so that the average number of local updates or micro-steps per sweep is of the order of the system size, the sweep itself is defined in terms of a fixed number of worm updates and not in terms of a fixed number of micro-steps. By allowing the insertion and removal of external sources/sinks, we have extended the configuration space explored by the Markov process, so that it generates both, configurations that contribute to the usual partition function $Z$, as well as configurations that contribute to partition functions of the form $Z_{2}\of{x,y}=\spartd{Z}{s^{*}_{x}}{s_{y}}$, which describe the behavior of the system in the presence of an external source $\phi$ at $x$ and an external sink $\phi^{*}$ at $y$ for all possible $\of{x,y}$. So the Markov process in fact samples configurations that contribute to the generalized partition function
\[
Z_{gen}\,=\,Z\,+\,\sum\limits_{x,y}\,Z_{2}\of{x,y}\ .\label{eq:genpartf}
\]
But we want for example,
\[
\avof{\phi\of{x}\phi^{*}\of{y}}\,=\,\frac{Z_{2}\of{x,y}}{Z}\ ,\label{eq:twopointf}
\]
and not 
\[
\avof{\phi\of{x}\phi^{*}\of{y}}_{gen}\,=\,\frac{Z_{2}\of{x,y}}{Z_{gen}}\ ,\label{eq:twopointfgen}
\]
and therefore, only if we define a sweep in terms of a fixed number of worm updates (which, since worm updates always start and end on closed-worm configurations, equals the number of configurations that contribute to $Z$) instead of a fixed number of micro-steps, we get the correct result without having to determine the correction factor $Z_{gen}/Z$. For observables which depend on configurations in $Z$ only, this correction factor is irrelevant. Also the masses obtained from fits to \eqref{eq:twopointf} and \eqref{eq:twopointfgen} would of course be identical. But for example the magnetic susceptibility obtained from \eqref{eq:twopointfgen}, i.e.
\[
\chi_{m,gen}\,=\,\frac{1}{V}\sum_{x,y}\,\avof{\phi\of{x}\phi^{*}\of{y}}_{gen}
\]
would in general be wrong.\\

The introduction of $Z_{gen}$ \eqref{eq:genpartf} looks like an unnecessary complication. Indeed, in the standard worm algorithm (see e.g. \cite{Wolff}), one simply uses reweighting of $Z_{2}\of{x,x}$ to estimate $Z$. Here we need $Z_{gen}$ because, in the remainder of this section, we will further generalize to
\[
Z_{gen}\,=\,Z\,+\,\sum\limits_{k}\sum\limits_{x} Z^{\ssof{k}}_{1}\of{x}\,+\,\sum\limits_{l}\sum\limits_{x,y} Z_{2}^{\ssof{l}}\of{x,y}\ ,
\] 
where
\[
Z^{\ssof{1}}_{1}\of{x}=\partd{Z}{s^{*}_{x}}=Z\,\avof{\phi\of{x}}\quad,\quad Z^{\ssof{2}}_{1}\of{x}=\partd{Z}{s_{x}}=Z\,\avof{\phi^{*}\of{x}}
\]
and
\[
Z^{\ssof{1,2,3,4}}_{2}\of{x,y}\,=\,\bcof{\spartd{Z}{s^{*}_{x}}{s_{y}}\,,\,\spartd{Z}{s^{*}_{x}}{s^{*}_{y}}\,,\,\spartd{Z}{s_{x}}{s^{*}_{y}}\,,\,\spartd{Z}{s_{x}}{s_{y}}}\ .
\]
This will allow us to measure also more general correlators like $\avof{\phi^{*}\of{x}\phi^{*}\of{y}}$ or $\avof{\phi\of{x}\phi\of{y}}$, as well as the corresponding condensates $\avof{\phi}$ and $\avof{\phi^{*}}$ \emph{during} the worm updates, instead of on closed-worm configurations only. \\

\subsection{General Correlator in Terms of External Sources: New Worm Moves}\label{ssec:gencorr}
To demonstrate how one can measure general correlators during the worm updates (without the use of reweighting), we use \eqref{eq:polarsource} to write the source terms in the partition function \eqref{eq:phifourfluxreppartf} in polar form:
\begin{multline}
Z\,=\,\sum\limits_{\cof{k,l,p,q}}\prod\limits_{x}\bcof{\bof{\prod\limits_{\nu}\frac{\of{\frac{\kappa}{2}}^{\abs{k_{x,\nu}}+2\,l_{x,\nu}}}{\of{\abs{k_{x,\nu}}+l_{x,\nu}}!\,l_{x,\nu}!}}\,\frac{\sof{\frac{r_{s}}{2}}^{\abs{p_{x}}+2\,q_{x}}\,\e^{\ii\,\theta_{s}\,p_{x}}\,\e^{2\,\mu\,k_{x,d}}}{\of{\abs{p_{x}}+q_{x}}!\,q_{x}!}\\
\delta\sof{p_{x}+\sum\limits_{\nu}\of{k_{x,\nu}-k_{x-\hat{\nu},\nu}}}\,W_{\lambda}\sof{\abs{p_{x}}+2\,q_{x}+\sum\limits_{\nu}\sof{\abs{k_{x,\nu}}+\abs{k_{x-\hat{\nu},\nu}}+2\,\ssof{l_{x,\nu}+l_{x-\hat{\nu},\nu}}}}}.\label{eq:phifourfluxreppartfsph}
\end{multline}
We can then for example define the correlator for \emph{radial excitations} by
\[
\avof{r\of{x}\,r\of{y}}-\avof{r\of{x}}\avof{r\of{y}}\,=\,\spartd{\log\of{Z}}{r_{s,x}}{r_{s,y}}\,=\,\frac{1}{Z}\spartd{Z}{r_{s,x}}{r_{s,y}}-\frac{1}{Z}\partd{Z}{r_{s,x}}\frac{1}{Z}\partd{Z}{r_{s,y}},\label{eq:radcorr}
\]
where
\begin{multline}
\partd{Z}{r_{s,z}}\,=\,{\color{fhl}\frac{1}{2}}\sum\limits_{\cof{k,l,p,q}}{\color{fhl}\of{\abs{p_{z}}+2\,q_{z}}}\prod\limits_{x}\bcof{\bof{\prod\limits_{\nu}\frac{\of{\frac{\kappa}{2}}^{\abs{k_{x,\nu}}+2\,l_{x,\nu}}}{\of{\abs{k_{x,\nu}}+l_{x,\nu}}!\,l_{x,\nu}!}}\,\frac{\sof{\frac{r_{s}}{2}}^{\abs{p_{x}}+2\,q_{x}{\color{fhl}-\delta_{x,z}}}\,\e^{\ii\,\theta_{s}\,p_{x}}\,\e^{2\,\mu\,k_{x,d}}}{\of{\abs{p_{x}}+q_{x}}!\,q_{x}!}\\
\delta\sof{p_{x}+\sum\limits_{\nu}\of{k_{x,\nu}-k_{x-\hat{\nu},\nu}}}\,W_{\lambda}\sof{\abs{p_{x}}+2\,q_{x}+\sum\limits_{\nu}\sof{\abs{k_{x,\nu}}+\abs{k_{x-\hat{\nu},\nu}}+2\,\ssof{l_{x,\nu}+l_{x-\hat{\nu},\nu}}}}}.\label{eq:radderiv1}
\end{multline}
The meaning of this correlator becomes clear when writing the original action \eqref{eq:phifouraction} in terms of spherical coordinates $\of{r_x,\,\theta_{x}}$:
\begin{multline}
S\fof{\phi}\,=\,\sum\limits_{x}\bcof{-\kappa\,\sum\limits_{\nu=1}^{d}\,r_{x}r_{x+\hat{\nu}}\cos\of{\theta_{x}-\theta_{x+\hat{\nu}}-2\,\ii\,\mu\,\delta_{\nu,d}}\\
+\,r_{x}^{2}\,+\,\lambda\of{r_{x}^{2}-1}^{2}\,-\,r_{s,x}r_{x}\cos\of{\theta_{x}-\theta_{s,x}}}.\label{eq:phifouractionsph}
\end{multline}
Taking the derivative of the partition function with respect to $r_{s,z}$ brings therefore down a factor $r_{z}\cos\of{\theta_{z}-\theta_{s}}$ under the integral, which for all $\theta_{z}$ measures the magnitude of the projection of the field on the direction specified by $\theta_{s,z}$. Analogously, taking the derivative of the partition function with respect $\theta_{s,z}$ would bring down a factor $r_{s,z}r_{z}\sin\of{\theta_{z}-\theta_{s,z}}$ which, after dividing by $r_{s,z}$, yields for all $\theta_{z}$ the magnitude of the projection of the field on the direction perpendicular to the direction specified by $\theta_{s,z}$.\\

If we write
\[
\abs{p_{z}}+2\,q_{z}\,=\,\frac{1}{2}\of{\abs{p_{z}}+p_{z}}+q_{z}+\frac{1}{2}\of{\abs{p_{z}}-p_{z}}+q_{z},
\]
we recognize, that by using again shifts of the form \eqref{eq:varshift1}, \eqref{eq:varshift2}, we can write \eqref{eq:radderiv1} as
\begin{multline}
\partd{Z}{r_{s,z}}\,=\,{\color{fhl}\frac{1}{2}}\sum\limits_{\sigma\in\pm 1}\,{\color{fhl}\e^{\ii\,\theta_{s}\,\sigma}}\,\sum\limits_{\cof{k,l,p,q}}\prod\limits_{x}\bcof{\bof{\prod\limits_{\nu}\frac{\of{\frac{\kappa}{2}}^{\abs{k_{x,\nu}}+2\,l_{x,\nu}}}{\of{\abs{k_{x,\nu}}+l_{x,\nu}}!\,l_{x,\nu}!}}\\
\frac{\sof{\frac{r_{s}}{2}}^{\abs{p_{x}}+2\,q_{x}}\,\e^{\ii\,\theta_{s}\,p_{x}}\,\e^{2\,\mu\,k_{x,d}}}{\of{\abs{p_{x}}+q_{x}}!\,q_{x}!}\,\delta\sof{p_{x}{\color{fhl}+\delta_{x,z}\,\sigma}+\sum\limits_{\nu}\of{k_{x,\nu}-k_{x-\hat{\nu},\nu}}}\\
W_{\lambda}\sof{\abs{p_{x}}+2\,q_{x}{\color{fhl}+\delta_{x,z}}+\sum\limits_{\nu}\sof{\abs{k_{x,\nu}}+\abs{k_{x-\hat{\nu},\nu}}+2\,\ssof{l_{x,\nu}+l_{x-\hat{\nu},\nu}}}}}\\
=\frac{1}{\sqrt{2}}\bof{\e^{-\ii\,\theta_{s}}\,\partd{Z}{s^{*}_{z}}+\e^{\ii\,\theta_{s}}\,\partd{Z}{s_{z}}}.\label{eq:radderiv2}
\end{multline}
Proceeding in the same way with the second derivative, we find:
\begin{multline}
\frac{1}{Z}\spartd{Z}{r_{s,z_1}}{r_{s,z_2}}\,=\,\frac{1}{{\color{fhl}4}\,Z}\sum\limits_{\sigma_1,\sigma_2\in\pm 1}\,{\color{fhl}\e^{\ii\,\theta_{s}\of{\sigma_1+\sigma_2}}}\,\sum\limits_{\cof{k,l,p,q}}\prod\limits_{x}\bcof{\bof{\prod\limits_{\nu}\frac{\of{\frac{\kappa}{2}}^{\abs{k_{x,\nu}}+2\,l_{x,\nu}}}{\of{\abs{k_{x,\nu}}+l_{x,\nu}}!\,l_{x,\nu}!}}\\
\frac{\sof{\frac{r_{s}}{2}}^{\abs{p_{x}}+2\,q_{x}}\,\e^{\ii\,\theta_{s}\,p_{x}}\,\e^{2\,\mu\,k_{x,d}}}{\of{\abs{p_{x}}+q_{x}}!\,q_{x}!}\,\delta\sof{p_{x}{\color{fhl}+\delta_{x,z_1}\,\sigma_1+\delta_{x,z_2}\,\sigma_2}+\sum\limits_{\nu}\of{k_{x,\nu}-k_{x-\hat{\nu},\nu}}}\\
W_{\lambda}\sof{\abs{p_{x}}+2\,q_{x}{\color{fhl}+\delta_{x,z_1}+\delta_{x,z_2}}+\sum\limits_{\nu}\sof{\abs{k_{x,\nu}}+\abs{k_{x-\hat{\nu},\nu}}+2\,\ssof{l_{x,\nu}+l_{x-\hat{\nu},\nu}}}}}\\
=\frac{1}{2}\bof{\frac{1}{Z}\spartd{Z}{s^{*}_{z_1}}{s^{\phantom{*}}_{z_2}}+\frac{1}{Z}\spartd{Z}{s^{\phantom{*}}_{z_1}}{s^{*}_{z_2}}+\frac{\e^{2\,\ii\,\theta_{s}}}{Z}\spartd{Z}{s_{z_1}}{s_{z_2}}+\frac{\e^{-2\,\ii\,\theta_{s}}}{Z}\spartd{Z}{s^{*}_{z_1}}{s^{*}_{z_2}}}.\label{eq:radcorr2}
\end{multline}
Of course, \eqref{eq:radcorr2} could have been obtained in a simpler way by using that
\[
s_{x}=\frac{r_{s,x}\e^{\ii\theta_{s,x}}}{\sqrt{2}}\quad,\qquad s^{*}_{x}=\frac{r_{s,x}\e^{-\ii\theta_{s,x}}}{\sqrt{2}},
\]
and therefore
\[
\partd{Z}{r_{s,z}}\,=\,\sum\limits_{x}\bof{\partd{s^{*}_{x}}{r_{s,z}}\partd{Z}{s^{*}_{x}}\,+\,\partd{s_{x}}{r_{s,z}}\partd{Z}{s_{x}}}\,=\,\frac{\e^{-\ii\theta_{s,z}}}{\sqrt{2}}\partd{Z}{s^{*}_{z}}\,+\,\frac{\e^{\ii\theta_{s,z}}}{\sqrt{2}}\partd{Z}{s_{z}},
\]
and so on, but it is instructive to repeat the argument with the shifts of the $p$- and $q$-variables which give rise to the Kronecker deltas in the local constraints and local weights on the second and third line of \eqref{eq:radcorr2}.\\

In a similar way, one finds the correlator for \emph{tangential} or \emph{angular excitations}:
\begin{multline}
\frac{1}{Z}\frac{1}{r_{s,z_1}r_{s,z_2}}\spartd{Z}{\theta_{s,z_1}}{\theta_{s,z_2}}\,=\\
\frac{1}{2}\bof{\frac{1}{Z}\spartd{Z}{s^{*}_{z_1}}{s^{\phantom{*}}_{z_2}}+\frac{1}{Z}\spartd{Z}{s^{\phantom{*}}_{z_1}}{s^{*}_{z_2}}-\frac{\e^{2\,\ii\,\theta_{s}}}{Z}\spartd{Z}{s_{z_1}}{s_{z_2}}-\frac{\e^{-2\,\ii\,\theta_{s}}}{Z}\spartd{Z}{s^{*}_{z_1}}{s^{*}_{z_2}}}.\label{eq:angcorr2}
\end{multline}

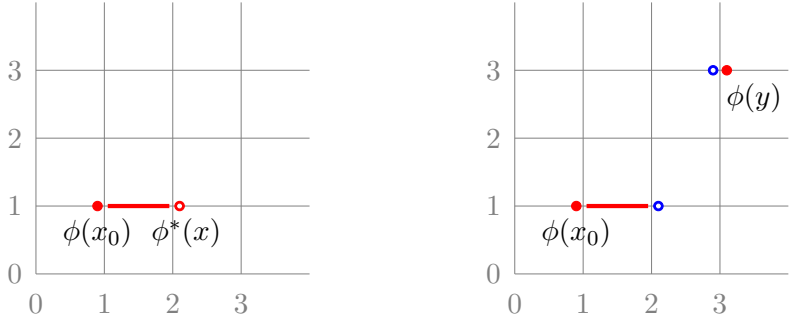
\begin{figure}[H]
\centering
\begin{minipage}[t]{\linewidth}
\centering
\begin{tikzpicture}[scale=0.9,nodes={inner sep=0}]
  \pgfpointtransformed{\pgfpointxy{1}{1}};
  \pgfgetlastxy{\vx}{\vy}
  \begin{scope}[node distance=\vx and \vy]
    \foreach \i in {0,...,3} {
        \draw [very thin,gray] (\i,0) -- (\i,4)  node[below] at (\i,-0.3) {$\i$};
    }
    \foreach \i in {0,...,3} {
        \draw [very thin,gray] (0,\i) -- (4,\i) node[left] at (-0.2,\i) {$\i$};
    }
    \draw [ultra thick,red!100] (1.05,1) -- (1.95,1);
    \node[draw,circle,inner sep=1.25,fill,color=red] at (0.9,1) {};
    \node[draw,circle,inner sep=1.25,fill,color=red] at (2.1,1) {};
    \node[draw,circle,inner sep=0.5,fill,color=white] at (2.1,1) {};
    \node at (0.9,0.6) {$\phi\ssof{x_{0}}$};
    \node at (2.2,0.6) {$\phi^{*}\ssof{x}$};
  \end{scope}
  \begin{scope}[node distance=\vx and \vy]
    
    \foreach \i in {0,...,3} {
        \draw [very thin,gray] (\i+7,0) -- (\i+7,4)  node[below] at (\i+7,-0.3) {$\i$};
    }
    \foreach \i in {0,...,3} {
        \draw [very thin,gray] (0+7,\i) -- (4+7,\i) node[left] at (7-0.2,\i) {$\i$};
    }
    \draw [ultra thick,red!100] (7+1.05,1) -- (7+1.95,1);
    \node[draw,circle,inner sep=1.25,fill,color=red] at (7+0.9,1) {};
    \node[draw,circle,inner sep=1.25,fill,color=blue] at (7+2.1,1) {};
    \node[draw,circle,inner sep=0.5,fill,color=white] at (7+2.1,1) {};

    \node[draw,circle,inner sep=1.25,fill,color=blue] at (7+2.9,3) {};
    \node[draw,circle,inner sep=0.5,fill,color=white] at (7+2.9,3) {};
    \node[draw,circle,inner sep=1.25,fill,color=red] at (7+3.1,3) {};
    \node at (7+0.9,0.6) {$\phi\ssof{x_{0}}$};
    \node at (7+3.5,2.6) {$\phi\ssof{y}$};
  \end{scope}
\end{tikzpicture}
\end{minipage}
\caption{New worm moves: for $\abs{s}>0$, starting from the configuration on the left-hand side, the (necessarily disconnected) head-changing worm-update consists of proposing to remove $\phi^{*}\of{x}$ from $x$, and to insert instead $\phi\of{y}$ at some random other site $y$. The resulting charge-imbalance in the system and at $x$ and $y$ is compensated by increasing both, $p_{x}$ and $p_{y}$ by 1 (blue circle), as depicted in the right-hand figure. The inverse move works analogously: propose to replace $\phi\of{y}$ by $\phi^{*}\of{x}$ and to simultaneously decrease $p_{x}$ and $p_{y}$ by 1.}
  \label{fig:worm3}
\end{figure}

The first two pieces inside the bracket on the last line of \eqref{eq:radcorr2} and on the right hand side of \eqref{eq:angcorr2} can directly be measured during the worm update described above, no matter whether $r_{s}$ is zero or not. The other two pieces are non-zero only if $r_{s}> 0$ and in order to measure them during the worm update, we have to extend the worm algorithm by an additional type of move which allows the worm to change the external source/sink at its head. This can be done by modifying the disconnected worm update described above in Fig. \ref{fig:worm2} in such a way, that instead of moving the external source $\phi^{*}$ from $x$ to some other site $y$ and shifting $p_{x}\rightarrow p_{x}+1$ and $p_{y}\rightarrow p_{y}-1$, one can remove $\phi^{*}\of{x}$ completely from the system, still compensate the absence of its charge at $x$ by shifting $p_{x}\rightarrow p_{x}+1$, but now, instead of inserting at $y$ again a $\phi^{*}$ and compensating for its charge by shifting $p_{y}\rightarrow p_{y}-1$, we insert a $\phi$ and shift $p_{y}\rightarrow p_{y}+1$ (see Fig. \ref{fig:worm3}). If this move is accepted, the worm samples from now on the last instead of the first piece from inside the brackets on the last lines of \eqref{eq:radcorr2} and \eqref{eq:angcorr2}. The move which changes the external charge at the head of the worm back from $\phi$ to $\phi^{*}$ works completely analogously, except that the $p$-variables have to be shifted in the opposite direction. The complex phases appearing in these expressions get always canceled by the complex phases coming from the shifts of the $p$-variables during the head-changing moves. Note also that with these additional head-changing moves, the worm can still be terminated only if its head and tail are located on the same site, and in addition, if the head's external charge is again the same as at the beginning of the worm.\\

The head-changing moves are completely general: if we had chosen for example the $\SU{2}$ principal chiral model instead of complex $\phi^4$, we could also have introduced moves that would allow us to sample correlators like $\savof{\pi^{-}\of{x}\pi^{0}\of{y}}$, $\savof{\pi^{+}\of{x}\sigma\of{y}}$, etc. during the worm \cite{Rindlisbacher1}. In \cite{Akerlund} we used this method to verify the existence of oscillatory two-point functions in the three-state Potts model if the real and imaginary parts of the field are coupled to different external fields.

\subsection{Measuring Condensates with a Worm: Why a Source Term is Needed}\label{ssec:measurecond}
\begin{figure}[H]
\centering
\begin{minipage}[t]{\linewidth}
\centering
\begin{tikzpicture}[scale=0.9,nodes={inner sep=0}]
  \pgfpointtransformed{\pgfpointxy{1}{1}};
  \pgfgetlastxy{\vx}{\vy}
  \begin{scope}[node distance=\vx and \vy]
    \foreach \i in {0,...,3} {
        \draw [very thin,gray] (\i,0) -- (\i,4)  node[below] at (\i,-0.3) {$\i$};
    }
    \foreach \i in {0,...,3} {
        \draw [very thin,gray] (0,\i) -- (4,\i) node[left] at (-0.2,\i) {$\i$};
    }
    \node[draw,circle,inner sep=1.25,fill,color=blue] at (0.9,1) {};
    \node[draw,circle,inner sep=1.25,fill,color=red] at (1.1,1) {};
    \node[draw,circle,inner sep=0.5,fill,color=white] at (1.1,1) {};
    \node at (1.3,0.6) {$\phi^{*}\ssof{x}$};
  \end{scope}
  \begin{scope}[node distance=\vx and \vy]
    
    \foreach \i in {0,...,3} {
        \draw [very thin,gray] (\i+6,0) -- (\i+6,4)  node[below] at (\i+6,-0.3) {$\i$};
    }
    \foreach \i in {0,...,3} {
        \draw [very thin,gray] (0+6,\i) -- (4+6,\i) node[left] at (6-0.2,\i) {$\i$};
    }
    \draw [ultra thick,red!100] (6+1.05,1) -- (6+1.95,1);
    \node[draw,circle,inner sep=1.25,fill,color=blue] at (6+0.9,1) {};
    \node[draw,circle,inner sep=1.25,fill,color=red] at (6+2.1,1) {};
    \node[draw,circle,inner sep=0.5,fill,color=white] at (6+2.1,1) {};
    \node at (6+2.3,0.6) {$\phi^{*}\ssof{x+\hat{\nu}}$};
  \end{scope}
  \begin{scope}[node distance=\vx and \vy]
    
    \foreach \i in {0,...,3} {
        \draw [very thin,gray] (\i+12,0) -- (\i+12,4)  node[below] at (\i+12,-0.3) {$\i$};
    }
    \foreach \i in {0,...,3} {
        \draw [very thin,gray] (0+12,\i) -- (4+12,\i) node[left] at (12-0.2,\i) {$\i$};
    }
    \draw [ultra thick,red!100] (12+1.05,1) -- (12+1.95,1);
    \node[draw,circle,inner sep=1.25,fill,color=blue] at (12+0.9,1) {};
    \node[draw,circle,inner sep=1.25,fill,color=blue] at (12+2.1,1) {};
    \node[draw,circle,inner sep=0.5,fill,color=white] at (12+2.1,1) {};
  \end{scope}
\end{tikzpicture}
\end{minipage}
\caption{The left-hand figure shows the situation where the insertion of a $\phi^{*}\of{x}$ at some site $x$ (indicated by the red circle) and a simultaneous shift, $p_{x}\rightarrow p_{x}-1$, (indicated by the filled blue dot) has been accepted. One can now propose either to remove $\phi^{*}$ again from the system and to shift $p_{x}$ back to its original value, or one can propose to move $\phi^{*}\of{x}$ to the site $x+\ohat{\nu}$ and to shift $k_{x,\nu}\rightarrow k_{x,\nu}+1$, as depicted in the middle figure. If the latter move is accepted one can again propose to either shift $\phi^{*}\of{x+\ohat{\nu}}$ further to a neighboring site of $x+\ohat{\nu}$, or to remove $\phi^{*}$ from the system and shifting $p_{x+\ohat{\nu}}\rightarrow p_{x+\ohat{\nu}}+1$, such that we would be left with the situation in the right-hand figure.}
  \label{fig:worm4}
\end{figure}
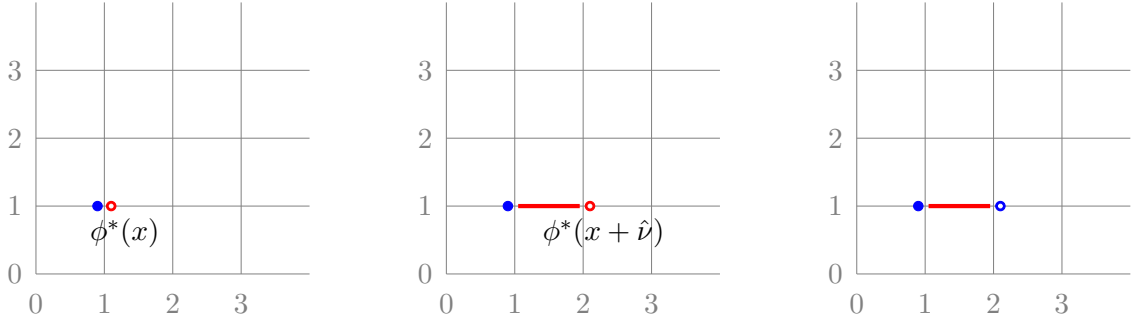

The partition function for our complex $\phi^4$ model is defined in terms of an Euclidean path integral which explicitly sums over all possible field configurations. This means that without an explicit breaking of the global $\U{1}$ symmetry (e.g. by a source term in the action), the condensates $\avof{\phi}$ and $\avof{\phi^{*}}$ have to vanish identically even in the ordered phase, as the sum of all the condensates corresponding to different $\U{1}$-degenerate vacua, is zero. In the context of "ordinary" Monte Carlo simulations, based on local Metropolis or heat-bath updates of the original field variables, $\phi_{x}$, this might not always be so obvious, as in such simulations, with increasing system size, the Markov chain usually has more and more problems to tunnel between different degenerate vacua. The reason for this is of course the local update strategy and the fact, that the different vacua should indeed become non-interacting in the thermodynamic limit. This means that any intermediate configuration required for the tunneling from one vacuum into another, becomes energetically more and more expensive with increasing system size and therefore more and more unlikely to be realized, so that the Markov chain tends to remain for a long time in the same vacuum, leading to apparently non-zero condensates. This is of course an illusion: it just reflects the fact that the algorithm is non-ergodic and therefore fails to generate a representative subset of configurations that contribute to the path integral.\\   
In the flux-representation \eqref{eq:phifourfluxreppartf} the situation is qualitatively different: a non-zero source term is mandatory for a non-zero condensate, even for very large systems! The reason for this is the following: as we have integrated out the complex phases of the $\phi$-field analytically in order to obtain the weights for the dual configurations in terms of flux variables, we also integrated out the global $\U{1}$ symmetry. Therefore every admissible flux-variable configuration, which could be understood to describe a configuration of particle world-lines, stands in fact for the superposition of the simultaneous realizations of these particle world-line configurations in each of the degenerate vacua. So, without adding source terms to the system, a spontaneous breaking of the global $\U{1}$ symmetry can only be observed in quantities which look the same in all degenerate vacua, or which are non-local in the sense that they depend on at least two sites
\footnote{To be more precise: as we have integrated out the complex phase of the $\phi$ field to arrive at the partition function \eqref{eq:phifourfluxreppartf}, we cannot even break the global $\U{1}$ symmetry by setting the source $s$ to a non-zero value. One manifestation of this is the fact that \eqref{eq:phifourfluxreppartf} does not depend on the phase $\theta_{s}$ of the external source but only on its magnitude $r_{s}=\sqrt{2}\abs{s}$, as shown at the end of Sec. \ref{ssec:phi4partf}. Another manifestation is that $\sabs{\avof{\phi}}=\sabs{\avof{\phi^{*}}}$ which can be seen from \eqref{eq:naivederiv1} and \eqref{eq:naivederiv2}, together with the fact that $\sum_{x}p_{x}=0$ for all valid configurations contributing to \eqref{eq:phifourfluxreppartf}. However, by specifying a non-zero value for $r_{s}$ we can measure the $\U{1}$ invariant $\sabs{\avof{\phi}}=\sabs{\avof{\phi^{*}}}$ in polar coordinates (see also \cite{Rindlisbacher1}).}: 
for example the value of the $k$-variable between two sites, which measures how well the fields on these two sites are "in phase"\footnote{Here "in phase" effectively means "aligned", but for the case where one sums over all possible orientations in which the fields can align in the internal space}, i.e. how dominant the "in phase" contribution to the integral over the individual phases of the fields on the two sites is. If the average value of the $k$-variables increases, this indicates that the system gets more and more ordered.\\
By looking at equations \eqref{eq:deriv1n} and \eqref{eq:deriv2n}, we can see that if $\abs{s}=0$ (which implies $p_{x}=0\,\forall x$), there is no way to satisfy the local delta function constraints in \eqref{eq:deriv1n} or \eqref{eq:deriv2n} simultaneously for all sites: one can move the defect (introduced by the $\pm 1$ in the delta function at $x$) around by changing a sequence of $k$-variables, but without the insertion of a second external source or sink at which this sequence of changed $k$-variables could terminate (which would then correspond to measuring a two-point function instead of a condensate), there will always be a vanishing delta function factor somewhere in the weight of the configuration. So there exist no configurations which could support a non-zero measurement of a condensate if the source is set to zero.\\
However, if $\abs{s}$ is non-zero, a second external source or sink is not required, as then the defect can be compensated by a shift of one of the $p$-variables in the system. It is precisely this property which we can use to efficiently measure condensates (and therefore also the disconnected piece of correlators) even in the case of small sources (see Fig. \ref{fig:worm4}): we propose for example to insert at some site $x$ a single external source or sink, $\phi$ or $\phi^{*}$, and to compensate the corresponding $\pm 1$ appearing in the delta function at $x$ by an appropriate shift $p_{x}\rightarrow p_{x}\mp 1$. If this insertion is accepted, we can either continue by proposing to remove the external source/sink again and to shift $p_{x}$ back to its original value, or we can propose to move the external source/sink to one of the neighboring sites of $x$, say $x+\hat{\nu}$ and to update $k_{x,\nu}$ in order to compensate for the resulting charge-displacement (completely analogous to the "connected move" in the "charged worm" described in the previous section). This procedure is repeated until the update which removes the external source/sink is accepted. Here, our algorithm is analogous to the \emph{open-worm} algorithm from \cite{Gattringer2}. The expectation values $\avof{\phi^{*}}$ and $\avof{\phi}$ are then obtained by measuring the fraction of the Monte Carlo time, for which the external source/sink is present in the system.\\

Yet another advantage of this method for measuring condensates is, that it provides a simple means of variance reduction. One can absorb the summation over the $q$-variables into the weight-function instead of sampling their values, i.e. one uses
\begin{multline}
\tilde{W}_{\lambda,r_{s}}\of{A,\abs{p}}\,=\,\sum_{q}\,\frac{\sof{\frac{r_{s}}{2}}^{\abs{p}+2\,q}}{\of{\abs{p}+q}!\,q!}\,W_{\lambda}\sof{\abs{p}+2\,q\,+\,A}\\
=\,\int\limits_{0}^{\infty}\idd{u}{}\,\frac{u^{A/2}\,\e^{-u-\lambda\of{u-1}^{2}}}{2}\,I_{\abs{p}}\of{r_{s}\sqrt{u}}\ ,\label{eq:summedweights}
\end{multline}
in the worm algorithm instead of $W_{\lambda}\sof{\abs{p}+2\,q+A}$ itself, where $A$ contains the sum over the $k$ and $l$-variables, and the $I_{\nu}\of{x}$ are modified Bessel functions of the first kind. The values of \eqref{eq:summedweights} can be pre-computed for a set of positive integers $A$ and $p$ and stored in a two-dimensional array\footnote{More precisely: one can even pre-compute directly the relevant weight ratios $\tilde{W}_{\lambda,r_{s}}\of{A,\abs{p}+1}/\tilde{W}_{\lambda,r_{s}}\of{A,\abs{p}}$ and $\tilde{W}_{\lambda,r_{s}}\of{A+1,\abs{p}}/\tilde{W}_{\lambda,r_{s}}\of{A,\abs{p}}$ .}. This speeds up the simulation and reduces the noise in the correlator and condensate measurements further.\\

In Appendix \ref{sec:wormupdate}, we provide pseudo-code describing our implementation of the concepts introduced in this and the previous sections, together with some comments and a derivation of the transition probabilities required to ensure detailed balance during the Monte Carlo simulation.\\

\subsection{Gain in Efficiency}
\begin{figure}[H]
\centering
\begin{minipage}[t]{0.33\linewidth}
\centering
$\scriptstyle \spartd{\log\of{Z}}{s^{*}_{x}}{s_{y}}$\\
\includegraphics[width=\linewidth]{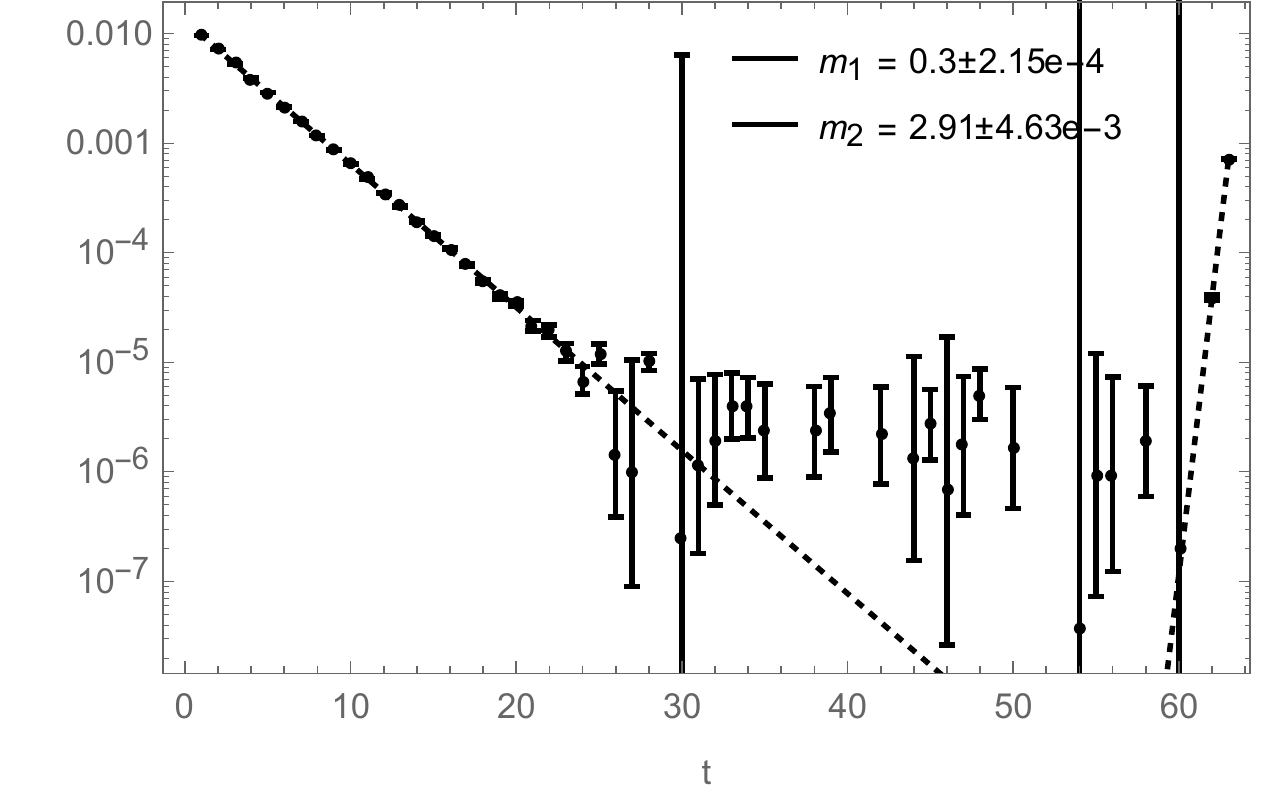}
\end{minipage}\hfill
\begin{minipage}[t]{0.33\linewidth}
\centering
$\scriptstyle \spartd{\log\of{Z}}{r_{s,x}}{r_{s,y}}$\\
\includegraphics[width=\linewidth]{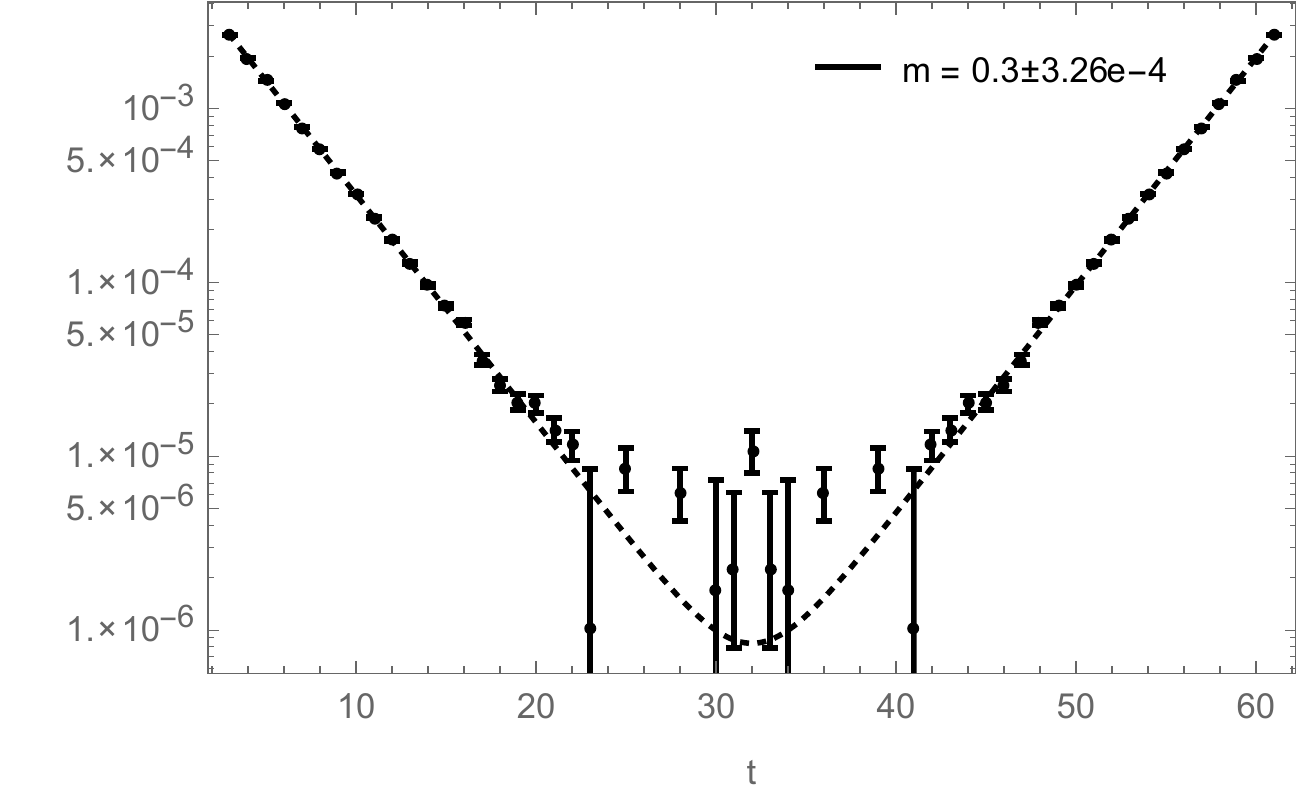}
\end{minipage}\hfill
\begin{minipage}[t]{0.33\linewidth}
\centering
$\scriptstyle \frac{1}{r_{s}^2}\spartd{\log\of{Z}}{\theta_{s,x}}{\theta_{s,y}}$\\
\includegraphics[width=\linewidth]{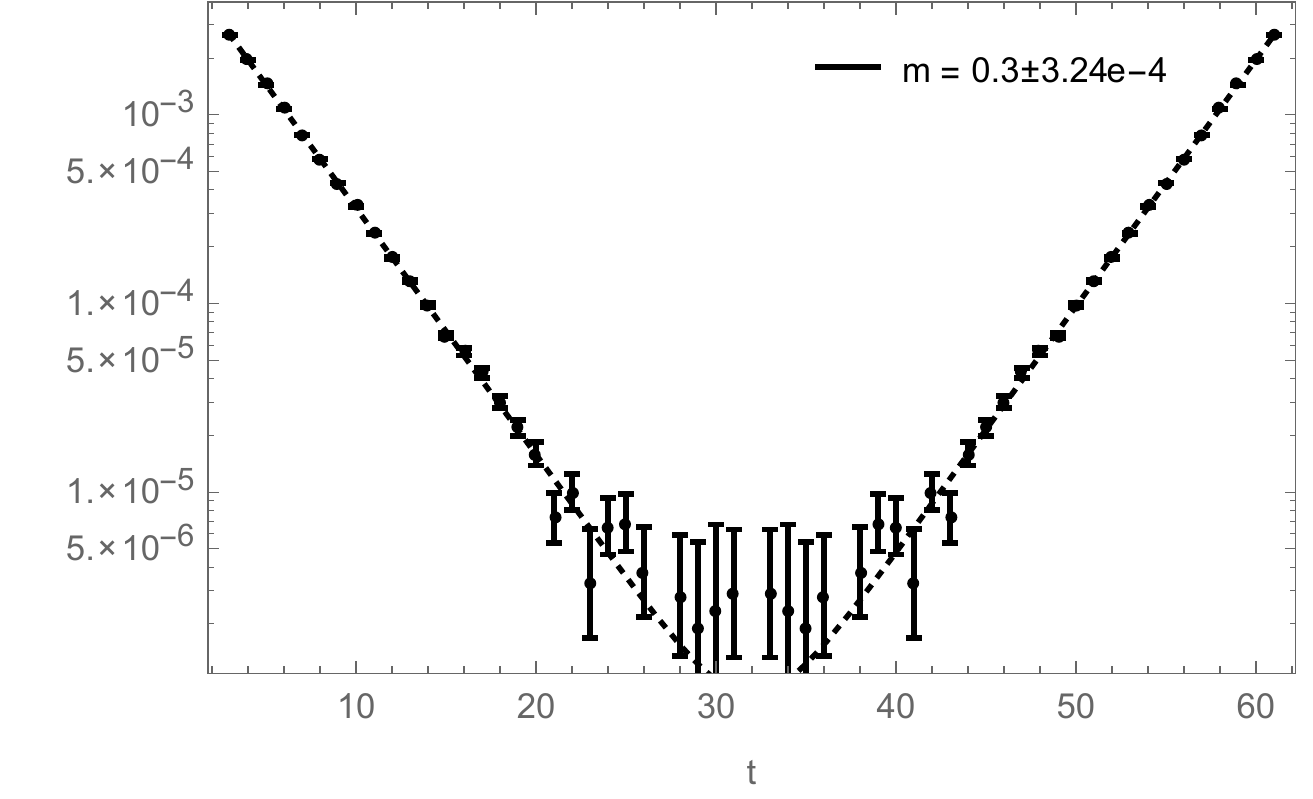}
\end{minipage}\\
\begin{minipage}[t]{0.33\linewidth}
\centering
\includegraphics[width=\linewidth]{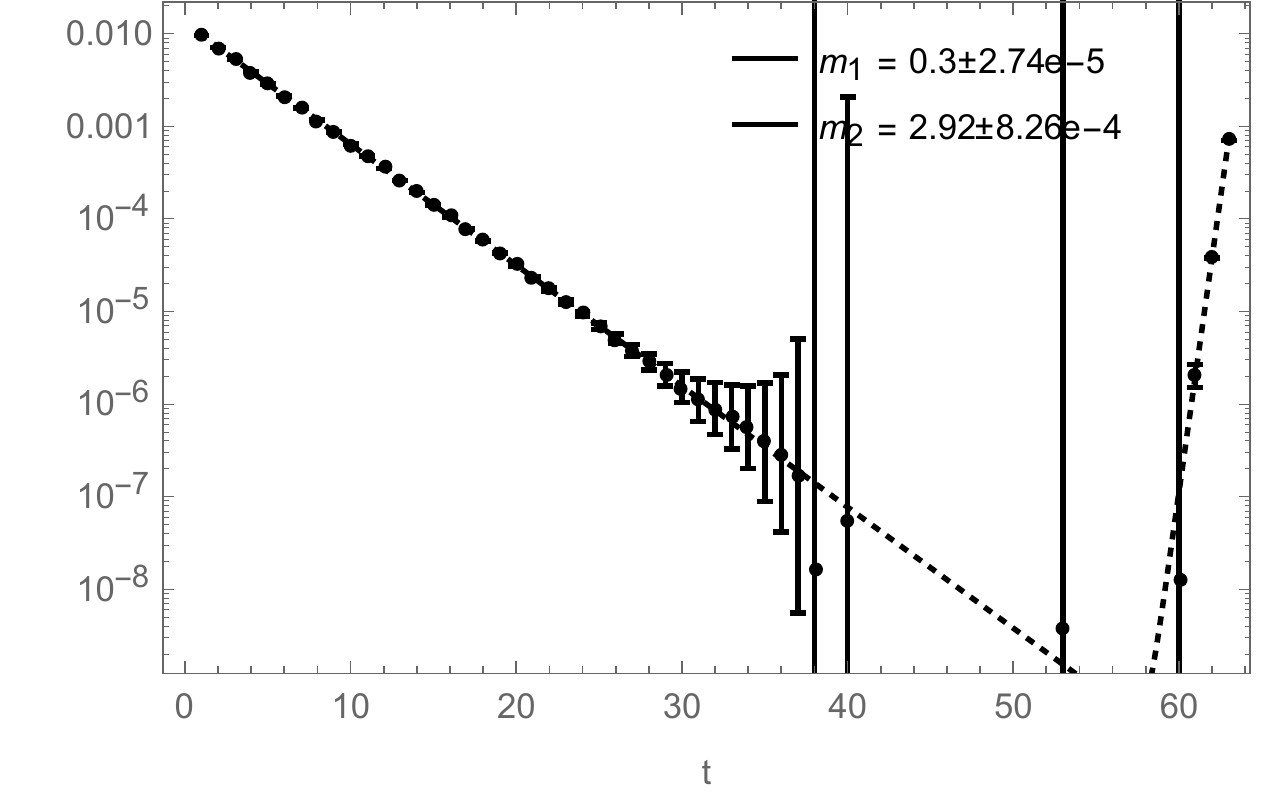}
\end{minipage}\hfill
\begin{minipage}[t]{0.33\linewidth}
\centering
\includegraphics[width=\linewidth]{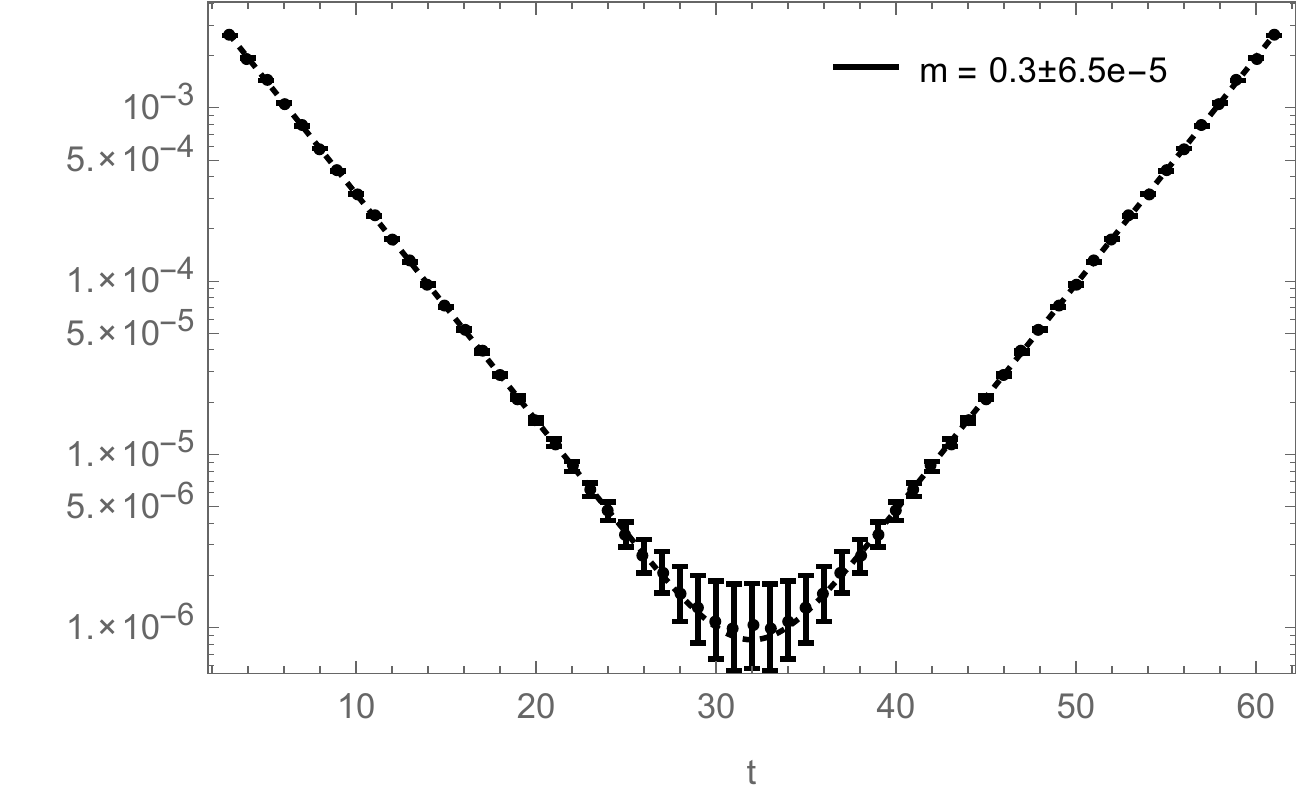}
\end{minipage}\hfill
\begin{minipage}[t]{0.33\linewidth}
\centering
\includegraphics[width=\linewidth]{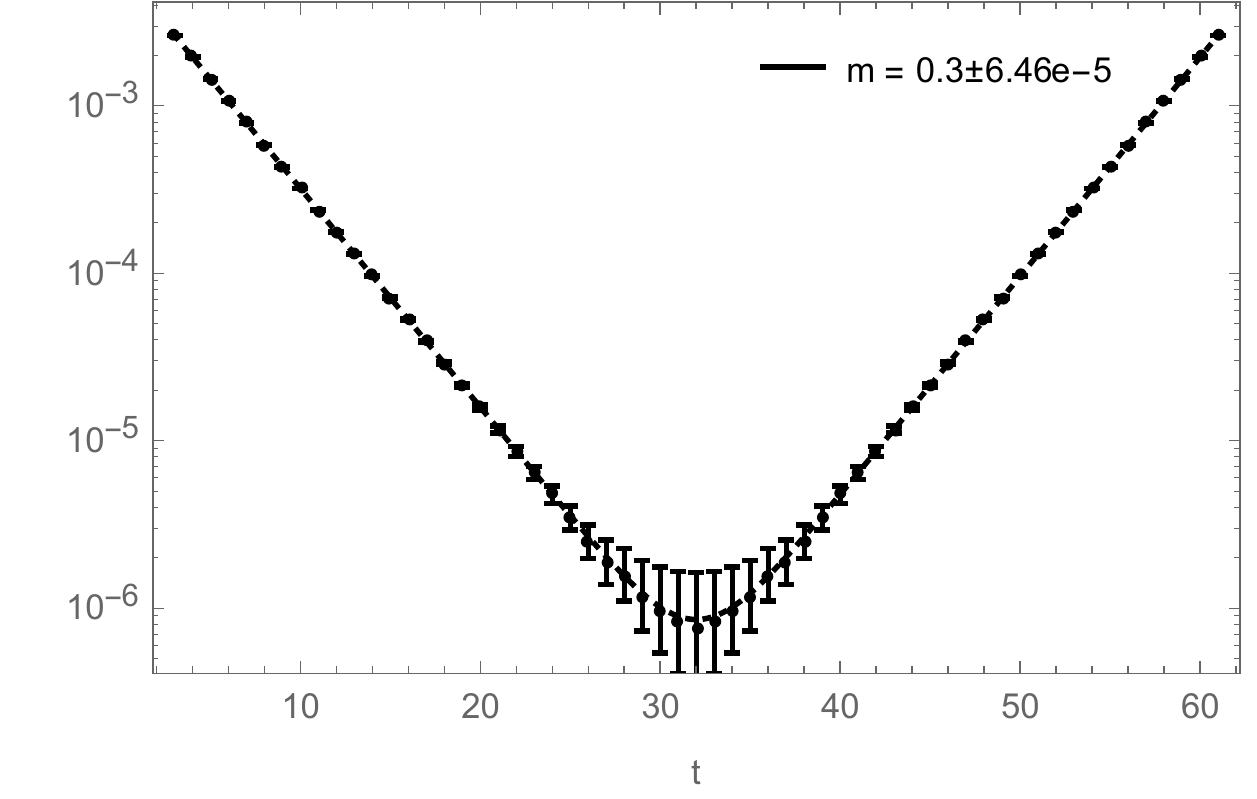}
\end{minipage}
\caption{We compare for a (2+1)-dimensional system with $N_{s}=8$, $N_{t}=64$, $\kappa=0.25$, $\lambda=7.5$, $r_s=0.01$ and $\mu=0.65$, the zero-momentum pieces of the indicated two-point functions, obtained with two different measurement methods: the correlators in the upper row are obtained with the naive approach, by making use of eqns. \eqref{eq:naivederiv1}, \eqref{eq:naivederiv2} and \eqref{eq:naivesderiv1} (and analogous ones), i.e. by measuring the $p$ and $q$ content of closed-worm configurations, while the ones in the bottom row are obtained by measuring the correlators and condensates during the worm, making use of eqns. \eqref{eq:deriv1n}, \eqref{eq:deriv2n} and \eqref{eq:sderivn} (and similar ones). Both data sets were obtained from the same simulation. For the top row correlators, measurements were taken after every sweep and for the bottom row correlators the histograms were stored and reset after every 1000 sweeps. The error bars were then obtained with the jack-knife procedure (note that the correlators are displayed on a log-scale, whereas each error-bar is on a linear scale, which is set by the log-re-scaling factor of the corresponding data point). The dotted lines correspond to exponential fits which lead to the mass-values displayed in the upper right corner of each plot. Note the reduction in the error promoted by the new method.}
  \label{fig:corr}
\end{figure}

\begin{figure}[H]
\centering
\begin{minipage}[t]{0.33\linewidth}
\centering
$\scriptstyle \spartd{\log\of{Z}}{s^{*}_{x}}{s_{y}}$\\
\includegraphics[width=\linewidth]{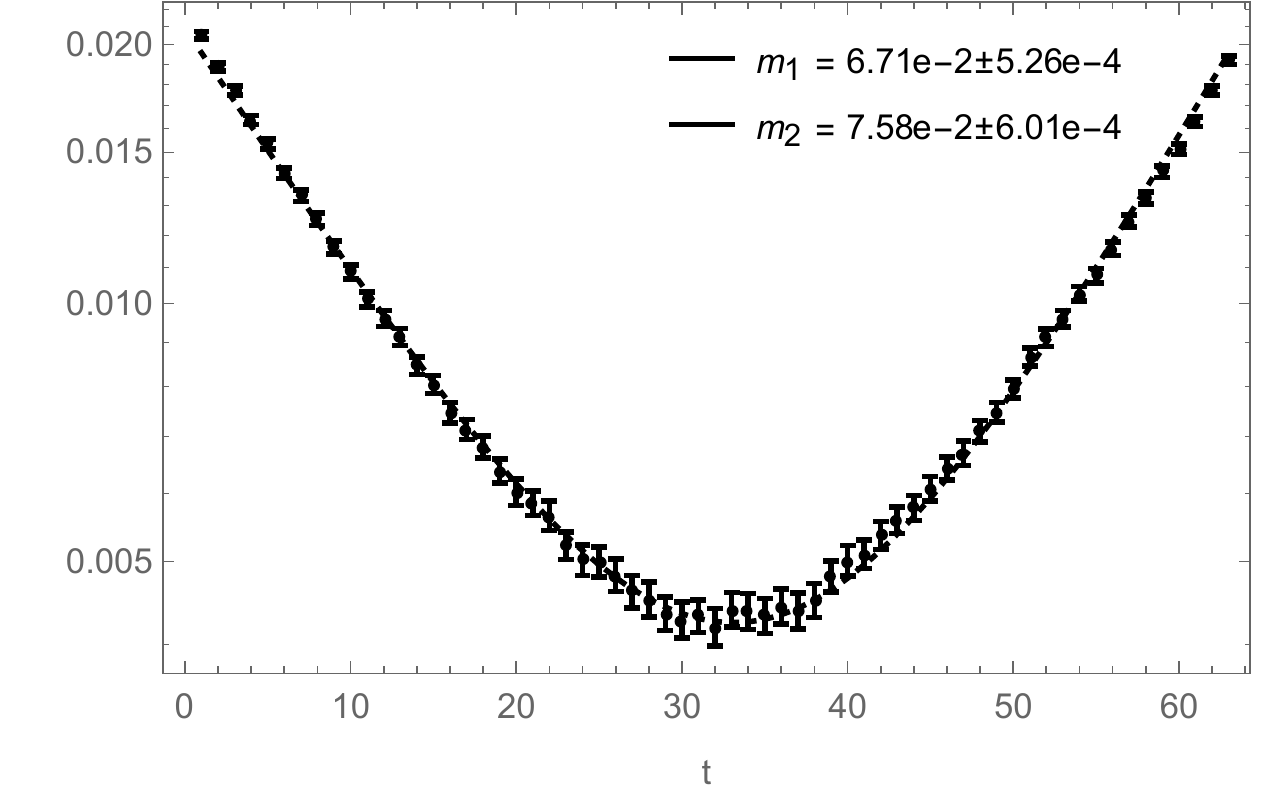}
\end{minipage}\hfill
\begin{minipage}[t]{0.33\linewidth}
\centering
$\scriptstyle \spartd{\log\of{Z}}{r_{s,x}}{r_{s,y}}$\\
\includegraphics[width=\linewidth]{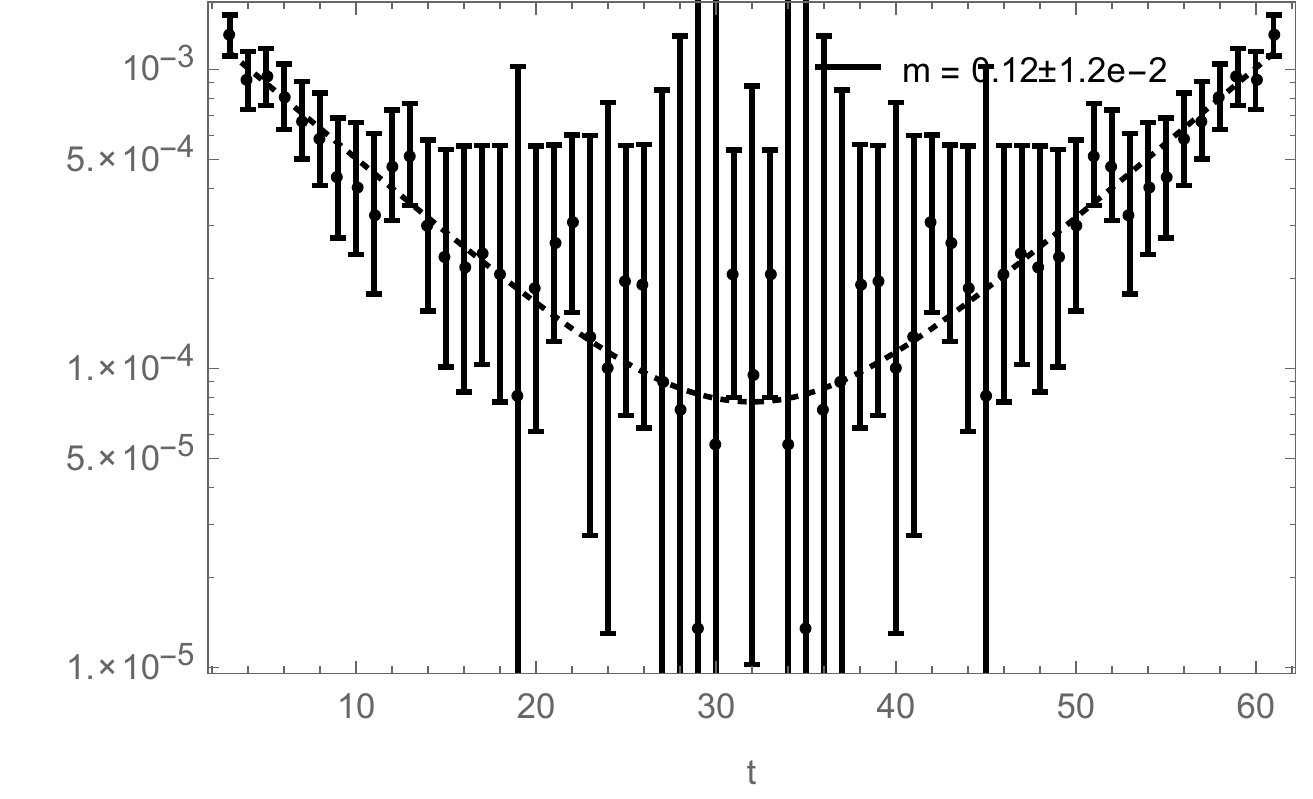}
\end{minipage}\hfill
\begin{minipage}[t]{0.33\linewidth}
\centering
$\scriptstyle \frac{1}{r_{s}^2}\spartd{\log\of{Z}}{\theta_{s,x}}{\theta_{s,y}}$\\
\includegraphics[width=\linewidth]{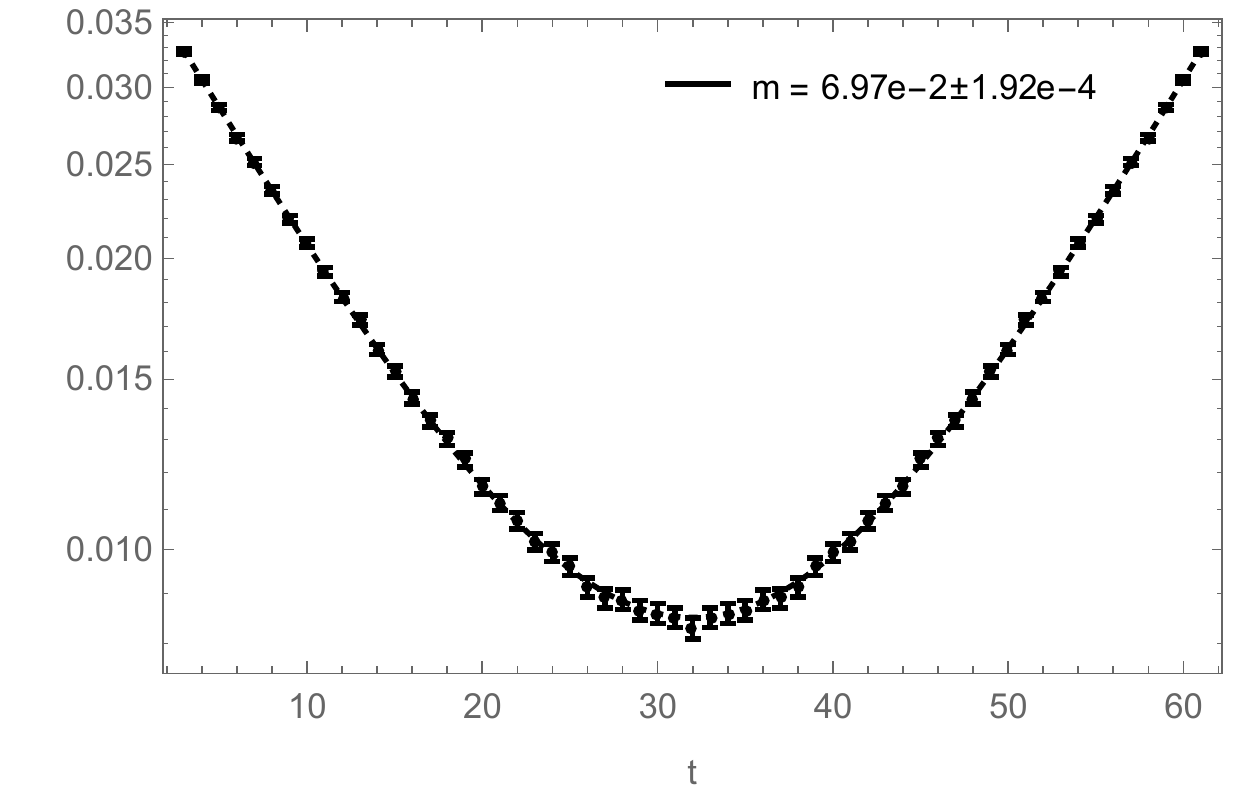}
\end{minipage}\\
\begin{minipage}[t]{0.33\linewidth}
\centering
\includegraphics[width=\linewidth]{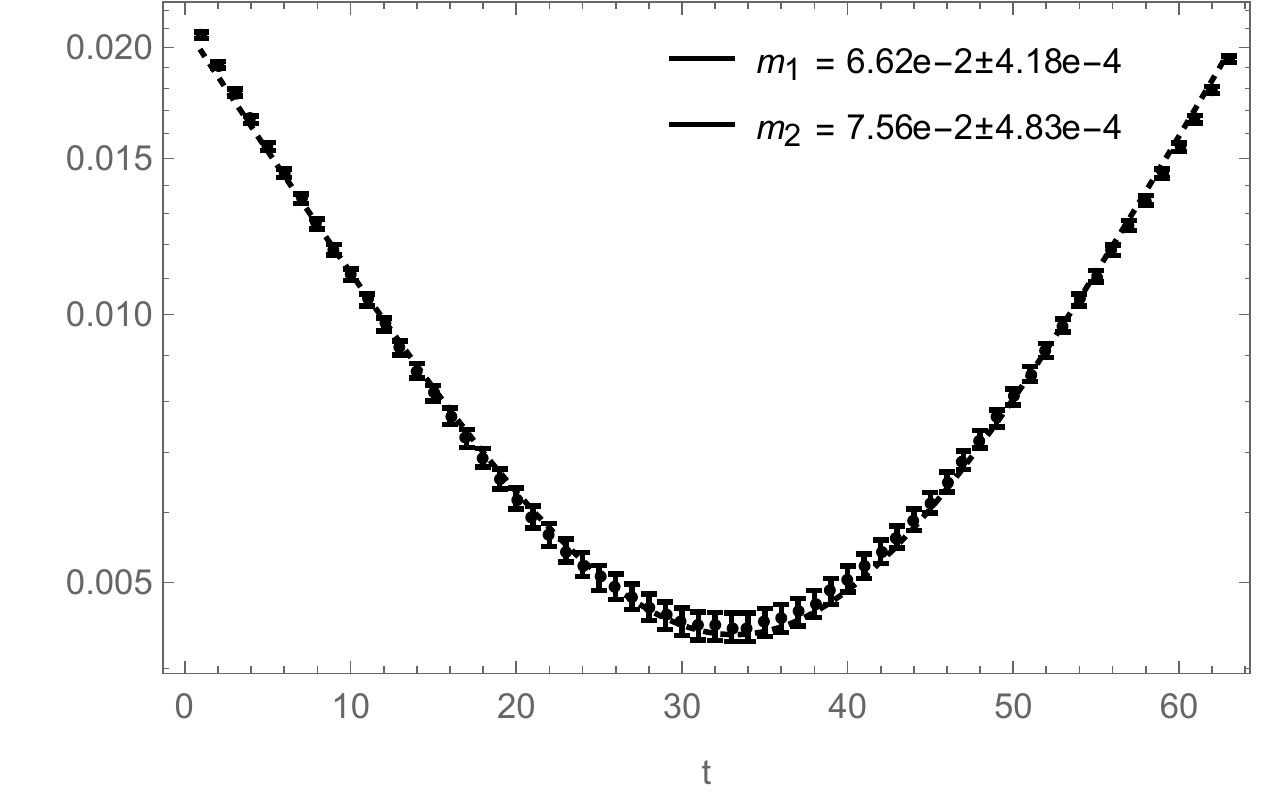}
\end{minipage}\hfill
\begin{minipage}[t]{0.33\linewidth}
\centering
\includegraphics[width=\linewidth]{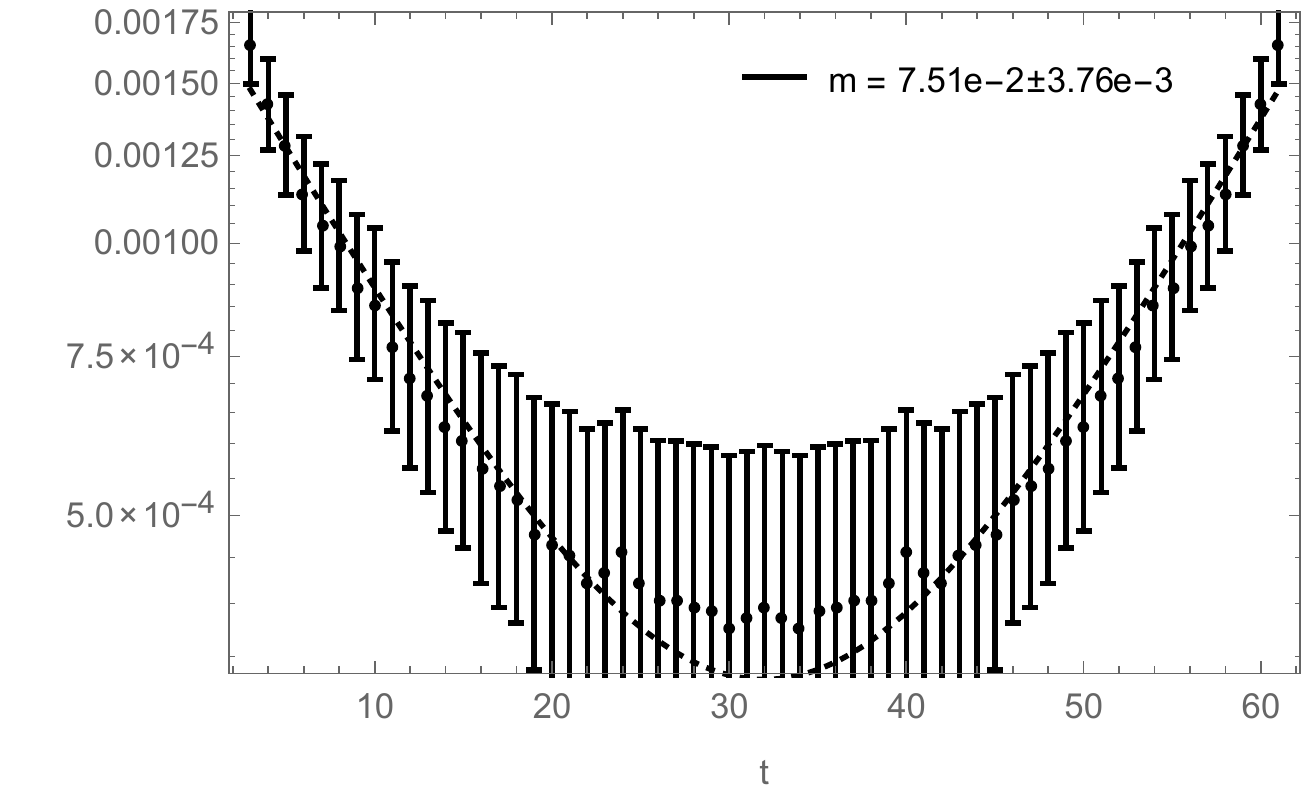}
\end{minipage}\hfill
\begin{minipage}[t]{0.33\linewidth}
\centering
\includegraphics[width=\linewidth]{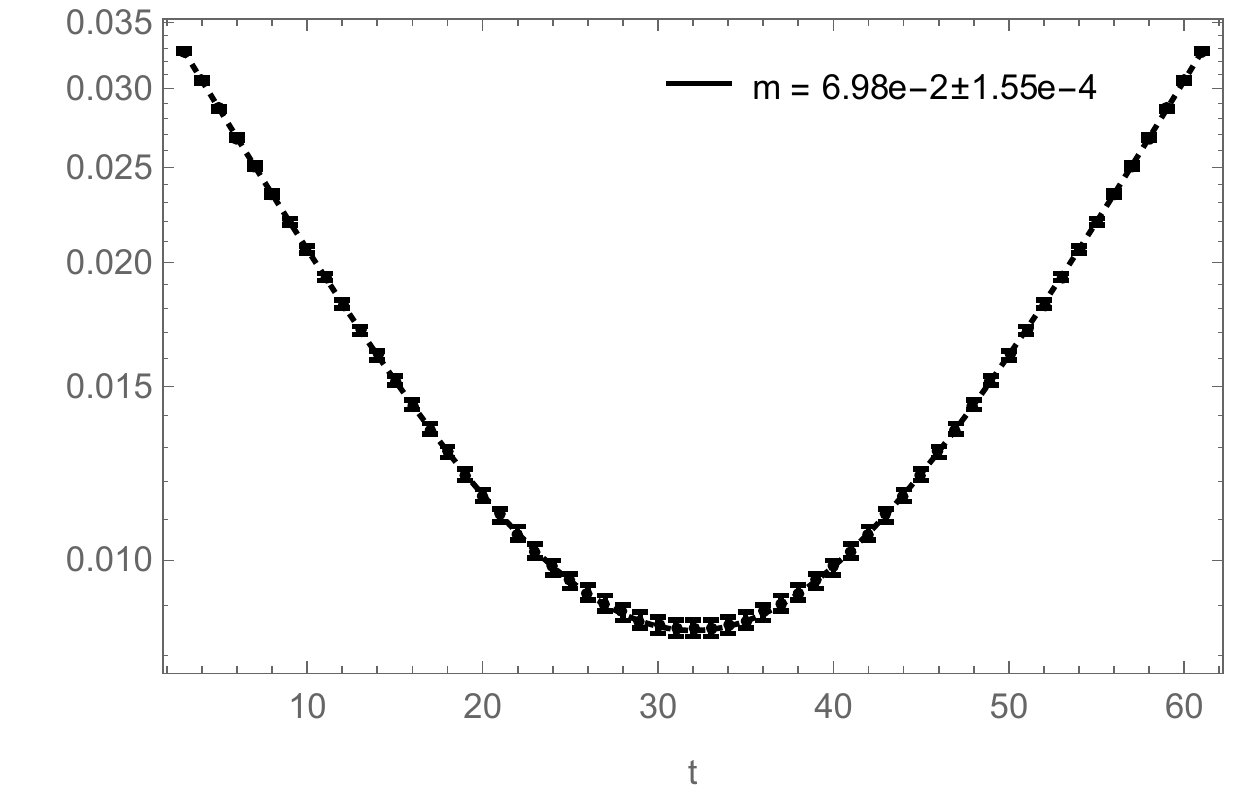}
\end{minipage}
\caption{Same as in figure \ref{fig:corr}, but for $\mu=1.2$, which is already in the symmetry broken phase. The correlators in the upper three plots were obtained with the standard measurement method on closed-worm configurations while the lower three figures were obtained with the improved method during the worm updates. Again the dotted lines correspond to exponential fits from which the mass-values are obtained which are displayed in the upper right corner of each plot. The relatively large, but homogeneous (note the log scale) errors visible in the two plots in the middle are mainly due to the subtraction of the large disconnected piece. Note the reduction in the error promoted by the new method.}
  \label{fig:corr2}
\end{figure}

In figures \ref{fig:corr} and \ref{fig:corr2} we compare results for the connected zero-momentum pieces of three different correlators, the charged one,
\[
\spartd{\log\of{Z}}{s^{*}_{x}}{s_{y}},\label{eq:chargedcorr1}
\]
the one for radial,
\[
\spartd{\log\of{Z}}{r_{s,x}}{r_{s,y}}\label{eq:radialcorr1}
\]
and the one for angular excitations,
\[
\frac{1}{r_{s}^2}\spartd{\log\of{Z}}{\theta_{s,x}}{\theta_{s,y}},\label{eq:angularcorr1}
\]
obtained with two different measurement techniques: once with the naive approach, based on measuring \eqref{eq:naivederiv1}, \eqref{eq:naivederiv2} and \eqref{eq:naivesderiv1} on \emph{closed-worm configurations}, and once with the improved method, by measuring the correlators and condensates during the worm-updates according to \eqref{eq:deriv1n}, \eqref{eq:deriv2n} and \eqref{eq:sderivn}, as well as \eqref{eq:radcorr2} and \eqref{eq:angcorr2}. The results from both methods were obtained during the same simulation and thus are directly comparable. For the naive approach, measurements were taken after every sweep (with a sweep defined as below eq. \eqref{eq:avenergy}). For the improved method, every $N_{sweeps}=1000$ sweeps, the histograms for the correlators and the condensates, accumulated during the worm-updates, were normalized by $1/(L_{sweep}\times N_{sweeps})$ (which is the Monte Carlo time over which the histograms were accumulated), stored and reset. The simulations had a length of $5\times 10^6$ sweeps (plus thermalization). The errors were then obtained with the jack-knife method. As can be seen, the correlators obtained with the improved method provide a much cleaner and less noisy signal than the ones obtained with the naive procedure (based on the values of the $p$ and $q$-variables in \emph{closed worm configurations}). The large relative errors in the correlators shown in Fig.~\ref{fig:corr2}, are due to the subtraction of the disconnected piece in \eqref{eq:chargedcorr1} and \eqref{eq:radialcorr1}, which is large in the ordered phase. Overall, the new method yields smaller errors on the fitted mass, by a factor $\sim 5$.\\

\begin{figure}[H]
\centering
\begin{minipage}[t]{0.49\linewidth}
\centering
\includegraphics[width=\linewidth]{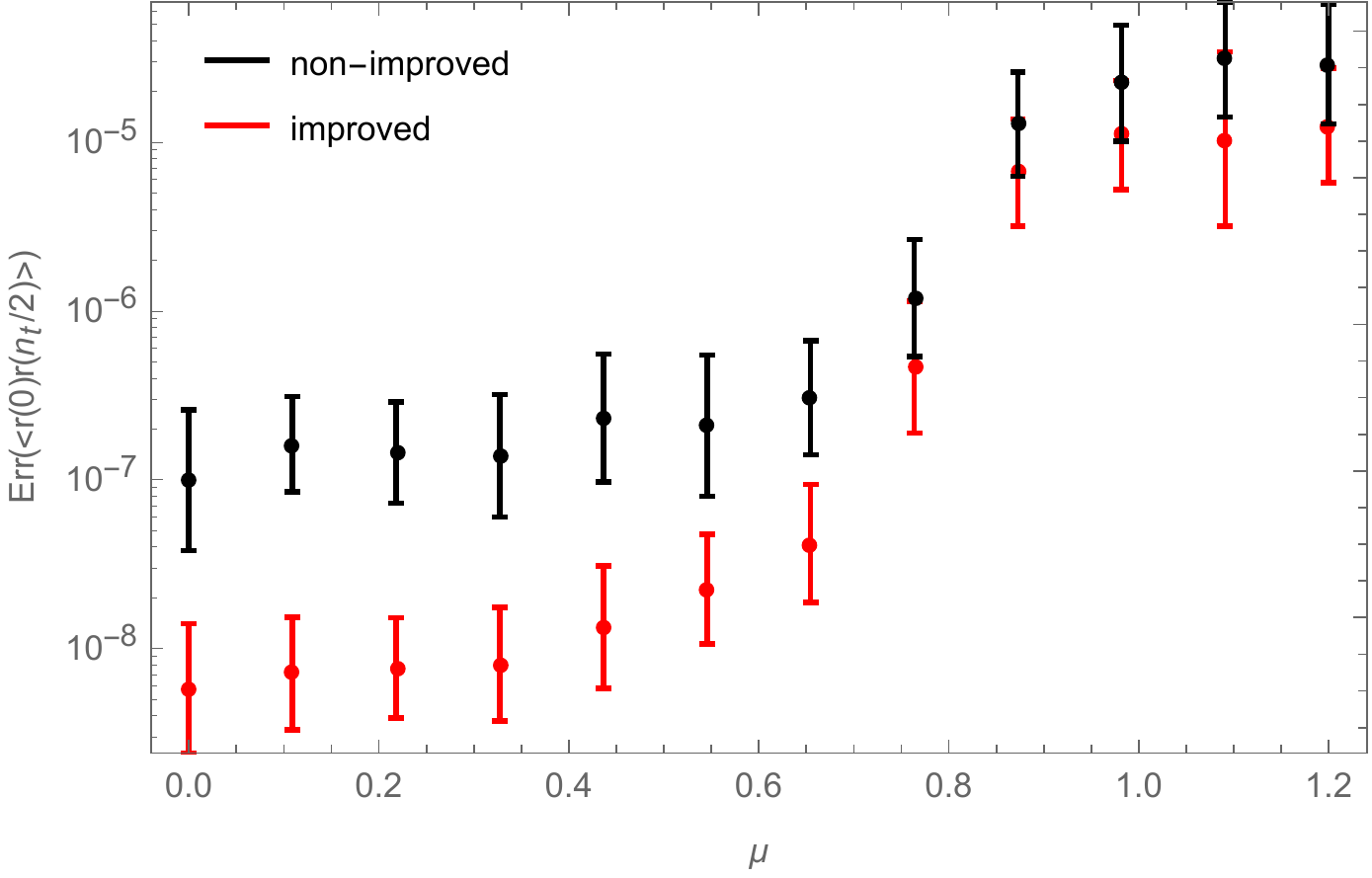}
\end{minipage}\hfill
\begin{minipage}[t]{0.49\linewidth}
\centering
\includegraphics[width=\linewidth]{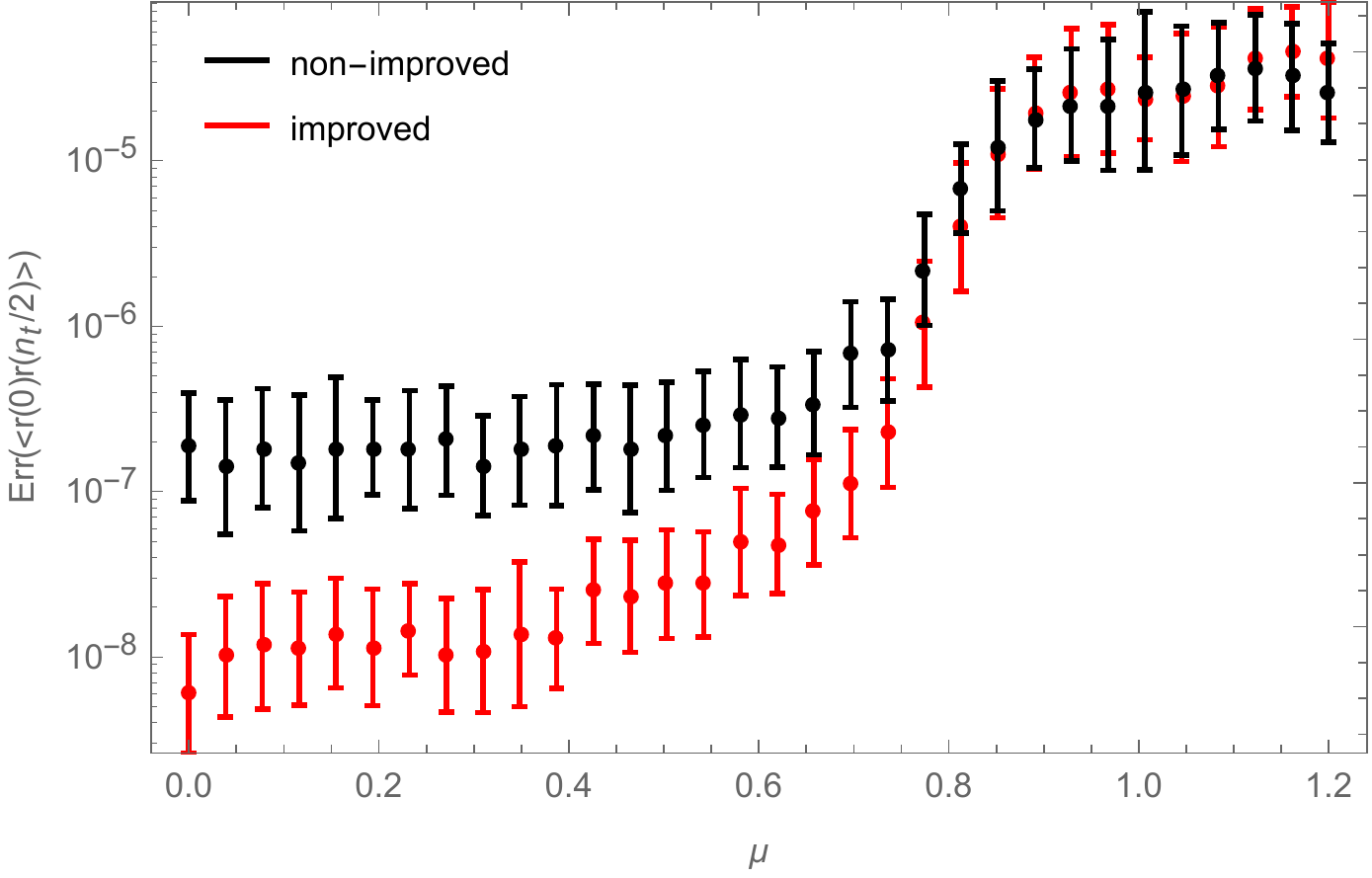}
\end{minipage}
\caption{The figure shows for a (2+1)-dimensional system with $N_{s}=8$, $\kappa=0.25$, $\lambda=7.5$, $r_s=0.01$ and $N_{t}=64$ (left) or $N_{t}=256$ (right) the \emph{error} in the zero-momentum piece of the correlator $\avof{r\of{0}r\of{t}}$ at distance $t=n_{t}/2$ as a function of $\mu$. The black data corresponds to the error in the measurements obtained with the non-improved method while the red data shows the error in the measurements obtained with the improved method.}
  \label{fig:correrror}
\end{figure}

\begin{figure}[H]
\centering
\begin{minipage}[t]{0.45\linewidth}
\centering
\includegraphics[width=\linewidth]{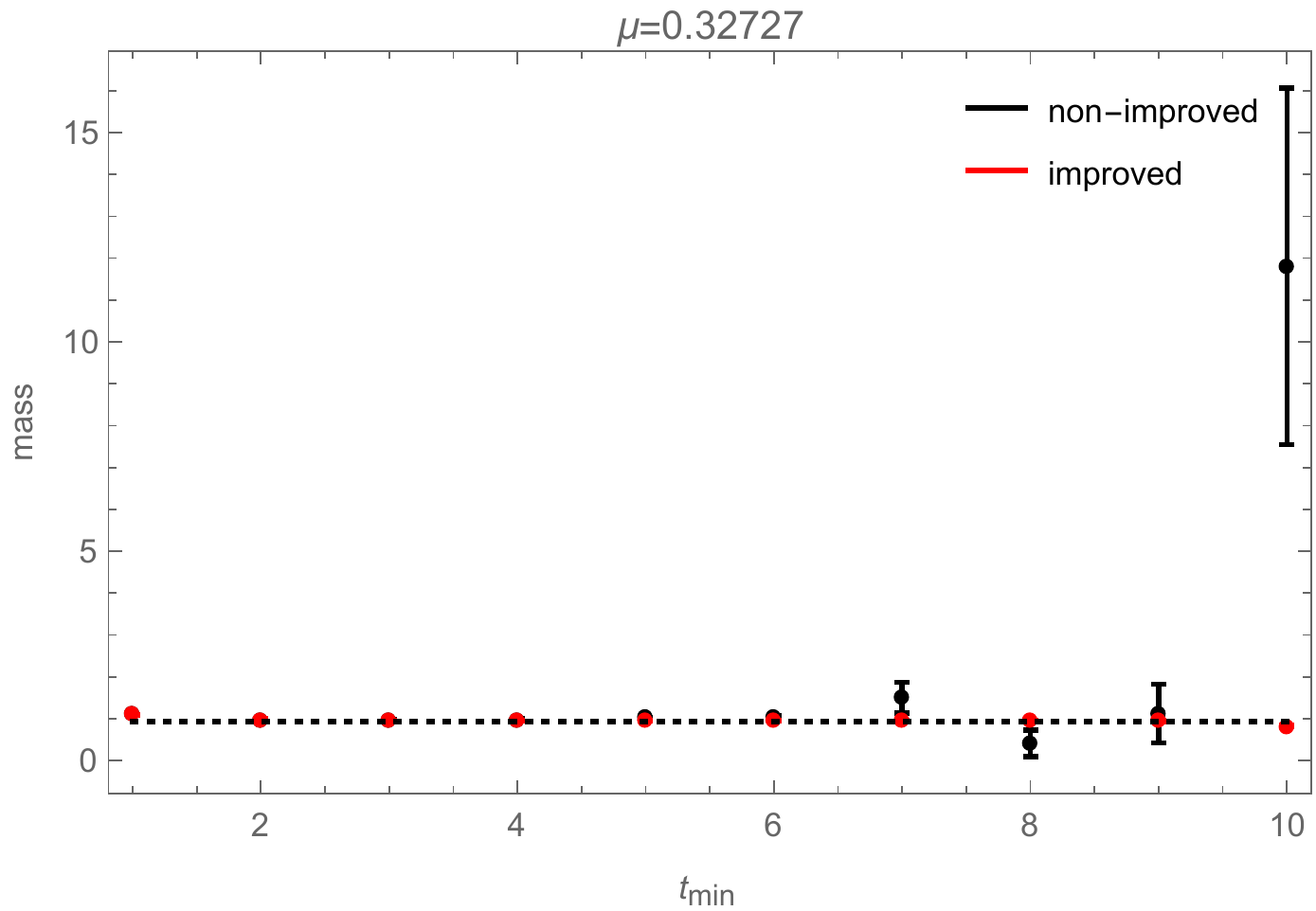}
\end{minipage}\hfill
\begin{minipage}[t]{0.45\linewidth}
\centering
\includegraphics[width=\linewidth]{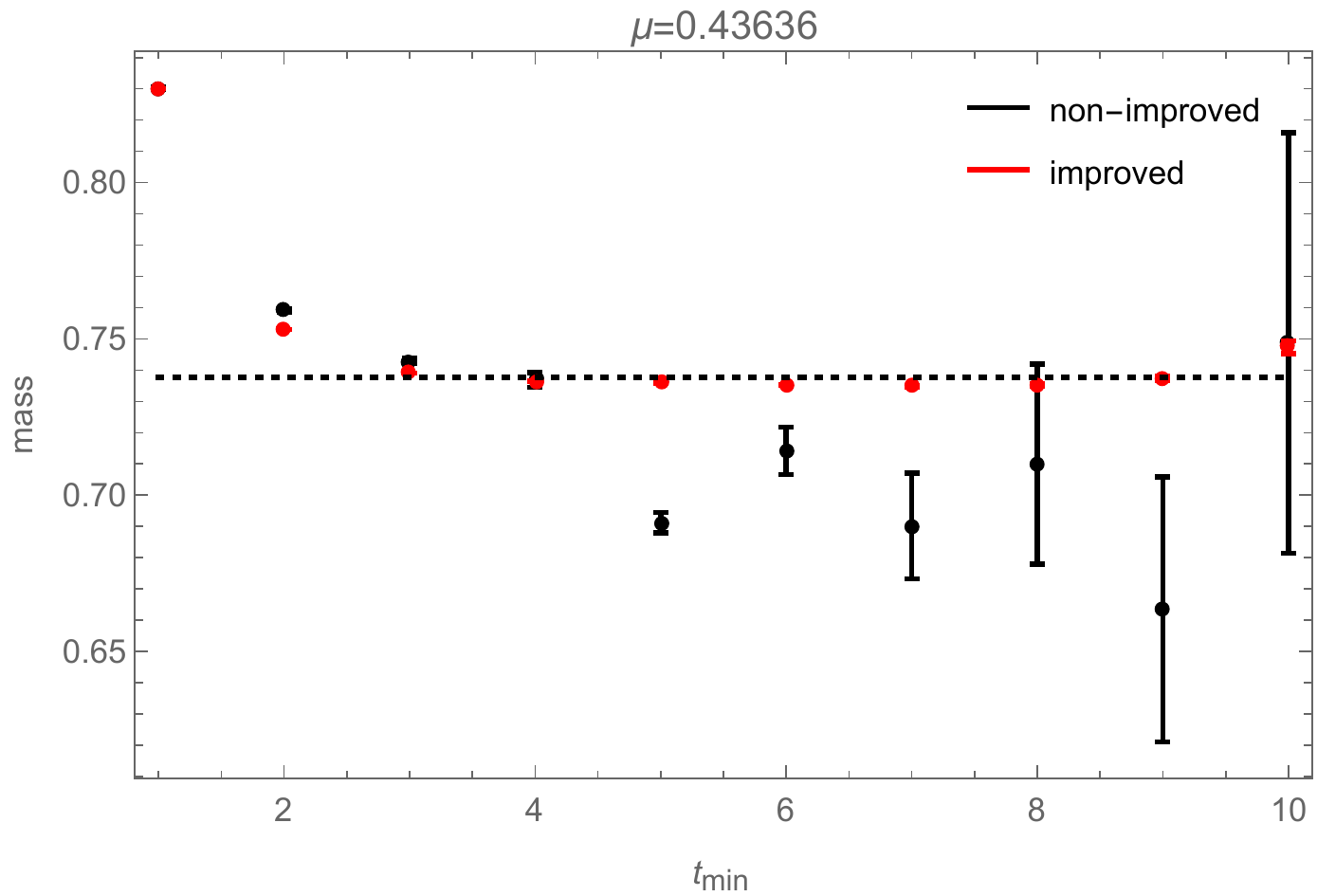}
\end{minipage}\\
\begin{minipage}[t]{0.45\linewidth}
\centering
\includegraphics[width=\linewidth]{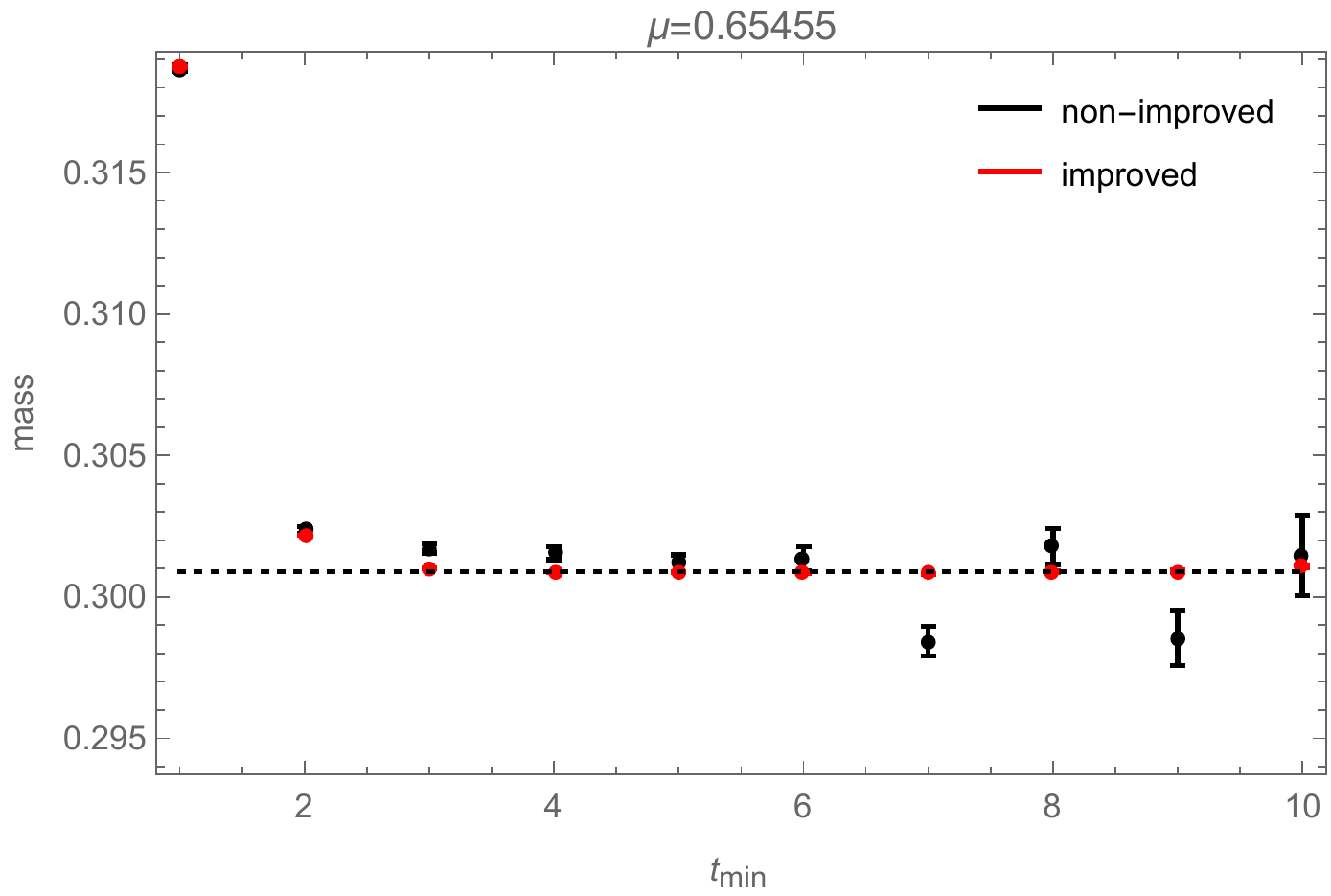}
\end{minipage}\hfill
\begin{minipage}[t]{0.45\linewidth}
\centering
\includegraphics[width=\linewidth]{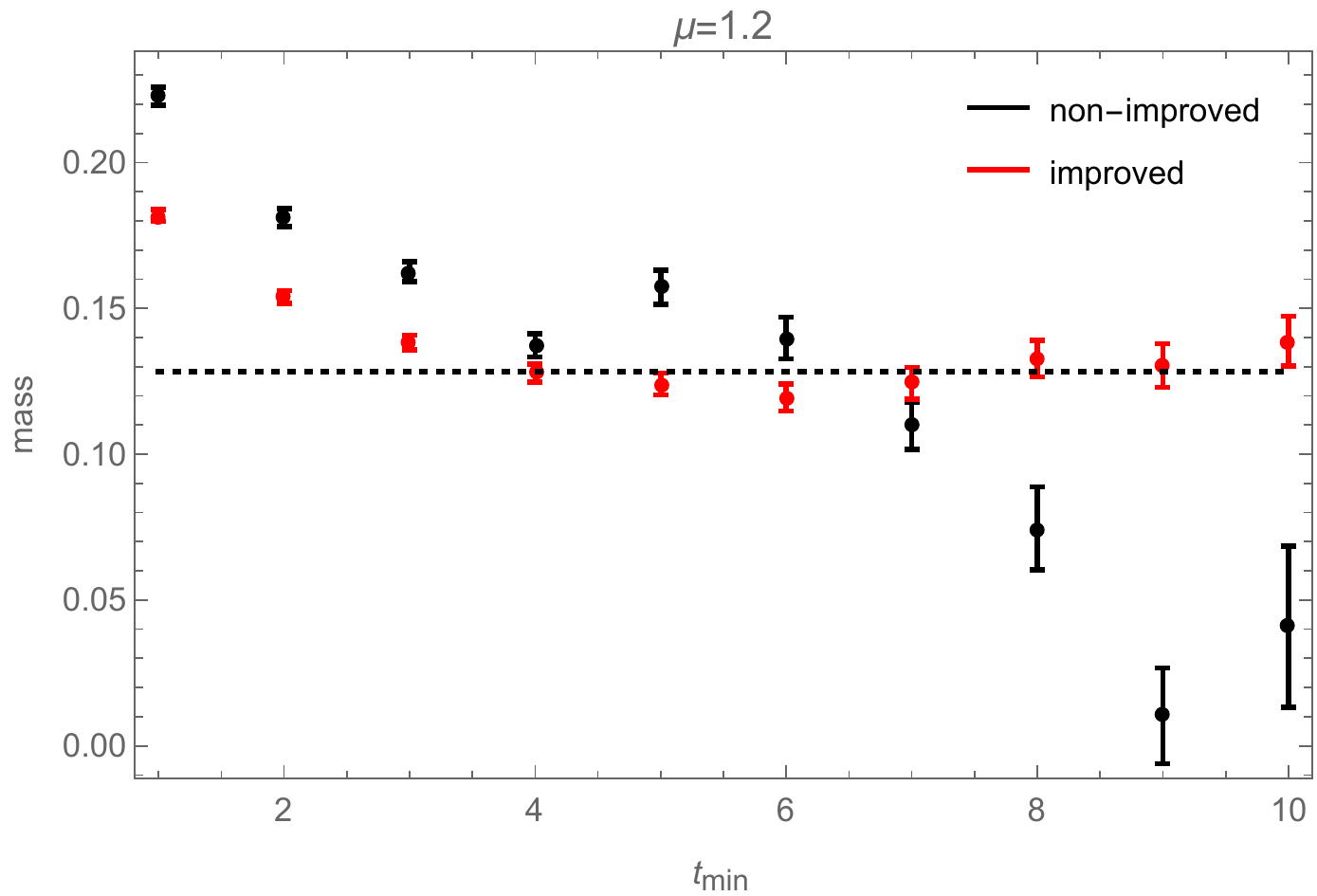}
\end{minipage}
\caption{The figure shows for a (2+1)-dimensional system with $N_{s}=8$, $N_{t}=64$, $\kappa=0.25$, $\lambda=7.5$, $r_s=0.01$ and for different values of $\mu$, the dependency of the mass $m$, obtained by fitting \eqref{eq:radfittingfunc} to the zero-momentum piece of the radial correlator $\avof{r\of{0}r\of{t}}$, on the fitting range $\fof{t_{min}, N_{t}-t_{min}}$. The red data results from fits to the correlation functions obtained with the improved method while the black data was obtained from fits to the naively measured correlators. A clear mass plateau (indicated by the dotted horizontal line) is visible with the improved method, and only with it. Note that the critical value of $\mu$ is $\mu_{c}\approx 0.8$.}
  \label{fig:massvsfitrange}
\end{figure}

In Fig.~\ref{fig:correrror} we show, for the two measurement methods, as a function of $\mu$, the \emph{errors} in the smallest component of the zero momentum piece of \eqref{eq:radialcorr1} (i.e. for $t=N_{t}/2$), before the disconnected piece is subtracted. In the disordered phase ($\mu<0.8$), the improved measurement reduces the error in this observable by about an order of magnitude. In the ordered phase on the other hand ($\mu>0.8$), the two methods seem to yield similar errors. The same is true for other components of the zero-momentum correlator, away from $t=N_{t}/2$.\\
If one is just interested in extracting a mass from \eqref{eq:radialcorr1}, the disconnected piece is irrelevant as it can be taken into account by simply adding a constant $C$ to the fitting function:
\[
f_{rad}\of{t}\,=\,C\,+\,A\,\cosh\of{m\of{t-N_{t}/2}}\ .\label{eq:radfittingfunc}
\]
Still, in Fig.~\ref{fig:massvsfitrange} we show how the reduced errors in the correlation functions, shown in Fig.~\ref{fig:correrror}, affect such a mass-fit: more precisely, Fig.~\ref{fig:massvsfitrange} shows how the fitted mass depends on the fitting range. The zero-momentum correlation functions measured during a simulation contain contributions from several states which have different masses. They all decay exponentially like $\sim\e^{- m_i\,t}$, which on a periodic lattice (and for neutral states) leads to terms like $\sim \cosh\of{m_{i}\of{t-N_{t}/2}}$. In order to extract the correct ground-state mass by fitting \eqref{eq:radfittingfunc} to the measured correlation function, one therefore either has to include additional $\cosh$-terms to take the excited states into account (which is usually difficult) or one has to restrict the fitting range to some interval $\fof{t_{min},N_{t}-t_{min}}$ in which the excited states have already died away sufficiently. Fig.~\ref{fig:massvsfitrange} now shows how the mass $m$, obtained by fitting \eqref{eq:radfittingfunc} (without excited states) to the correlation function, changes as function of $t_{min}$. For the improved correlator (black), it can be seen that the fitted mass approaches a clearly visible plateau with increasing $t_{min}$. For the non-improved correlator (red), the plateau would of course also be there, but due to the bigger noise in this correlator measurement, the ground state signal gets quickly lost with increasing $t_{min}$, sometimes even before the excited states become negligible, such that the plateau cannot be observed.\\

\subsection{Alternatives and Possible Improvements}\label{ssec:altimp}
Despite the obvious improvement in the signal-to-noise ratio provided by the method described above in Sec. \ref{ssec:gencorr} for measuring different channels of general two-point functions during different stages of a generalized worm-algorithm, this measuring at different stages of the algorithm has the potential draw-back that the different channels of the two-point function are not necessarily determined with the same accuracy: with the current algorithm, the statistics in the measurement of any component of the two-point function is proportional to its expectation value. This is fine as long as there are not too many channels in the two-point functions or if one is only interested in the dominant propagating modes. However, if one is for example interested in the full mass spectrum of the theory, including sub-dominant modes/channels, this can be problematic. A naive attempt to deal with this problem would be to use reweighting to obtain always measurements for all channels during all stages of the worm-algorithm: if the worm is for example currently sampling $\avof{\phi^{*}\of{x}\phi\of{y}}$, then measurements for $\avof{\phi\of{x}\phi\of{y}}$, $\avof{\phi^{*}\of{x}\phi^{*}\of{y}}$ and $\avof{\phi\of{x}\phi^{*}\of{y}}$ can be obtained by appropriate reweighting factors, virtually replacing the external sources at head and tail of the worm (and appropriately shifting some $p$-variables) and vice versa. It is however clear that there will in general be a big overlap problem between the different channels of the correlator.\\
The proportionality between statistics and true expectation value for the components of two-point functions which are sampled during the evolution of the worm, already leads to problems when dealing with just the standard two-point function, $\avof{\phi^{*}\of{x}\phi\of{y}}$, if one is interested in very low temperatures or if the masses are all large, such that the correlator decays very quickly. To overcome this problem it was proposed in \cite{Wolff1} to use an additional weight $\rho\of{x-y}$ in the sampling of the two-point function $\avof{\phi^{*}\of{x}\phi\of{y}}$, where $\rho\of{x-y}$ can be thought of as an initial guess for $\avof{\phi^{*}\of{x}\phi\of{y}}$, such that the algorithm has to sample only the ratio
\[
\frac{\avof{\phi^{*}\of{x}\phi\of{y}}}{\rho\of{x-y}}\ ,
\]
which, for a good choice of $\rho\of{x-y}$, should vary much less as function of $\of{x-y}$ than $\avof{\phi^{*}\of{x}\phi\of{y}}$ itself, leading to an exponential improvement in the signal-to-noise ratio. This $\rho\of{x-y}$ could of course be generalized to a set of functions $\rho_{f_1,f_2}\of{x-y}$, where $f_{1}$ and $f_{2}$ label the different possible choices for the external charges at head and tail of the worm. Furthermore, one could then use a method similar to the Wang-Landau sampling \cite{Landau1} in order to find the optimal $\rho_{f_1,f_2}\of{x-y}$, leading to equal statistics for all components and channels of the two-point function.

\subsection{Effect of Source Term}\label{ssec:effectofsource}
\begin{figure}[h!]
\centering
\begin{minipage}[t]{0.49\linewidth}
\centering
\includegraphics[width=\linewidth]{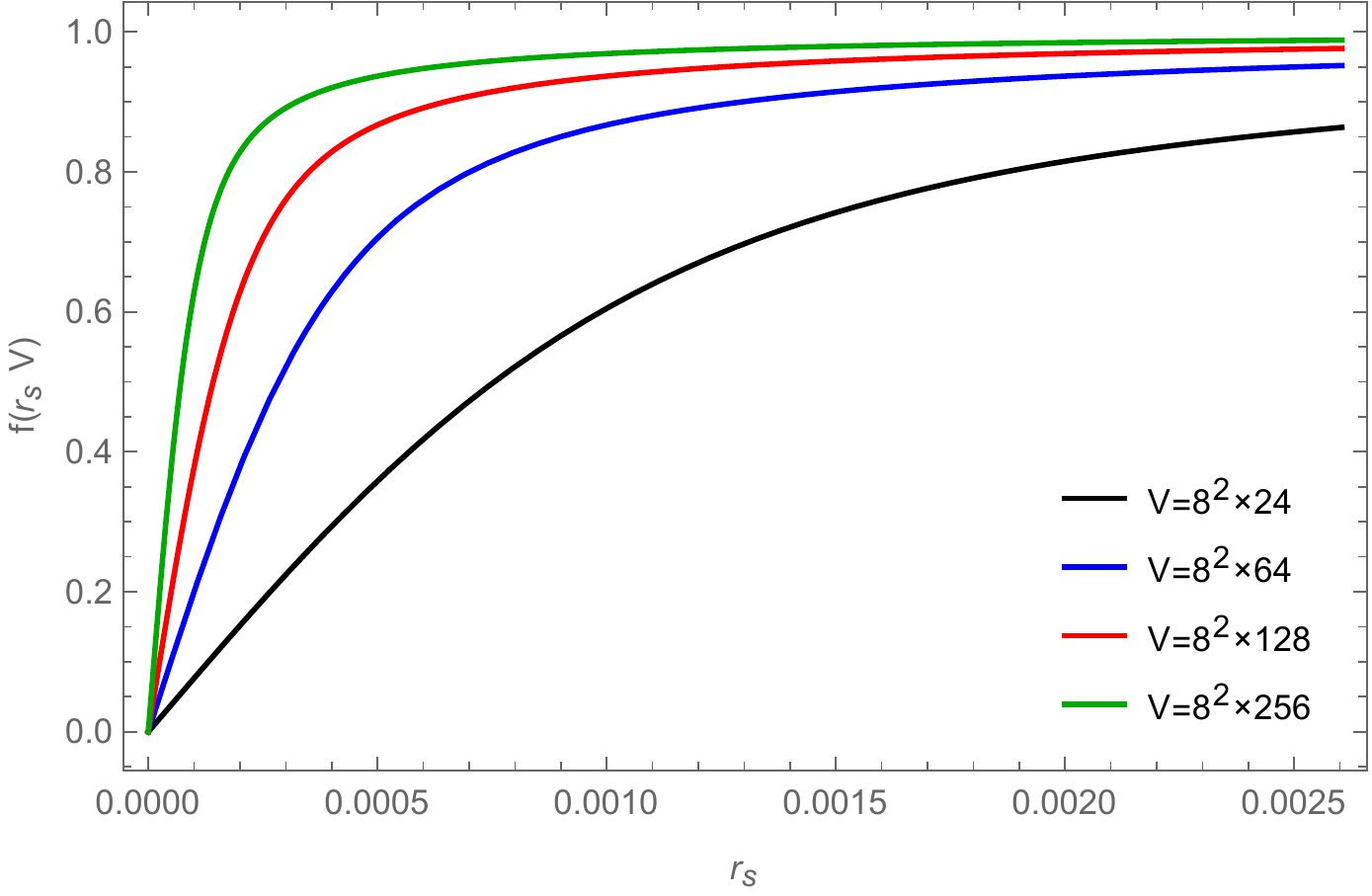}
\end{minipage}\hfill
\begin{minipage}[t]{0.49\linewidth}
\centering
\includegraphics[width=\linewidth]{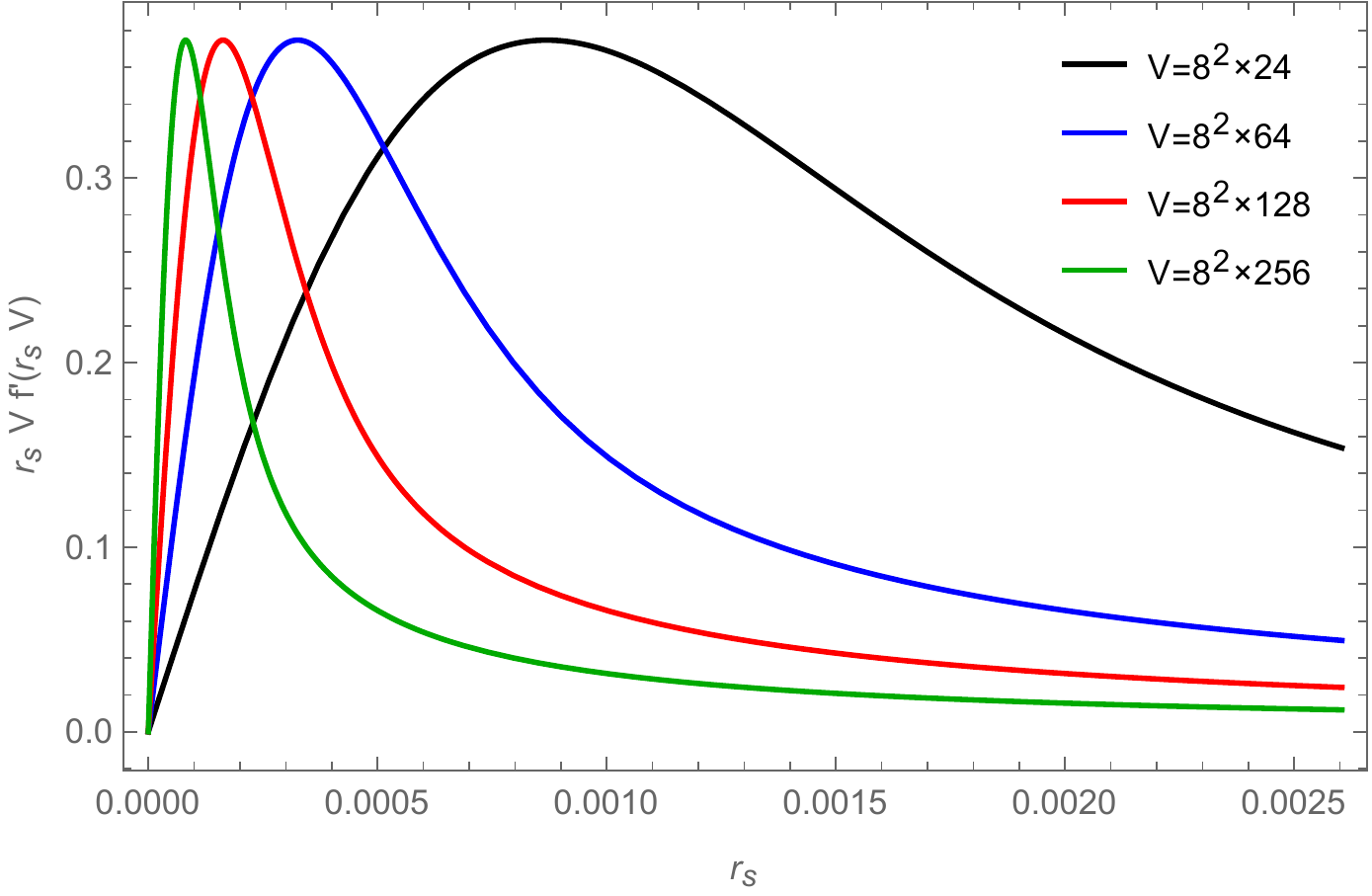}
\end{minipage}
\caption{The figure illustrates the behavior of \eqref{eq:approxordradex} (left) and \eqref{eq:approxordradsusc} (right) as a function of $r_{s}$ for four different volumes $V=8^2\cdot 24,\,8^2\cdot 64,\,8^2\cdot 128,\,8^2\cdot 256$, where we have set $\obar{r}=1$ and in the axes labels we have used the abbreviation $f\of{z}=I_{1}\of{z}/I_{0}\of{z}$. The peaks in the curves on the right-hand side mark the value of $r_{s}$ where $f\of{r_{s} V}$ stops to be proportional to $r_{s}$, and are obtained when $r_{s}\approx 4/\of{3\,V}$.}
\label{fig:besselfuncratio}
\end{figure}

As already mentioned in Sec. \ref{ssec:measurecond}, the Euclidean path integral that defines the partition function for our theory sums over all possible field configurations and thereby integrates out the global $\U{1}$ symmetry. In the dual formulation in terms of flux variables this feature of the partition function is manifest as during the dualization process, the integration over the field $\phi_{x}$ has been carried out exactly. Spontaneous symmetry breaking can therefore only be observed in our simulations either when looking at non-local observables as for example two-point functions or the charge density, $\frac{1}{2\,V}\partd{\log\of{Z}}{\mu}=\avof{\frac{1}{V}\sum_{x}k_{x,d}}$, or after the introduction of a non-vanishing source and the choice of an appropriate coordinate system for the internal space, in which one can then observe spontaneous symmetry breaking also in condensates.\\
In this section we now address the question of how big the source has to be chosen and how one can extract results for the theory at zero source from simulations carried out in the presence of a non-zero source. For this discussion, we will use the partition function \eqref{eq:phifourpartf} in terms of the original field variables in polar form: 
\[
Z=\int\DD{r,\theta}\e^{\sum\limits_{x}\scof{\kappa\,\sum\limits_{\nu=1}^{d}\,r_{x}r_{x+\hat{\nu}}\cos\of{\theta_{x}-\theta_{x+\hat{\nu}}-2\,\ii\,\mu\,\delta_{\nu,d}}-r_{x}^{2}-\lambda\of{r_{x}^{2}-1}^{2}+r_{s}r_{x}\cos\of{\theta_{x}-\theta_{s}}}}\ .\label{eq:origpartfsph}
\]
In these variables the condensate for radial excitations \eqref{eq:radderiv2} then reads
\begin{multline}
\avof{r}\,=\,\frac{1}{V}\partd{\log\of{Z}}{r_{s}}\,=\,\frac{1}{V\,Z}\int\DD{r,\theta}\sof{{\color{fhl}\sum\limits_{x}r_{x}\cos\of{\theta_{x}-\theta_{s}}}}\\
\cdot\e^{\sum\limits_{x}\scof{\kappa\,\sum\limits_{\nu=1}^{d}\,r_{x}r_{x+\hat{\nu}}\cos\of{\theta_{x}-\theta_{x+\hat{\nu}}-2\,\ii\,\mu\,\delta_{\nu,d}}-r_{x}^{2}-\lambda\of{r_{x}^{2}-1}^{2}+r_{s}r_{x}\cos\of{\theta_{x}-\theta_{s}}}}\ ,\label{eq:origcondrad}
\end{multline}
and the cosine in the observable under the integral makes it clear that this will only yield the expected result in the ordered phase if $r_{s}$ is sufficiently large in order to give dominant weight to the $\phi$-configurations where $\theta_{x}\sim\theta_{s}$ when integrating over the phases $\theta_{x}$. On the other hand, we would like $r_{s}$ to be as small as possible in order to reduce the associated bias. To get an approximate answer on how large $r_{s}$ should be chosen, we can proceed as follows: we know that in the ordered phase, the exponential of the interaction term ("interaction" in the sense of coupling fields on neighboring sites),
\[
\e^{\kappa\sum\limits_{x}\sum\limits_{\nu=1}^{d}r_{x}r_{x+\hat{\nu}}\cos\of{\theta_{x}-\theta_{x+\hat{\nu}}-\,2\,\ii\,\mu\,\delta_{\nu,d}}}\ ,\label{eq:nnintterm}
\]
dominates over the entropy in the configuration space of the $\theta_{x}$ variables and the dominant contribution to the integral over these variables will come from configurations where all $\theta_{x}$ are parallel or, as we called it earlier, "in phase", i.e. we can assume that in the ordered phase, the partition function \eqref{eq:origpartfsph} behaves like
\begin{multline}
Z_{ord}=\int\DD{r}\int\idd{\theta}{}\e^{\sum\limits_{x}\scof{\kappa\,\sum\limits_{\nu=1}^{d}\,r_{x}r_{x+\hat{\nu}}\cosh\of{2\,\mu\,\delta_{\nu,d}}-r_{x}^{2}-\lambda\of{r_{x}^{2}-1}^{2}+r_{s}r_{x}\cos\of{\theta-\theta_{s}}}}\\
=\int\DD{r}\e^{\sum\limits_{x}\scof{\kappa\,\sum\limits_{\nu=1}^{d}\,r_{x}r_{x+\hat{\nu}}\cosh\of{2\,\mu\,\delta_{\nu,d}}-r_{x}^{2}-\lambda\of{r_{x}^{2}-1}^{2}}}I_{0}\of{r_{s}\sum_{x}r_{x}}\ ,\label{eq:approxorderedpartf}
\end{multline}
with $I_{\nu}\of{x}$ being the modified Bessel functions of the first kind. If we now evaluate \eqref{eq:origcondrad} by using \eqref{eq:approxorderedpartf}, we find
\begin{multline}
\avof{r}\sim\frac{1}{V}\partd{\log\of{Z_{ord}}}{r_{s}}\,=\,\frac{1}{Z_{ord}}\int\DD{r}\frac{\of{\sum_{x}r_{x}}}{V}\frac{I_{1}\of{r_{s}\sum_{x}r_{x}}}{I_{0}\of{r_{s}\sum_{x}r_{x}}}\\
\cdot\e^{\sum\limits_{x}\scof{\kappa\,\sum\limits_{\nu=1}^{d}\,r_{x}r_{x+\hat{\nu}}\cosh\of{2\,\mu\,\delta_{\nu,d}}-r_{x}^{2}-\lambda\of{r_{x}^{2}-1}^{2}}}I_{0}\of{r_{s}\sum\limits_{x}r_{x}}\approx \bar{r}\,\frac{I_{1}\of{r_{s}\,V\,\bar{r}}}{I_{0}\of{r_{s}\,V\,\bar{r}}}\ ,\label{eq:approxordradex}
\end{multline}
where $\bar{r}$ is the mean field solution to \eqref{eq:approxordradex} and $V$ is the system volume. The ratio of Bessel functions in \eqref{eq:approxordradex} vanishes if $r_{s}\,V\,\bar{r}$ is zero and asymptotically approaches $1$ for $\of{r_{s}\,V\,\bar{r}\rightarrow \infty}$ as shown in the left panel of Fig.~\ref{fig:besselfuncratio} as a function of $r_{s}$. 

In the thermodynamic limit $\of{V\rightarrow\infty}$, a non-zero value of $r_{s}$ leads to
\[
\frac{I_{1}\of{r_{s}\,V\,\bar{r}}}{I_{0}\of{r_{s}\,V\,\bar{r}}}\underset{\of{V\rightarrow\infty}}{\longrightarrow} 1\ ,
\]
no matter how small $r_{s}$ is. However, in practice we can simulate only finite systems and the thermodynamic limit has to be obtained by extrapolation (finite volume scaling). If we are interested in results for $\avof{r}$ in an infinite system at $r_{s}=0$, \eqref{eq:approxordradex} reminds us that one first has to take the infinite volume limit before sending $\of{r_{s}\rightarrow 0}$; otherwise the result will always be zero, since for every finite volume there is a minimal value for $r_{s}$ below which $\of{r_{s}\,V\,\bar{r}}$ will be in the region where $\frac{I_{1}\of{r_{s}\,V\,\bar{r}}}{I_{0}\of{r_{s}\,V\,\bar{r}}}$ is proportional to $r_{s}$. The value of $r_{s}$ at which this change of behavior of $\frac{I_{1}\of{r_{s}\,V\,\bar{r}}}{I_{0}\of{r_{s}\,V\,\bar{r}}}$ happens is given by the peak in the quantity
\[
r_{s}\,\partd{}{r_{s}}\bof{\frac{I_{1}\of{r_{s}\,V\,\bar{r}}}{I_{0}\of{r_{s}\,V\,\bar{r}}}}\label{eq:approxordradsusc}
\]
when plotted as a function of $r_{s}$ and we can define the volume dependent lower bound for $r_{s}$:
\[
r_{s}^{lb}\of{V}\,=\,\max_{r_{s}>0}\bof{r_{s}\,\partd{}{r_{s}}\bof{\frac{I_{1}\of{r_{s}\,V\,\bar{r}}}{I_{0}\of{r_{s}\,V\,\bar{r}}}}}\ ,
\]
which scales like
\[
r_{s}^{lb}\of{V}\propto\frac{1}{V}\ ,
\]
as could have been expected since $r_{s}$ couples to an extensive quantity in the action \eqref{eq:phifouractionsph}. In the full theory $r_{s}^{lb}\of{V}$ would be determined by the maximum in the susceptibility
\[
\chi_{r}\,=\,\frac{r_{s}}{V}\partdm{\log\of{Z}}{r_{s}}{2}\label{eq:suscdef}
\] 
as a function of $r_{s}$. The values of $r_{s}$ used for simulations that should yield results to be used in a scaling analysis to extract information on an infinite system at zero source, should therefore be chosen such that $r_{s}\gg r_{s}^{lb}\of{V_{min}}$, where $V_{min}$ is the volume of the smallest system involved in the analysis.\\

Deep in the disordered phase on the other hand, entropy dominates over the exponential of the interaction term \eqref{eq:nnintterm} and we can approximate the partition function \eqref{eq:origpartfsph} as follows:
\begin{multline}
Z_{dis}=\int\DD{r}\prod\limits_{x}\bof{\int\idd{\theta_{x}}{}\e^{-r_{x}^{2}-\lambda\of{r_{x}^{2}-1}^{2}+r_{s}r_{x}\cos\of{\theta_{x}-\theta_{s}}}}\\
=\int\DD{r}\prod\limits_{x}\bof{\e^{-r_{x}^{2}-\lambda\of{r_{x}^{2}-1}^{2}}I_{0}\of{r_{s}r_{x}}}=\bof{\int\idd{r}{}\e^{-r^{2}-\lambda\of{r^{2}-1}^{2}}I_{0}\of{r_{s}r}}^{V}\ .\label{eq:approxdisorderedpartf}
\end{multline}
For the condensate we then find
\[
\avof{r}\sim\frac{1}{V}\partd{\log\of{Z_{dis}}}{r_{s}}\,=\,\frac{\int\idd{r}{}\,r\,I_{1}\of{r_{s}r}\e^{-r^{2}-\lambda\of{r^{2}-1}^{2}}}{\int\idd{r}{}I_{0}\of{r_{s}r}\e^{-r^{2}-\lambda\of{r^{2}-1}^{2}}}\sim\bar{r}\,\frac{I_{1}\of{r_{s}\bar{r}}}{I_{0}\of{r_{s}\bar{r}}}\approx \frac{r_{s}\bar{r}^{2}}{2}+\order{r_{s}^{3}}\ .
\]
In the disordered phase the condensate is therefore expected to behave always like $\avof{r}\propto r_{s}$ unless $r_{s}\gg 1$ as there is now no volume factor in the argument of the Bessel functions. Due to this direct proportionality between condensate and source, it also follows that the susceptibility $\chi_{r}$ defined in eq. \eqref{eq:suscdef} should (to lowest order in $r_{s}$) be identical to the condensate.\\

\begin{figure}[h!]
\centering
\begin{minipage}[t]{0.49\linewidth}
\centering
\includegraphics[width=\linewidth]{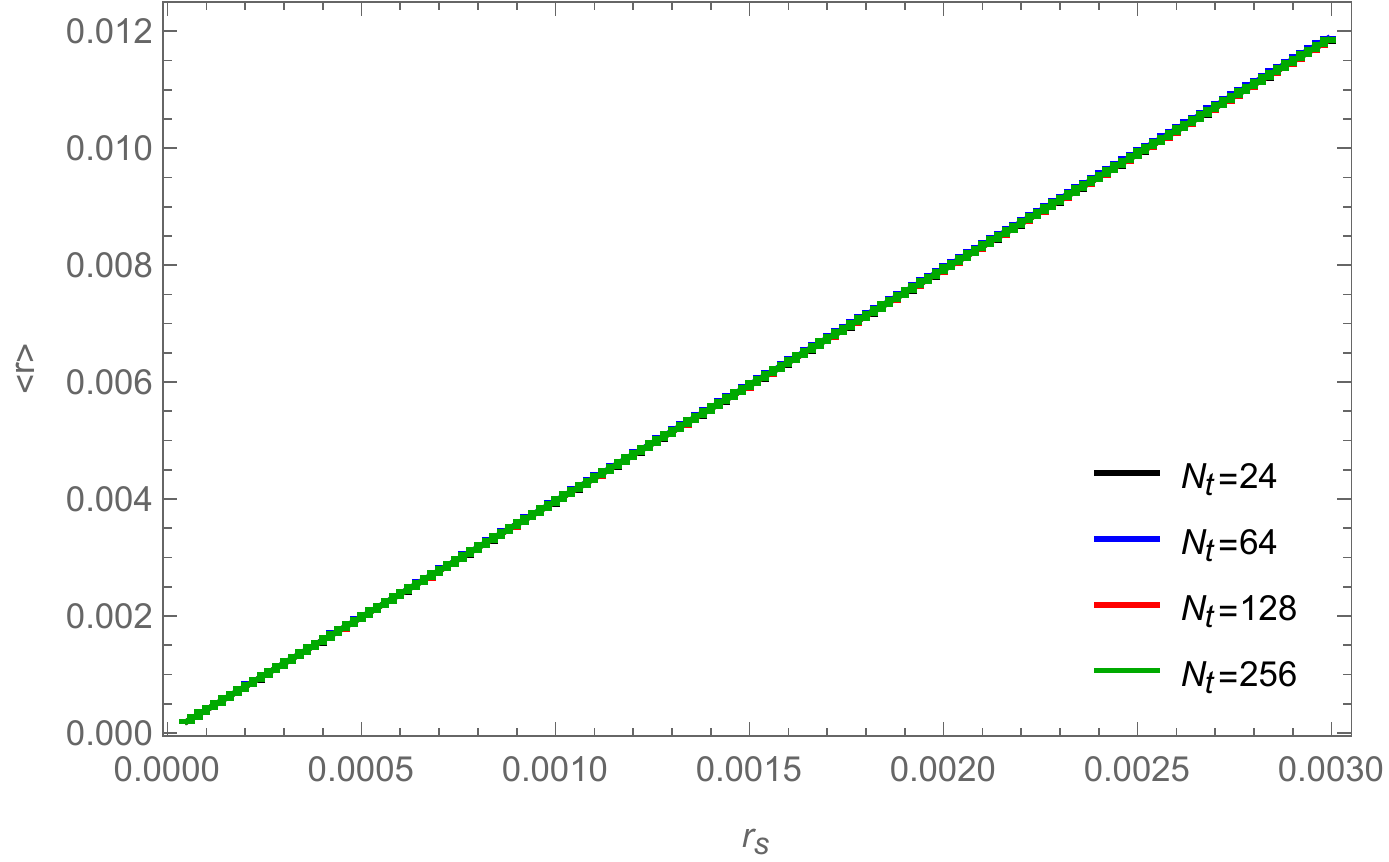}
\end{minipage}\hfill
\begin{minipage}[t]{0.49\linewidth}
\centering
\includegraphics[width=\linewidth]{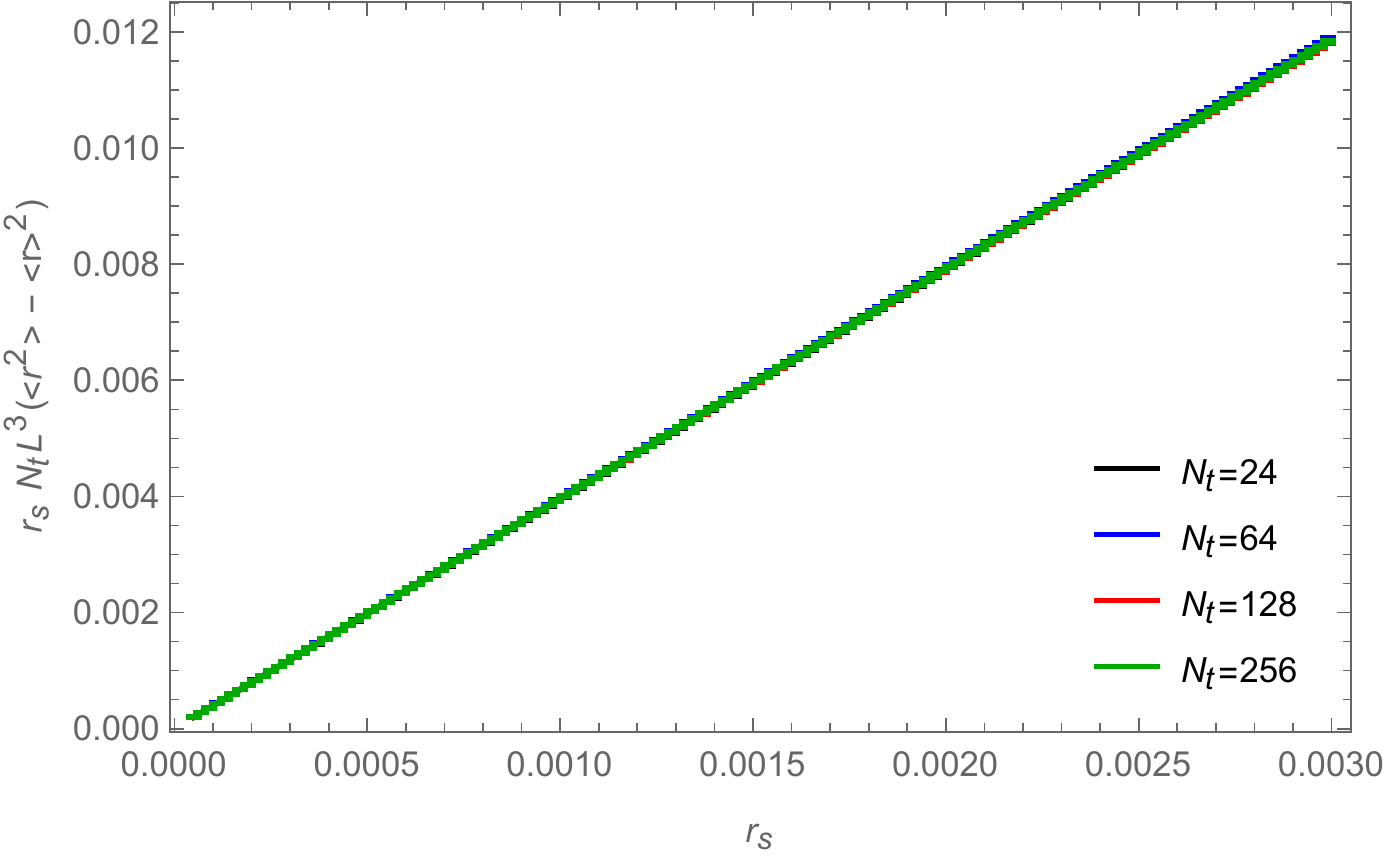}
\end{minipage}\\
\begin{minipage}[t]{0.49\linewidth}
\centering
\includegraphics[width=\linewidth]{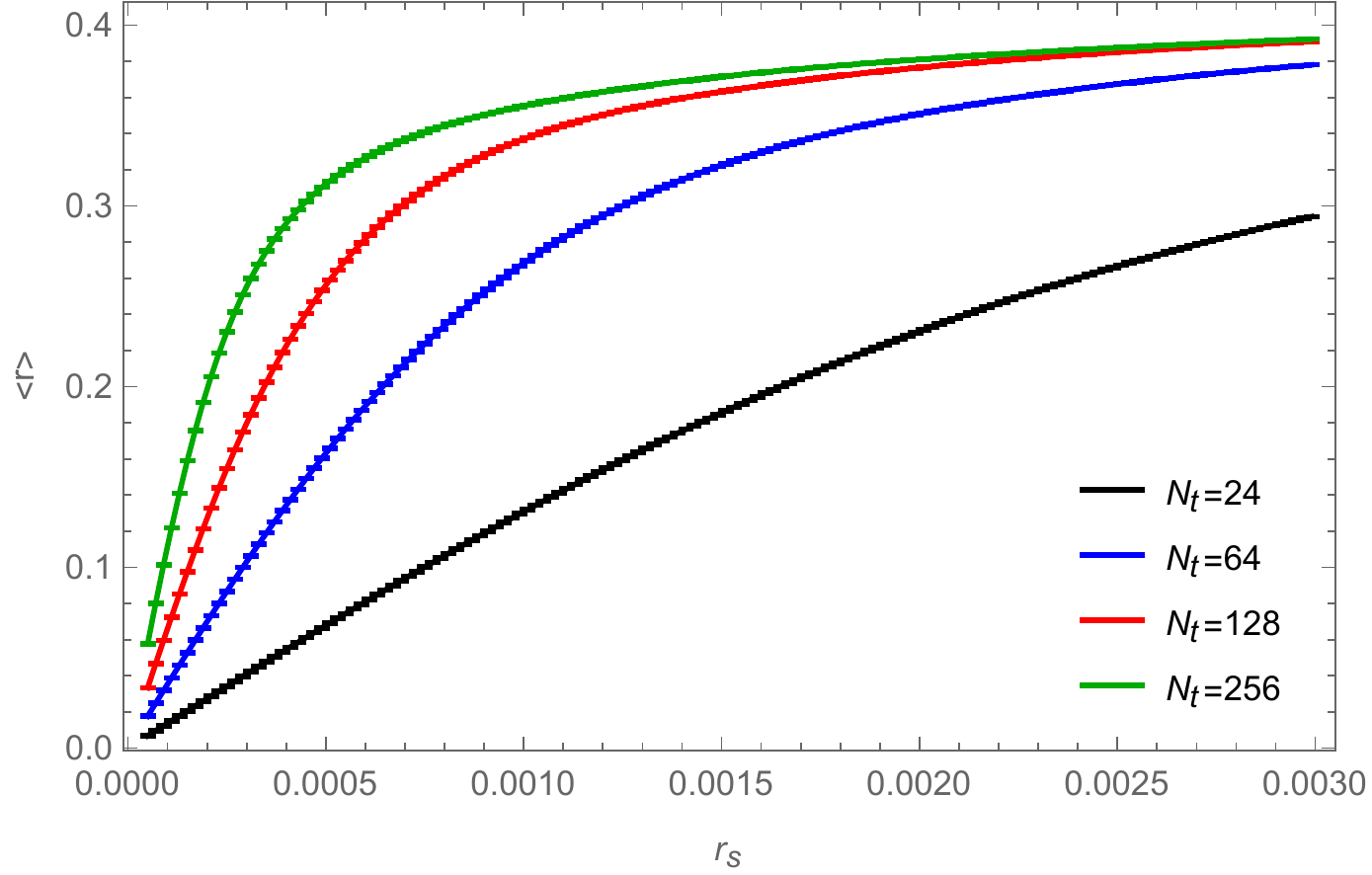}
\end{minipage}\hfill
\begin{minipage}[t]{0.49\linewidth}
\centering
\includegraphics[width=\linewidth]{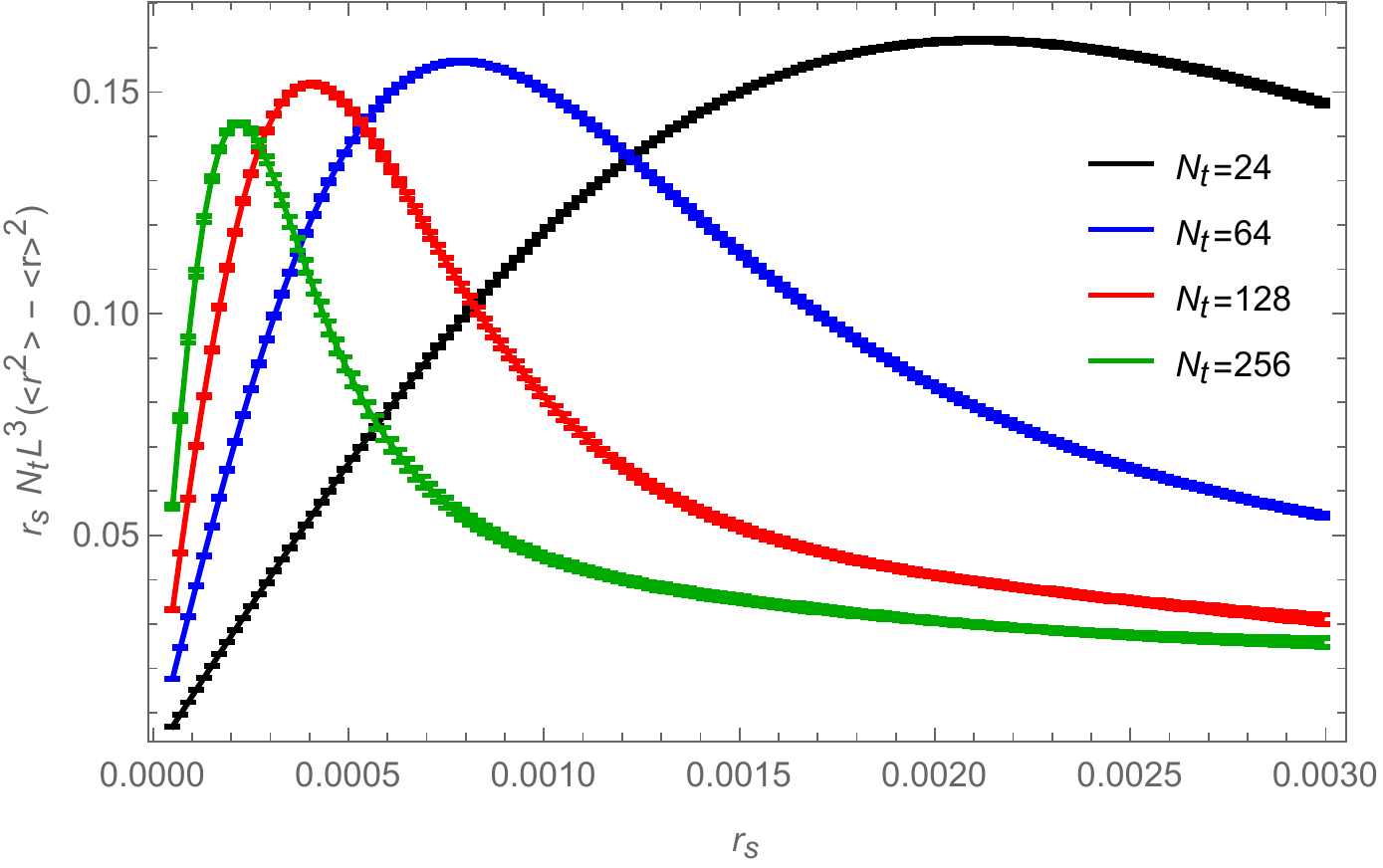}
\end{minipage}\\
\begin{minipage}[t]{0.49\linewidth}
\centering
\includegraphics[width=\linewidth]{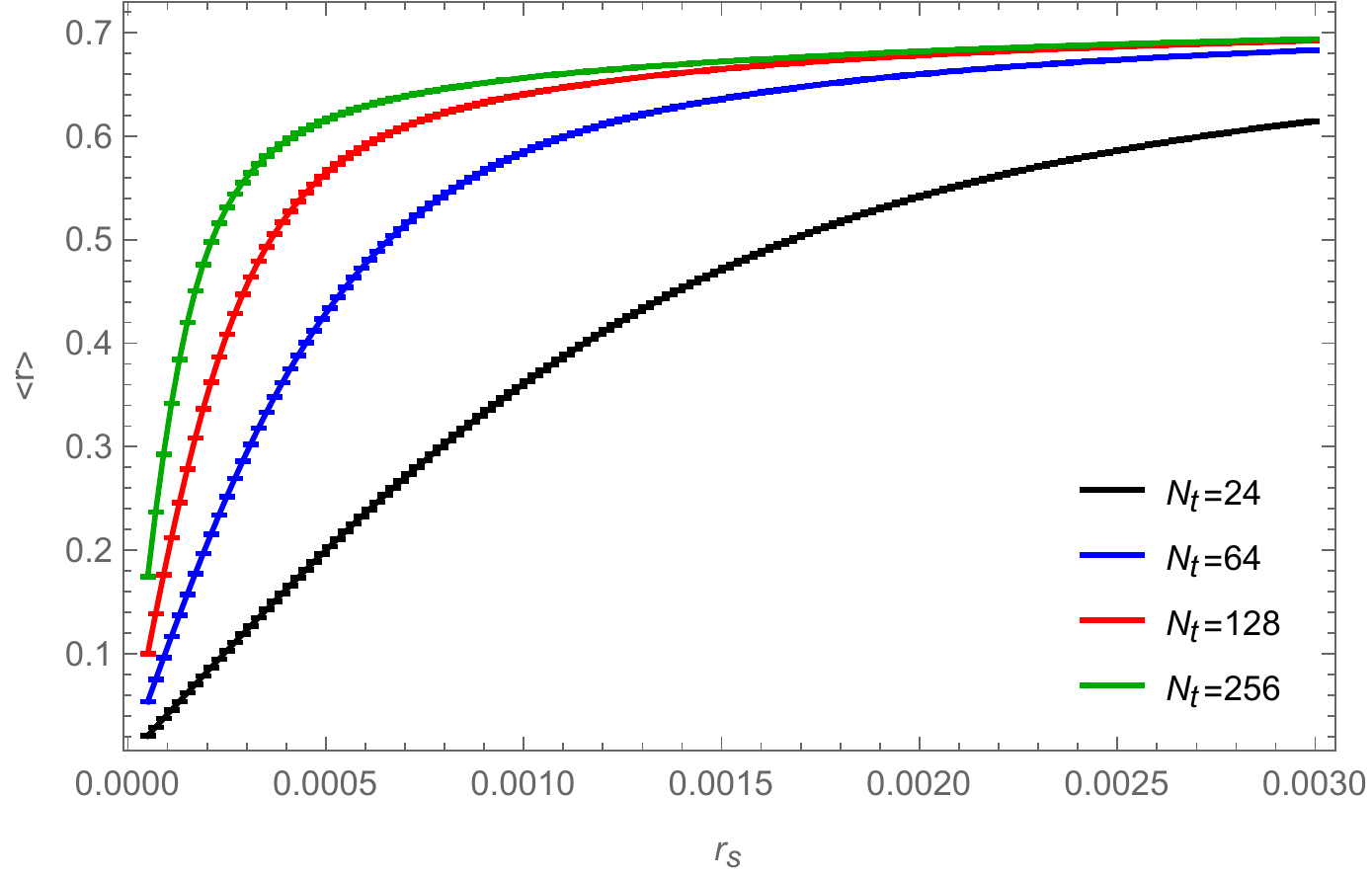}
\end{minipage}\hfill
\begin{minipage}[t]{0.49\linewidth}
\centering
\includegraphics[width=\linewidth]{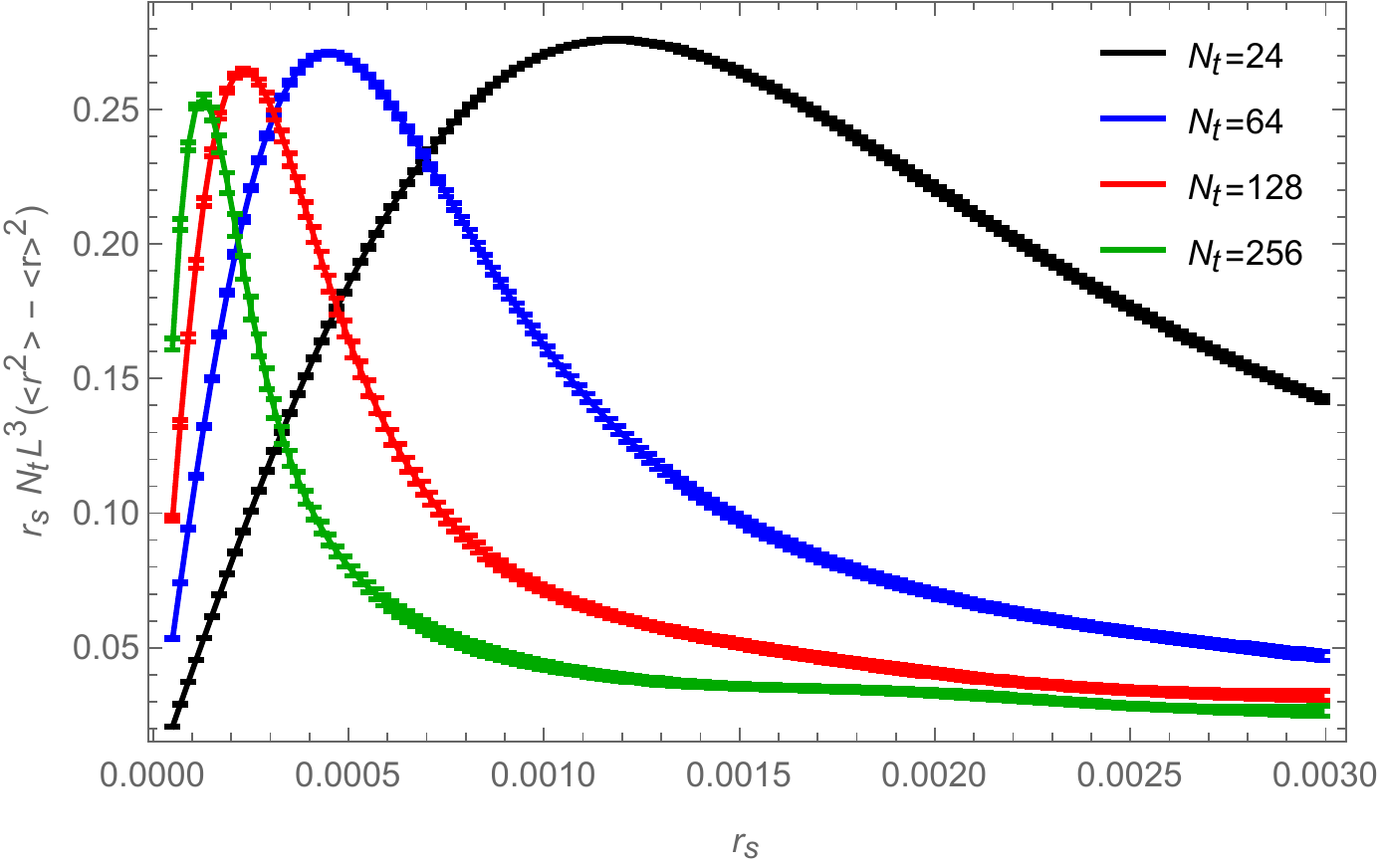}
\end{minipage}
\caption{The figure shows for a (2+1)-dimensional system with $N_{s}=8$, $\kappa=0.25$ and $\lambda=7.5$ the condensate $\avof{r}$ (left) and the corresponding susceptibility (right) as a function of the magnitude of the source $r_{s}$ for $N_{t}=24,64,182,256$ in the disordered phase at $\mu=0.686$ (top), just in the ordered phase at $\mu=0.857$ (middle) and deep in the ordered phase at $\mu=1.2$ (bottom). The continuous curves were obtained by multi-histogram reweighting.}
  \label{fig:condvssource}
\end{figure}

In Fig.~\ref{fig:condvssource} we show for a (2+1)-dimensional system with $\kappa=0.25$, $\lambda=7.5$, $N_{s}=8$, $N_{t}=24,\,64,\,128,\,256$ and for three different values of the chemical potential $\mu$ the condensate $\avof{r}$ (left) and the corresponding susceptibility $\chi_{r}$ (right) as a function of the source $r_{s}$. The top figures correspond to $\mu=0.686$, which is in the disordered phase, where condensate and susceptibility are both proportional to the source $r_{s}$ and independent of the system size. The other figures correspond to $\mu=0.857$ (middle) and $\mu=1.2$ (bottom) which are both in the ordered phase (since $\mu_{c}\approx 0.8$) where the condensate develops a plateau as soon as $r_{s}$ reaches a volume-dependent minimal value.\\

It should be mentioned at this point that if one is only interested in the magnitude of the condensate, an alternative to the introduction of a source for measuring it would be to use that
\[
\avof{\phi^{*}\of{x}\phi\of{y}}\underset{\abs{x-y}\rightarrow \infty}{\longrightarrow} \abs{\avof{\phi}}^{2}\ .\label{eq:chargedcorr3}
\]
The same equation also holds for the correlator of radial excitations \eqref{eq:radcorr2} (where for $\abs{s}=0$ only the first two terms in \eqref{eq:radcorr2} are non-zero):
\[
\avof{r\of{x}r\of{y}}\underset{\abs{x-y}\rightarrow \infty}{\longrightarrow} \avof{r}^{2}\,=\,2\,\abs{\avof{\phi}}^{2}\ . \label{eq:radexcorr}
\]
The latter is particularly convenient: as we are usually simulating periodic finite systems, zero-momentum correlation functions always have a minimum, which is the point where the deviation of the value of the correlation function from the square of the corresponding condensate is minimal. For the zero-momentum piece of \eqref{eq:radexcorr} the minimum is always located at $N_{t}/2$, while for the zero-momentum piece of \eqref{eq:chargedcorr3}, its location can change as function of $\mu$. However, since for vanishing source $r_{s}=\sqrt{2}\abs{s}=0$ only the first two terms of \eqref{eq:radcorr2} can be measured, while the other two terms (the ones that cannot be measured) would contribute similarly to the constant piece of the zero-momentum correlator, one has to keep in mind that the measured background should be multiplied by a factor $2$ in order to get the correct value for the squared condensate $\avof{r}^{2}$.\\ 

\begin{figure}[h!]
\centering
\includegraphics[width=0.6\linewidth]{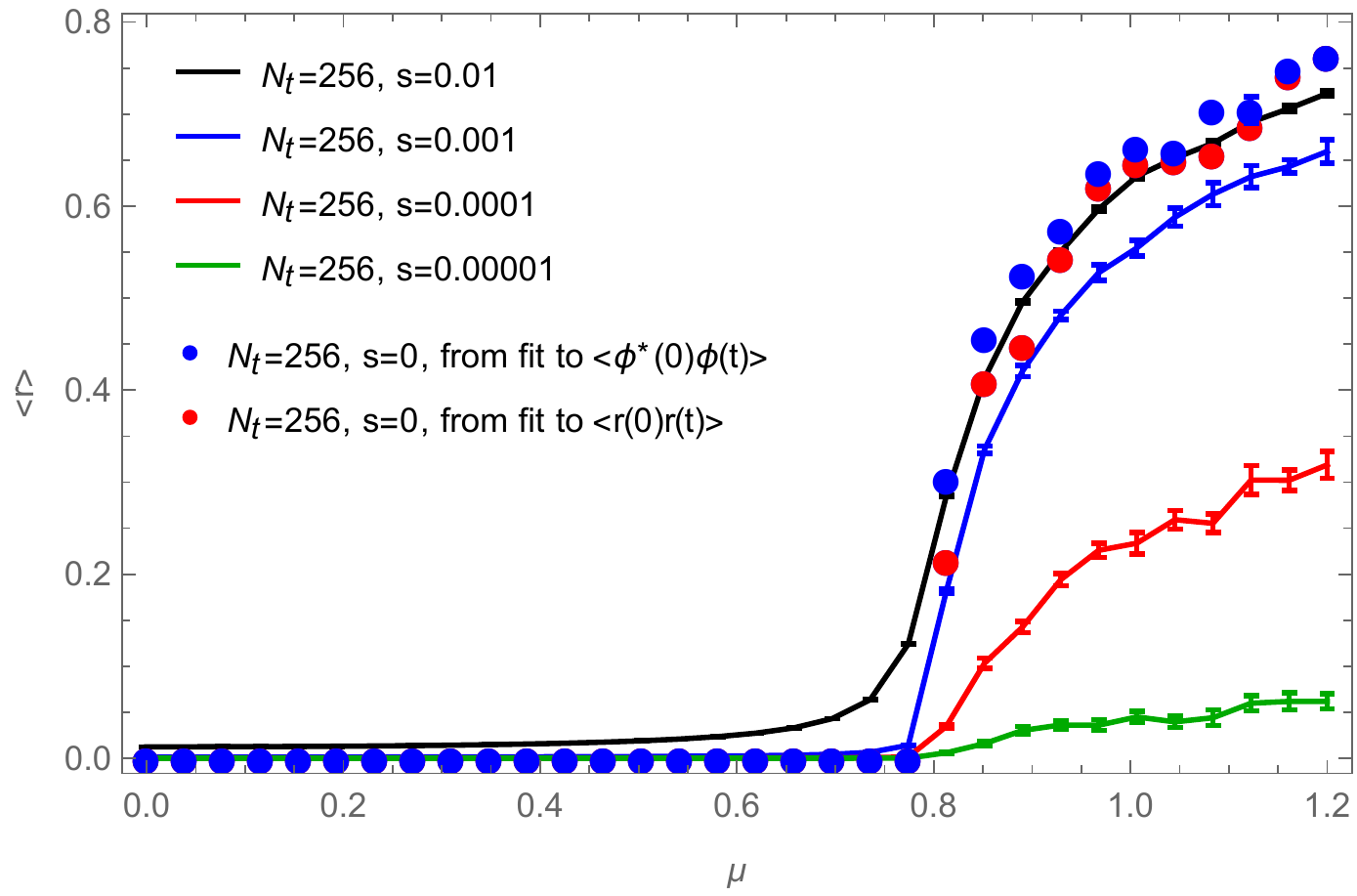}
\caption{The figure shows for a (2+1)-dimensional system with $N_{s}=8$, $N_{t}=256$, $\kappa=0.25$ and $\lambda=7.5$ the condensate $\avof{r}$ measured in three different ways: the data points which are joined by different lines correspond to direct measurements of the condensate for different values of the source $r_s=0.01, 0.001, 0.0001, 0.00001$, the blue and red dots were obtained by fitting \eqref{eq:condfitfunc} to the correlators for charged \eqref{eq:chargedcorr2} and radial \eqref{eq:radcorr2} excitations respectively for $r_s=0$.}
\label{fig:condfromcorr}
\end{figure}

\begin{figure}[h!]
\centering
\begin{minipage}[t]{0.49\linewidth}
\centering
\includegraphics[width=\linewidth]{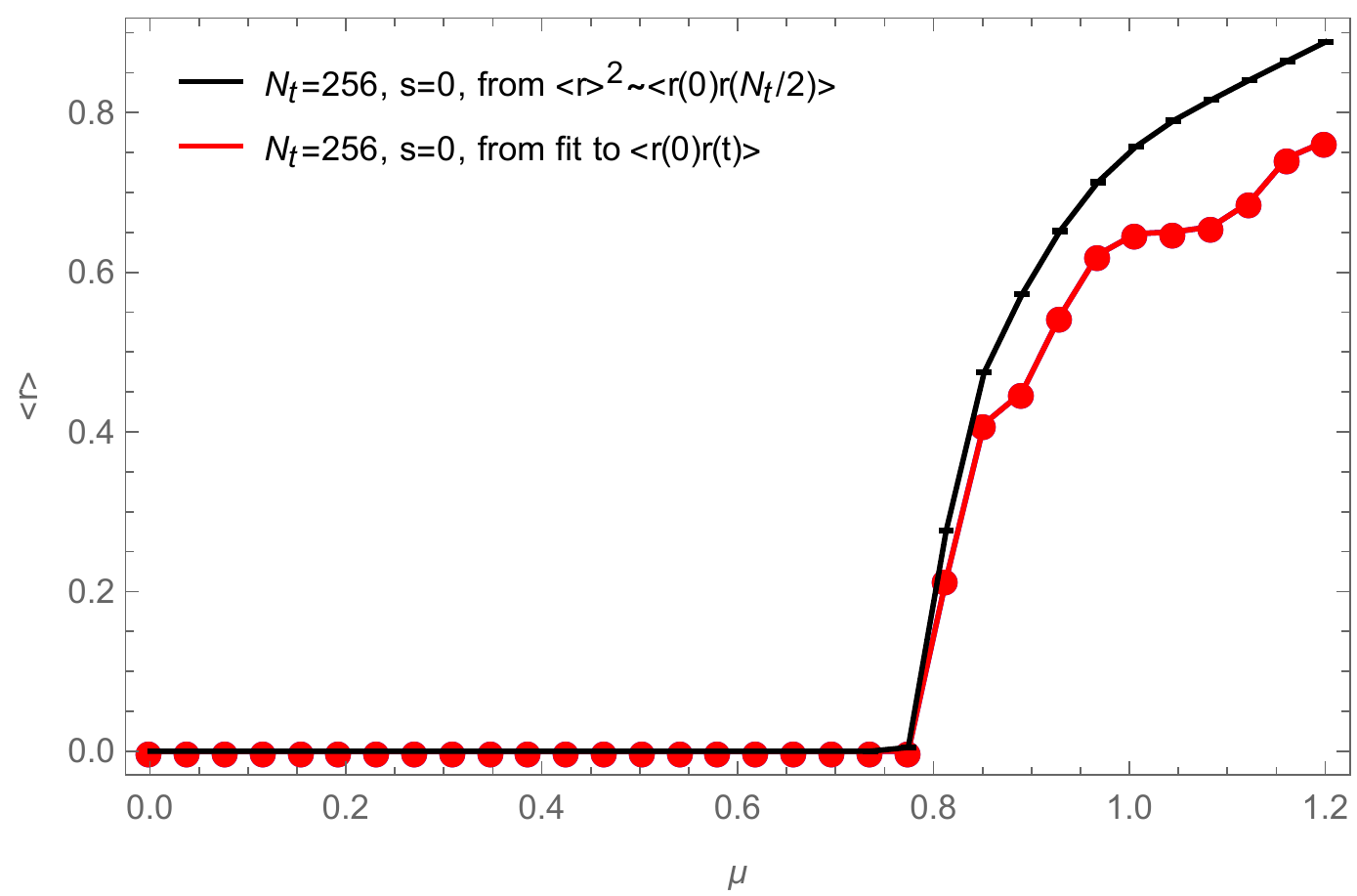}
\end{minipage}\hfill
\begin{minipage}[t]{0.49\linewidth}
\centering
\includegraphics[width=\linewidth]{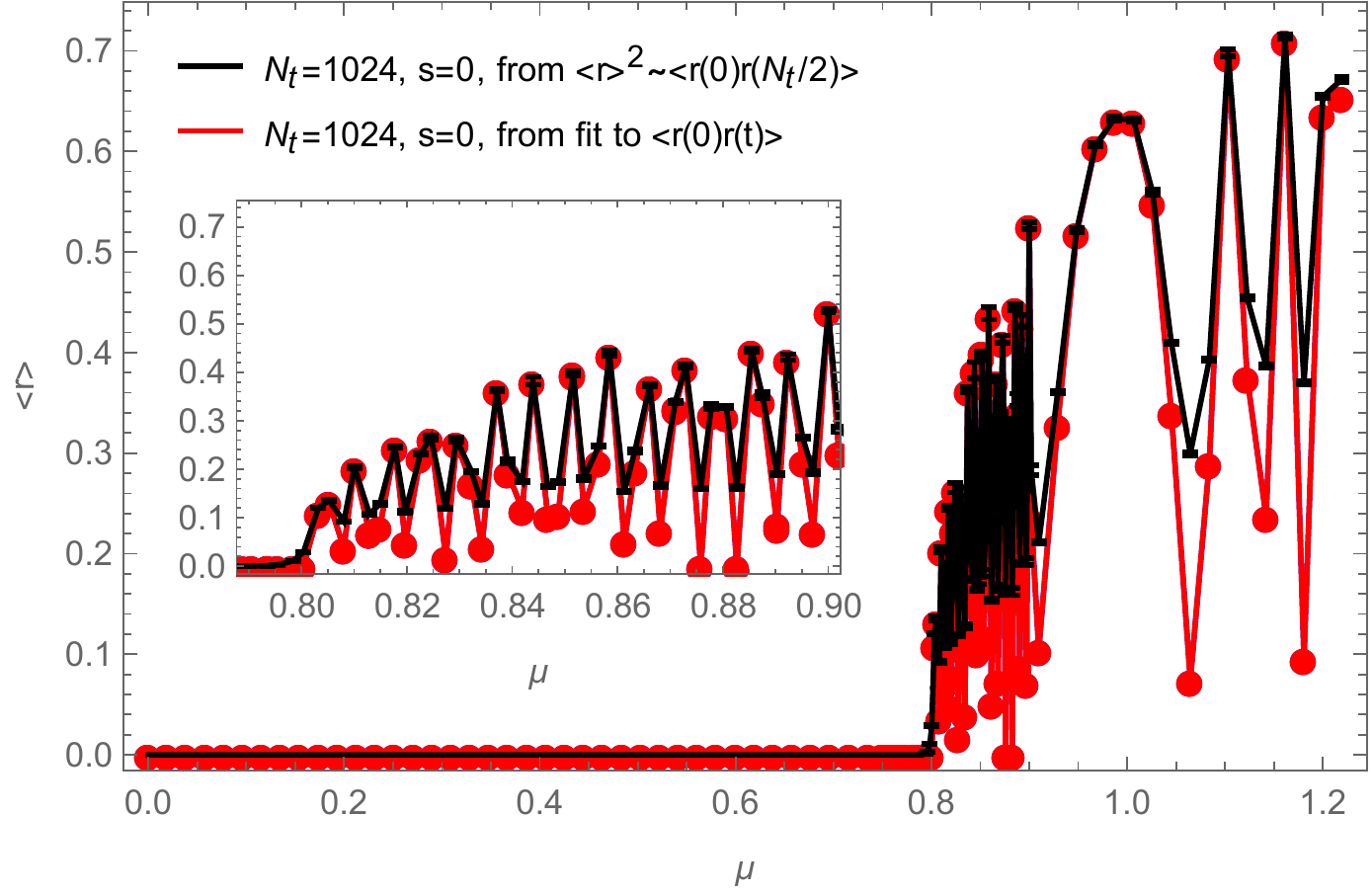}
\end{minipage}
\caption{The figure shows for two (2+1)-dimensional systems with $N_{s}=8$, $\kappa=0.25$, $\lambda=7.5$ and $r_s=0$ and $N_{t}=256$ (left) and $N_{t}=1024$ (right) the condensate $\avof{r}$ as a function of $\mu$, determined once by identifying $\avof{r}^2$ with twice (see explanation in text below eq.\eqref{eq:radexcorr}) the value of the zero-momentum piece of the measured \eqref{eq:radcorr2} (black), and once by fitting to the same correlator the function \eqref{eq:condrfitfunc} (red). As can be seen, for $N_{t}=256$ (left), the black and the red curve do not agree very well: the black curve shows too large values as the temperature is still too large and the zero-momentum correlator at $N_{t}/2$ deviates too much from the constant background. The red curve from the fit shows however already an oscillatory signal. For $N_{t}=1024$ (right), the agreement between the two curves is much better and the oscillations as a function of $\mu$ are very pronounced. We have increased the number of sampling points between $\mu=0.8$ and $\mu=0.9$ in order to resolve also some of the fast oscillations (see inset). Between $\mu=0.8$ and $\mu=1.2$, the charge density increases from zero to slightly above one, which happens in $N_{s}^{d-1}=8^2=64$ small steps. Each valley in the condensate corresponds to one of these small steps.}
  \label{fig:condcomp2}
\end{figure}

Another possibility would be to extract $\avof{\abs{\phi}}$ or $\avof{r}$ from a fit of the form
\[
f\of{t}\,=\,C_{1}^2\,+\,A\,\of{\e^{-m_1\,t}\,+\,e^{-m_2\,\of{N_t\,-\,t}}}\label{eq:condfitfunc}
\]
or
\[
f\of{t}\,=\,\frac{1}{2}\,C_{2}^2\,+\,A\,\cosh\of{m\,\of{t-N_{t}/2}}\label{eq:condrfitfunc}
\]
to the zero-momentum pieces of the correlators in \eqref{eq:sderivn} and \eqref{eq:radcorr2}, respectively, so that $\avof{\abs{\phi}}\,=\,C_{1}$ and $\avof{r}\,=\,C_{2}$, where the factor $1/2$ in front of the first term in \eqref{eq:condrfitfunc} is again due to the fact that for $r_{s}=0$ only the first two terms of \eqref{eq:radcorr2} can be measured. In Fig.~\ref{fig:condfromcorr}, we show an example of such measurements and compare them to results obtained by measuring the condensate directly, as described in Sec.~\ref{ssec:measurecond}, for four different, non-zero values of the source $r_{s}$. Note that the slight oscillations in the data coming from the systems with vanishing source are not just due to systematic errors in the fitting process, but are a manifestation of the phenomenon described in \cite{Chandrasekharan}, which is due to the quantization of the charge, visible in a finite system at low temperature (i.e. large $N_{t}$) (see also Fig. \ref{fig:condcomp2}, right). In Fig. \ref{fig:condcomp2}, we compare for two systems with different temporal extents, the result for the condensate $\avof{r}$ obtained with the fitting method with the one obtained by extracting the condensate from the minimum (at $t=N_{t}/2$) of the zero-momentum piece of \eqref{eq:radcorr2}. The fitting method is more involved but works better for not too large $N_{t}$, while for sufficiently large $N_{t}$ the two methods seem to work equally well.   

\section{Summary}\label{sec:summary}
Using the complex $\phi^4$ model as an example, we first reviewed the relation between the worm-algorithm and the charged correlator, how the former can be derived from the latter and how the standard worm can be thought of as sampling the charged correlator.\\
We then showed how the worm-algorithm can be generalized in order to sample also more general correlators and condensates and explained why this requires the inclusion of a non-vanishing source term in the action. In general, all correlators probing different internal space components of the field can be sampled during the worm Monte Carlo evolution by generalizing the standard worm-algorithm along the lines we presented.\\
In the Appendix we explain in some detail how our algorithm works and how the transition probabilities are computed in order to satisfy detailed balance.

\begin{appendices}
\section{Worm Algorithm}\label{sec:wormupdate}
In this appendix we describe our implementation of the generalized worm-algorithm introduced in the text. We explain why the code is organized in this particular way and show how the transition probabilities are obtained as to satisfy detailed balance.\\

\subsection{Organization of Program}\label{ssec:progorg}
The main components of our program are described in Alg. \ref{alg:wormupdate} in terms of pseudo-code. They consist of 3 main routines: 
\begin{itemize}
\item a routine "WORM\_UPDATE(x0,...)", which updates the $k$ and $l$ (and if the source is non-zero, also the $p$ and $q$) variables while sampling the full two-point function, starting and ending at "x0",
\item a routine "COND\_WORM\_UPDATE(x0,...)", which, if the source is non-zero, also updates the $k$, $l$, $p$ and $q$ variables while sampling the $\phi$ and $\phi^{*}$ condensates, starting at "x0" and ending at some other site (which will then be used as new value for "x0"),
\item and the routine "SWEEP()", which, for a fixed number of times, randomly executes either "WORM \_UPDATE(x0,...)" or "COND\_WORM\_UPDATE(x0,...)" (provided the source is non-zero), or picks a new random location for "x0".
\end{itemize}

In "SWEEP()", the probability for choosing next a new random location for "x0" is always 1/2. If $\abs{s}$ (or equivalently $r_{s}$) is set to zero, the probability for executing next "WORM\_UPDATE(x0,...)" is also 1/2. If $\abs{s}$ is non-zero, "WORM\_UPDATE(x0,...)" and "COND\_WORM\_UPDATE(x0,...)" share the 1/2 probability for being executed next and are therefore executed with probability 1/4 each. The two worms can always start in two different ways: for "WORM\_UPDATE(x0,...)" we can either have $\phi$ as the head and $\phi^{*}$ as the tail of the worm or vice versa, and similarly for "COND\_WORM \_UPDATE(x0,...)", we can either attempt to start by inserting a $\phi$ or a $\phi^{*}$ (together with an appropriate shift of a $p$-variable, i.e. the insertion of an appropriate dynamical monomer of opposite charge).\\

The $l$ and $q$ variables are updated along with the $k$ and $p$ variables in "WORM\_UP-DATE(x0,...)" and "COND\_WORM\_UPDATE(x0,...)": at every step of the worm the type of variable ($k$, $l$, $p$ or $q$ variable) to be updated next is selected at random, in such a way that the probability of selecting an $l$ variable ($q$ variable) is the same as the probability of selecting a $k$ variable ($p$ variable). In for example \cite{Gattringer} another updating strategy was used, where only the $k$ variables were updated during the worm evolution while the $l$ variables were updated in separate sweeps, executed periodically after a fixed number of worms. This can become problematic as soon as the system develops long-range order and the average worm-length becomes of the order of the system size: the worm then evolves for a long time in a fixed $l$-background, which slows down de-correlation and gives rise to larger errors in the two-point functions measured during the worm evolution. Furthermore, alternating between different types of updates in a fixed order strictly speaking breaks detailed balance (although this is usually harmless). Our updating scheme avoids both of these problems: the $l$ and $q$ variables do not form a fixed background while the worm evolves and detailed balance is satisfied at every step of the simulation.\\  

Note that if one is interested in two-point functions only, one could in principle avoid the removal and re-insertion of the external source/sink pair (and the uniform sampling of $x0$) whenever the worm closes, and instead just do importance sampling with respect to the location of the pair, which would be slightly more efficient as it would require less calls to the random-number generator. Measurements of observables which have to be obtained on closed-worm configurations would then have to be corrected by a reweighting factor, compensating for the presence of the external source/sink pair at $x0$.\\

Note also that "WORM\_UPDATE(x0,...)" and "COND\_WORM\_UPDATE(x0,...)" could in principle be combined into a single worm, which, provided $\abs{s}$ is non-zero, would not only sample the location and type of the head of the worm, but also sample the state of the worms tail: whether it should consist of an external source or sink as in "WORM\_UPDATE(x0,...)" or if it should be replaced by an appropriate shift of the $p$-variable (i.e. by a dynamical monomer) at the location of the tail as is the case in "COND\_WORM\_UPDATE(x0,...)". The reason why we decided to use separate worms is that our full simulation program is implemented to work for arbitrary linear and non-linear $\On{N}$ sigma models (complex $\phi^4$ corresponds to the linear $\On{2}$-case and the also mentioned $\SU{2}$ chiral-effective model to the non-linear $\On{4}$-case) with arbitrary source terms, in which case the implementation with two different worms for sampling two- and one-point functions turned out to be simpler.

\subsection{Detailed Balance and Transition Probabilities}\label{ssec:detailedbalance}

\begin{figure}[h!]
\centering
\begin{minipage}[t]{\linewidth}
\centering
\begin{tikzpicture}[scale=0.8,nodes={inner sep=0}]
  \pgfpointtransformed{\pgfpointxy{1}{1}};
  \pgfgetlastxy{\vx}{\vy}
  \clip (-4.5,-3) rectangle (15,14);
  \begin{scope}[node distance=\vx and \vy]
    \node at (0,3.6) {$x$};
  \end{scope}
  \begin{scope}[node distance=\vx and \vy]
    \node[draw,circle,inner sep=1.25,fill,color=red] at (3.9,4) {};
    \node[draw,circle,inner sep=1.25,fill,color=red] at (4.1,4) {};
    \node[draw,circle,inner sep=0.5,fill,color=white] at (4.1,4) {};
    \node at (4,3.6) {$x$};
  \end{scope}
  \begin{scope}[node distance=\vx and \vy]
    \draw[ultra thick,red!100] (7+1.05,4) -- (7+1.95,4);
    \node[draw,circle,inner sep=1.25,fill,color=red] at (7+0.9,4) {};
    \node[draw,circle,inner sep=1.25,fill,color=red] at (7+2.1,4) {};
    \node[draw,circle,inner sep=0.5,fill,color=white] at (7+2.1,4) {};
    \node at (7+0.9,3.6) {$x$};
    \node at (7+2.2,3.6) {$x+\hat{\nu}$};
  \end{scope}
  
  \begin{scope}[node distance=\vx and \vy]
    \node at (12+0.9,4) {$\cdots$};
  \end{scope}
  
  \begin{scope}[node distance=\vx and \vy]
    \node at (-4,3.6) {$y$};
  \end{scope}
  \begin{scope}[node distance=\vx and \vy]
    \draw[<-] (-3.3,4.1) to[bend left] node[pos=0.5,above]{$\sfrac{1}{2(V-1)}$} (-0.7,4.1);
    \draw[->] (-3.3,3.9) to[bend right] node[pos=0.5,below]{$\sfrac{1}{2(V-1)}$} (-0.7,3.9);
  \end{scope}  
  
%  \begin{scope}[node distance=\vx and \vy]
%    \node[draw,circle,inner sep=1.25,fill,color=red] at (3.9,1) {};
%    \node[draw,circle,inner sep=1.25,fill,color=red] at (4.1,1) {};
%    \node[draw,circle,inner sep=0.5,fill,color=white] at (4.1,1) {};
%    \node at (4,0.6) {$y$};
%  \end{scope}  
  \begin{scope}[node distance=\vx and \vy]
    \node[draw,circle,inner sep=1.25,fill,color=red] at (5.5+0.9,0) {};
    \node[draw,circle,inner sep=1.25,fill,color=blue] at (5.5+1.1,0) {};
    \node[draw,circle,inner sep=0.5,fill,color=white] at (5.5+1.1,0) {};
    \node[draw,circle,inner sep=1.25,fill,color=blue] at (5.5+1.9,1) {};
    \node[draw,circle,inner sep=1.25,fill,color=red] at (5.5+2.1,1) {};
    \node[draw,circle,inner sep=0.5,fill,color=white] at (5.5+2.1,1) {};
    \node at (5.5+1,-0.4) {$y$};
    \node at (5.5+2,0.6) {$z$};
  \end{scope}
  \begin{scope}[node distance=\vx and \vy]
    \node[draw,circle,inner sep=1.25,fill,color=blue] at (-0.1,7) {};
    \node[draw,circle,inner sep=1.25,fill,color=red] at (0.1,7) {};
    \node[draw,circle,inner sep=0.5,fill,color=white] at (0.1,7) {};
    \node at (0,6.6) {$x$};
  \end{scope}
  
  \begin{scope}[node distance=\vx and \vy]
    \draw[->] (-0.5,7) -- node[pos=0.5,above]{$\sfrac{1}{2}$} (-2,7);
    \node[left] at (-2.3,7) {\scriptsize l- or q-update};
  \end{scope}  
  
    \begin{scope}[node distance=\vx and \vy]
    \node at (0,13) {$x+\hat{\mu}$};
  \end{scope}
  
  \begin{scope}[node distance=\vx and \vy]
    \draw[ultra thick,red!100] (2.3,10) -- (3.1,10);
    \node[draw,circle,inner sep=1.25,fill,color=blue] at (2.1,10) {};
    \node[draw,circle,inner sep=1.25,fill,color=red] at (3.3,10) {};
    \node[draw,circle,inner sep=0.5,fill,color=white] at (3.3,10) {};
    \node at (2.1,9.6) {$x$};
    \node at (3.3,9.6) {$x+\hat{\mu}$};
  \end{scope}
  
  \begin{scope}[node distance=\vx and \vy]
    \draw[->] (2.6,10.7) -- node[pos=0.5,left]{$\sfrac{1}{2}$} (2.6,11.8);
    \node at (2.7,12.2) {\scriptsize l- or q-update};
  \end{scope}
  
  \begin{scope}[node distance=\vx and \vy]
    \draw[<-] (1.6,10.4) to[bend left] node[pos=0.5,below,left]{$\sfrac{1}{4\,\text{nc}}$} (0.2,12.4);
    \draw[->] (1.8,10.6) to[bend right] node[pos=0.75,above,right]{$\sfrac{1}{4}$} (0.4,12.6);
  \end{scope}
  
  \begin{scope}[node distance=\vx and \vy]
    \draw[<-] (0.2,7.5) to[bend left] node[pos=0.75,above,left]{$\sfrac{1}{8\,\text{d}}$} (1.6,9.5);
    \draw[->] (0.4,7.4) to[bend right] node[pos=0.5,below,right]{$\sfrac{1}{8\,\text{d}}$} (1.8,9.4);
  \end{scope}
  
  \begin{scope}[node distance=\vx and \vy]
    \node at (7.6,10) {$\cdots$};
  \end{scope}
    
  \begin{scope}[node distance=\vx and \vy]
    \draw[<-] (4.1,10.1) to[bend left] node[pos=0.5,above]{$\sfrac{1}{8\,\text{d}}$} (6.7,10.1);
    \draw[->] (4.1,9.9) to[bend right] node[pos=0.5,below]{$\sfrac{1}{8\,\text{d}}$} (6.7,9.9);
  \end{scope}
  
  \begin{scope}[node distance=\vx and \vy]
    \node at (11+0.4,0.5) {$\cdots$};
  \end{scope}
    
  \begin{scope}[node distance=\vx and \vy]
    \draw[<-] (-0.1,4.5) to[bend left] node[pos=0.5,left]{$\sfrac{1}{4}$} (-0.1,6.1);
    \draw[->] (0.1,4.5) to[bend right] node[pos=0.5,right]{$\sfrac{1}{4\,\text{nc}}$} (0.1,6.1);
  \end{scope}
  
  \begin{scope}[node distance=\vx and \vy]
    \draw[<-] (0.7,4.1) to[bend left] node[pos=0.5,above]{$\sfrac{1}{2}$} (3.3,4.1);
    \draw[->] (0.7,3.9) to[bend right] node[pos=0.5,below]{$\sfrac{1}{4\,\text{nc}}$} (3.3,3.9);
  \end{scope}
  
%  \begin{scope}[node distance=\vx and \vy]
%    \draw[<-] (0.7,1.1) to[bend left] node[pos=0.5,above]{$\sfrac{1}{2}$} (3.3,1.1);
%    \draw[->] (0.7,0.9) to[bend right] node[pos=0.5,below]{$\sfrac{1}{4\,\text{nc}}$} (3.3,0.9);
%  \end{scope}
  
  \begin{scope}[node distance=\vx and \vy]
    \draw[->] (8.4,4.7) -- node[pos=0.5,left]{$\sfrac{1}{2}$} (8.4,5.7);
    \node at (8.5,6.1) {\scriptsize l- or q-update};
  \end{scope}

  \begin{scope}[node distance=\vx and \vy]
    \draw[<-] (7,-1.5) -- node[pos=0.5,left]{$\sfrac{1}{2}$} (7,-0.5);
    \node at (7.0,-1.9) {\scriptsize l- or q-update};
  \end{scope}
  
  \begin{scope}[node distance=\vx and \vy]
    \draw[<-] (4.7,4.1) to[bend left] node[pos=0.5,above]{$\sfrac{1}{2\,\text{mnu}}$} (7.3,4.1);
    \draw[->] (4.7,3.9) to[bend right] node[pos=0.5,below]{$\sfrac{1}{2\,\text{mnu}}$} (7.3,3.9);
  \end{scope}
  
  \begin{scope}[node distance=\vx and \vy]
    \draw[<-] (4.4,3.1) to[bend left] node[pos=0.5,above,right]{$\sfrac{1}{2\,\text{mnu}}$} (6.4,1.1);
    \draw[->] (4.2,2.9) to[bend right] node[pos=0.5,below,left]{$\sfrac{1}{2\,\text{mnu}}$} (6.2,0.9);
  \end{scope}
  
  \begin{scope}[node distance=\vx and \vy]
    \draw[<-] (9.7,4.1) to[bend left] node[pos=0.5,above]{$\sfrac{1}{2\,\text{mnu}}$} (12.3,4.1);
    \draw[->] (9.7,3.9) to[bend right] node[pos=0.5,below]{$\sfrac{1}{2\,\text{mnu}}$} (12.3,3.9);
  \end{scope}
  
  \begin{scope}[node distance=\vx and \vy]
    \draw[<-] (8.2,0.6) to[bend left] node[pos=0.5,above]{$\sfrac{1}{2\,\text{mnu}}$} (10.8,0.6);
    \draw[->] (8.2,0.4) to[bend right] node[pos=0.5,below]{$\sfrac{1}{2\,\text{mnu}}$} (10.8,0.4);
  \end{scope}
  
  \begin{scope}[node distance=\vx and \vy]
    \draw[black,dotted] (-3.5,6.3) to[out=330,in=320,looseness=1] (7.5,12) to[out=140,in=150,looseness=1] node[pos=0.1,fill=white]{COND\_WORM\_UPDATE()} (-3.5,6.3);
    \draw[black,dotted] (2,-2.5) to[out=330,in=340,looseness=1] (13.5,7.5) to[out=160,in=150,looseness=1] node[pos=0.1,fill=white]{WORM\_UPDATE()} (2,-2.5);
  \end{scope} 
\end{tikzpicture}
\end{minipage}
\caption{The figure illustrates the move-choice probabilities for the different moves which are possible at different stages of the worm algorithm. Theses move-choice probabilities are necessary to determine the correct transition probabilities, satisfying detailed balance.}
  \label{fig:mcprobabs}
\end{figure}
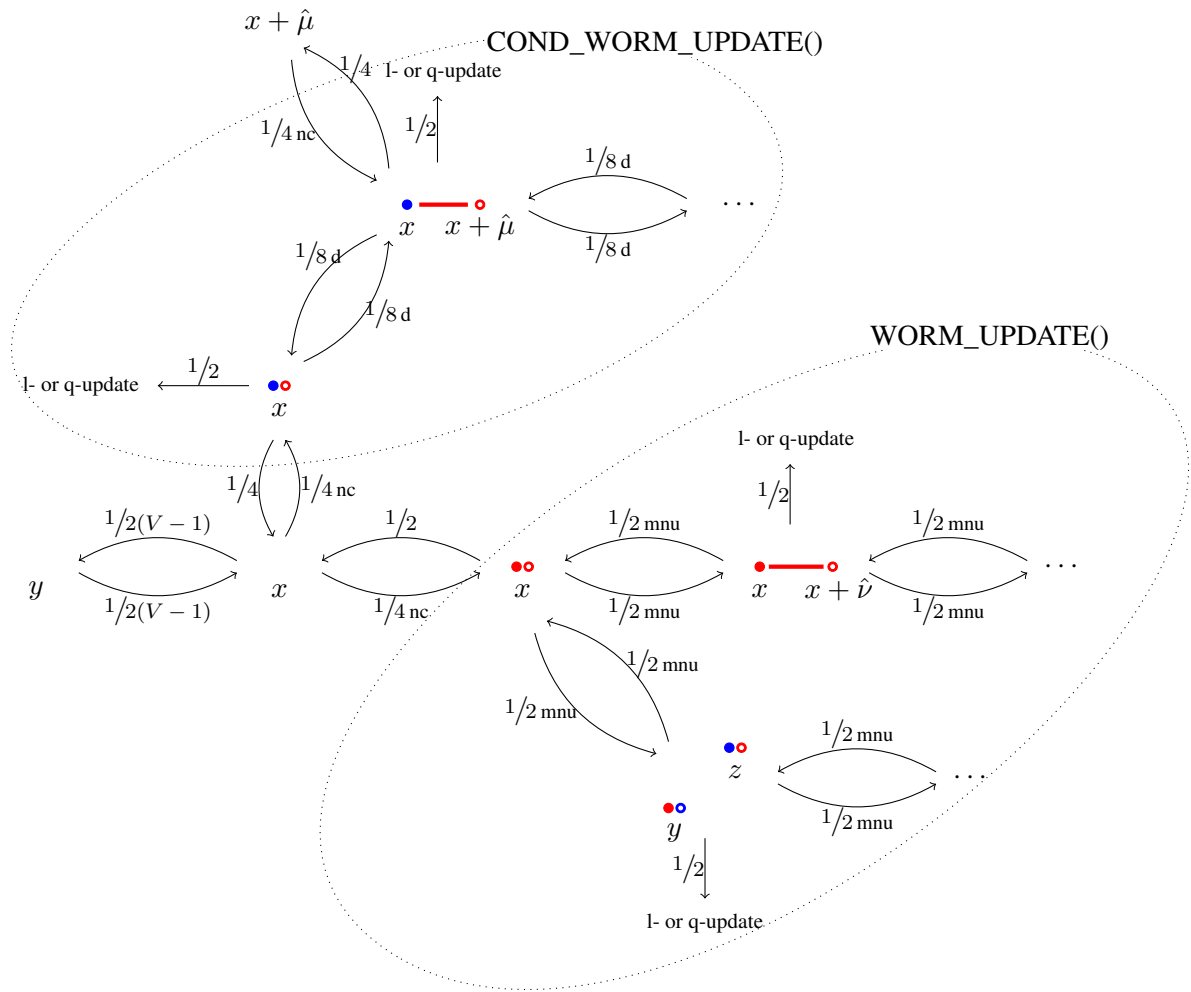

In order to ensure that our Markov chain really samples the partition, one- and two-point function, we have to ensure detailed balance between each pair of successive configurations $C$ and $C'$, i.e. configurations contributing to the partition, one- or two-point function, which can be turned into each other by a single update. The detailed balance equation takes the general form
\[
w\of{C}P\of{C\rightarrow C'}\,=\,w\of{C'}P\of{C'\rightarrow C}\ ,
\]
where $w\of{C}$ is the weight of the configuration $C$ and $P\of{C\rightarrow C'}$ is the transition probability for going from configuration $C$ to $C'$. However, as during different stages of our simulation, the number of possible moves changes, it can happen, that the move $C\rightarrow C'$ is chosen with a different probability than the inverse move $C'\rightarrow C$. We must then factor out from the transition probabilities the different \emph{move-choice probabilities}, $p$, and write
\[
P\of{C\rightarrow C'}\,=\,p\of{C\rightarrow C'}\,P_r\of{C\rightarrow C'}\ ,
\] 
and similarly
\[
P\of{C'\rightarrow C}\,=\,p\of{C'\rightarrow C}\,P_r\of{C'\rightarrow C}\ ,
\] 
and then use only the \emph{reduced transition probabilities}, $P_{r}$,  for the Metropolis acceptance test, i.e.
\[
P_r\of{C\rightarrow C'}\,=\,\min\bof{1,\frac{p\of{C'\rightarrow C}\,w\of{C'}}{p\of{C\rightarrow C'}\,w\of{C}}}\ ,\label{eq:redtransprobab1}
\]
and
\[
P_r\of{C'\rightarrow C}\,=\,\min\bof{1,\frac{p\of{C\rightarrow C'}\,w\of{C}}{p\of{C'\rightarrow C}\,w\of{C'}}}\ .\label{eq:redtransprobab2}
\]
The reason for this is, that in order for the Metropolis test to take place, the corresponding move has to be chosen already and the probability for this to happen should therefore not be taken into account a second time for the final decision whether the transition to the new configuration should take place or not. This is of course always the case, but if the move choice probabilities are the same for all the moves that are possible during the simulation, in particular for all pairs of move and inverse move, the move choice probabilities just cancel out in \eqref{eq:redtransprobab1} and \eqref{eq:redtransprobab2}.\\
The diagram in figure \ref{fig:mcprobabs} illustrates the different move-choice probabilities which occur at different stages of the simulation and in order to demonstrate how the corresponding reduced transition probabilities are determined, we show the derivation of these probabilities for the moves occurring in "WORM\_UPDATE(...)". In what follows we use the abbreviation $A_{x}=\abs{p_{x}}+2\,q_{x}+\sum\limits_{\nu}\sof{\abs{k_{x,\nu}}+\abs{k_{x-\hat{\nu},\nu}}+2\,\ssof{l_{x,\nu}+l_{x-\hat{\nu},\nu}}}$, and for the variable $n_{c}$ we have that $n_{c}=1$ if $r_{s}=\sqrt{2}\abs{s}=0$, and $n_{c}=2$ otherwise\footnote{The value of $n_{c}$ is different for $r_{s}=\sqrt{2}\abs{s}=0$ and $r_{s}=\sqrt{2}\abs{s}\neq 0$ because "WORM\_UPDATE(x0,...)" is executed with different probabilities within "SWEEP()" in these two cases, as explained at the beginning of Sec. \ref{ssec:progorg}.}.
\begin{itemize}
\item The detailed balance equation for starting the worm at site $x_{0}$, i.e. inserting an external source/sink pair at this point, reads:\\
\[
W_{\lambda}\sof{A_{x_0}}\,\frac{1}{4\,n_{c}}\,P_{r}\,=\,\frac{1}{4}\,W_{\lambda}\sof{A_{x_0}+2}\frac{1}{2}\,P'_{r},
\]
from which it follows that:
\[
P_{r}\,=\,\min\bof{1,\frac{n_{c}\,W_{\lambda}\sof{A_{x_0}+2}}{2\,W_{\lambda}\sof{A_{x_0}}}},
\]
\item If the head of the worm consists of a $\phi^{*}$, we define $\Delta=1$, and if the head is a $\phi$, we set $\Delta=-1$ instead. 
\begin{itemize}
\item For moving the head from site $x$ to its nearest-neighbor in positive direction, say to $x'=x+\hat{\nu}$, and changing $k_{x,\nu}\rightarrow k_{x,\nu}+\Delta=: k'_{x,\nu}$, the detailed balance equation reads:\\
\begin{multline}
\frac{\of{\frac{\kappa}{2}}^{\sabs{k_{x,\nu}}+2 l_{x,\nu}}\e^{2\,\mu\,k_{x,\nu}\,\delta_{\nu,d}}}{\of{\sabs{k_{x,\nu}}+l_{x,\nu}}!l_{x,\nu}!}\,W_{\lambda}\sof{A_{x}}\,W_{\lambda}\sof{A_{x'}}\,\frac{1}{2\,\of{2\,d+\,n_{c}}}\,P_{r}\,=\\
\frac{\of{\frac{\kappa}{2}}^{\sabs{k'_{x,\nu}}+2 l_{x,\nu}}\e^{2\,\mu\,k'_{x,\nu}\,\delta_{\nu,d}}}{\of{\sabs{k'_{x,\nu}}+l_{x,\nu}}!l_{x,\nu}!}\,W_{\lambda}\sof{A_{x}-1+\abs{k'_{x,\nu}}-\abs{k_{x,\nu}}}\\
\cdot\,W_{\lambda}\sof{A_{x'}+1+\abs{k'_{x,\nu}}-\abs{k_{x,\nu}}}\,\frac{1}{2\,\of{2\,d+\,n_{c}}}\,P'_{r},
\end{multline}
from which it follows that if $\abs{k'_{x,\nu}}>\abs{k_{x,\nu}}$:
\[
P_{r}\,=\,\min\bof{1,\frac{\frac{\kappa}{2}\e^{2\,\mu\,\Delta\,\delta_{\nu,d}}}{\abs{k'_{x,\nu}}+l_{x,\nu}}\frac{W_{\lambda}\sof{A_{x'}+2}}{W_{\lambda}\sof{A_{x'}}}},
\]
and if $\sabs{k'_{x,\nu}}<\sabs{k_{x,\nu}}$:
\[
P_{r}\,=\,\min\bof{1,\frac{\abs{k_{x,\nu}}+l_{x,\nu}}{\frac{\kappa}{2}\e^{-2\,\mu\,\Delta\,\delta_{\nu,d}}}\frac{W_{\lambda}\sof{A_{x}-2}}{W_{\lambda}\sof{A_{x}}}}.
\]
\item When moving the head from site $x$ to a nearest-neighbor in a negative direction, e.g. $x'=x-\hat{\nu}$ and shifting $k_{x',\nu}\rightarrow k_{x',\nu}-\Delta=: k'_{x',\nu}$, the corresponding detailed balance equation becomes:\\
\begin{multline}
\frac{\of{\frac{\kappa}{2}}^{\sabs{k_{x',\nu}}+2 l_{x',\nu}}\e^{2\,\mu\,k_{x',\nu}\delta_{nu,d}}}{\of{\sabs{k_{x',\nu}}+l_{x',\nu}}!l_{x',\nu}!}\,W_{\lambda}\sof{A_{x}}\,W_{\lambda}\sof{A_{x'}}\,\frac{1}{2\,\of{2\,d+\,n_{c}}}\,P_{r}\,=\\
\frac{\of{\frac{\kappa}{2}}^{\sabs{k'_{x',\nu}}+2 l_{x',\nu}}\e^{2\,\mu\,k_{x',\nu}\,\delta_{nu,d}}}{\of{\sabs{k'_{x',\nu}}+l_{x',\nu}}!l_{x',\nu}!}\,W_{\lambda}\sof{A_{x}-1+\sabs{k'_{x',\nu}}-\sabs{k_{x',\nu}}}\\
\cdot\,W_{\lambda}\sof{A_{x'}+1+\sabs{k'_{x',\nu}}-\sabs{k_{x',\nu}}}\,\frac{1}{2\,\of{2\,d+\,n_{c}}}\,P'_{r},
\end{multline}
such that for $\sabs{k'_{x',\nu}}>\sabs{k_{x',\nu}}$:
\[
P_{r}\,=\,\min\bof{1,\frac{\frac{\kappa}{2}\e^{-2\,\mu\,\Delta\,\delta_{\nu,d}}}{\sabs{k'_{x',\nu}}+l_{x',\nu}}\frac{W_{\lambda}\sof{A_{x'}+2}}{W_{\lambda}\sof{A_{x'}}}},
\]
and if $\sabs{k'_{x',\nu}}<\sabs{k_{x',\nu}}$:
\[
P_{r}\,=\,\min\bof{1,\frac{\sabs{k_{x',\nu}}+l_{x',\nu}}{\frac{\kappa}{2}\e^{2\,\mu\,\Delta\,\delta_{\nu,d}}}\frac{W_{\lambda}\sof{A_{x}-2}}{W_{\lambda}\sof{A_{x}}}}.
\]
\item Moving the head from site $x$ to another random site $x'$ and shifting $p_{x}\rightarrow p_{x}+\Delta=:p'_{x}$, $p_{x'}\rightarrow p_{x'}-\Delta=:p'_{x'}$ leads to the detailed balance equation,
\begin{multline}
\frac{\of{\frac{r_{s}}{2}}^{\abs{p_{x}}+2 q_{x}}}{\of{\abs{p_{x}}+q_{x}}!q_{x}!}\frac{\of{\frac{r_{s}}{2}}^{\abs{p_{x'}}+2 q_{x'}}}{\of{\abs{p_{x'}}+q_{x'}}!q_{x'}!}\,W_{\lambda}\sof{A_{x}}\,W_{\lambda}\sof{A_{x'}}\,\frac{1}{2\,\of{2\,d+\,n_{c}}}\,P_{r}\\
=\,\frac{\of{\frac{r_{s}}{2}}^{\abs{p'_{x}}+2 q_{x}}}{\of{\abs{p'_{x}}+q_{x}}!q_{x}!}\frac{\of{\frac{r_{s}}{2}}^{\abs{p'_{x'}}+2 q_{x'}}}{\of{\abs{p'_{x'}}+q_{x'}}!q_{x'}!}\\
\cdot\,W_{\lambda}\sof{A_{x}-1+\abs{p'_{x}}-\abs{p_{x}}}\,W_{\lambda}\sof{A_{x'}+1+\abs{p'_{x'}}-\abs{p_{x'}}}\,\frac{1}{2\,\of{2\,d+\,n_{c}}}\,P'_{r},\label{eq:detbalanced1}
\end{multline}
such that
\[
P_{r}\,=\,\min\sof{1,R_{r,x}\,R_{r,x'}},\label{eq:trprobabd1}
\]
where 
\[
R_{r,x}\,=\,
\begin{cases}
\frac{r_{s}}{\abs{p'_{x}}+q_{x}} &\mbox{if } \abs{p'_{x}}>\abs{p_{x}} \\ 
\frac{\abs{p_{x}}+q_{x}}{\frac{r_{s}}{4}}\frac{W_{\lambda}\sof{A_{x}-2}}{W_{\lambda}\sof{A_{x}}} & \mbox{if } \abs{p'_{x}}<\abs{p_{x}}.
\end{cases}\label{eq:prx1}
\]
and
\[ 
R_{r,x'}\,=\,
\begin{cases}
\frac{\frac{r_{s}}{4}}{\sabs{p'_{x'}}+q_{x'}}\frac{W_{\lambda}\sof{A_{x'}+2}}{W_{\lambda}\sof{A_{x'}}} &\mbox{if } \sabs{p'_{x'}}>\sabs{p_{x'}} \\ 
\frac{\sabs{p_{x'}}+q_{x'}}{r_{s}} & \mbox{if } \sabs{p'_{x'}}<\sabs{p_{x'}}.
\end{cases}\label{eq:prx2}
\]
\item Removing the head from site $x$ and inserting it again at $x'$ with opposite external charge leads again to \eqref{eq:detbalanced1}, \eqref{eq:trprobabd1}, \eqref{eq:prx1} and \eqref{eq:prx2}, but this time with $p_{x}\rightarrow p_{x}+\Delta=:p'_{x}$ and $p_{x'}\rightarrow p_{x'}+\Delta=:p'_{x'}$.
\end{itemize}

\item Finally, if head and tail are again located both on site $x_{0}$ and consist of opposite external charges, one can propose to terminate the worm, for which the detailed balance equation is given by\\
\[
\frac{1}{4}\,W_{\lambda}\sof{A_{x_0}}\frac{1}{2}\,P_{r}\,=\,W_{\lambda}\sof{A_{x_0}-2}\frac{1}{4\,n_{c}}\,P'_{r},
\]
and therefore:
\[
P_{r}\,=\,\min\bof{1,\frac{2\,W_{\lambda}\sof{A_{x_0}-2}}{n_{c}\,W_{\lambda}\sof{A_{x_0}}}}.
\]
\end{itemize}

The reduced transition probabilities for the moves in "COND\_WORM\_UPDATE(...)" can be obtained in a a completely analogous way.

\clearpage

\newsavebox{\smlmat}% Box to store smallmatrix content
\savebox{\smlmat}{\scriptsize $\begin{pmatrix} \phi\of{x_0}\phi\of{x_0+\ohat{d}\cdot t} & \phi\of{x_0}\phi^{*}\of{x_0+\ohat{d}\cdot t}\\ \phi^{*}\of{x_0}\phi\of{x_0+\ohat{d}\cdot t} & \phi^{*}\of{x_0}\phi^{*}\of{x_0+\ohat{d}\cdot t} \end{pmatrix}$}
\begin{algorithm}[H]
\caption{\small The following pseudo-code describes our simulation program. The routine "SWEEP()" makes use of a worm-update routine "WORM\_UPDATE()" which samples the two-point functions, allows for monomer insertions and head-changes of the worm (provided sr>0). If sr>0, it makes also use of a second worm-update routine "COND\_WORM\_UPDATE()" which samples the condensates. The input parameters are as follows:\\
\&sites is a reference to the two-dimensional array "sites", where:\\
sites[$i_x$][0]=$\abs{p_{x}}+2\,q_{x}+\sum\limits_{\nu}\sof{\abs{k_{x,\nu}}+\abs{k_{x-\ohat{\nu},\nu}}+2\,\ssof{l_{x,\nu}+l_{x-\ohat{\nu},\nu}}}$,\\
sites[$i_x$][1]=sites[x][0] (for a $\On{N}$ model with $N>2$ sites[$i_x$][1] would be different from sites[$i_x$][0]),\\
sites[$i_x$][2]=$p_{x}$, sites[$i_x$][3]=$q_{x}$,\\
\&bonds is a reference to the 3-dimensional array "bonds", where for nu<d, with $\nu$=nu+1:\\
bonds[$i_x$][nu][0]=$k_{x,\nu}$,\\
bonds[$i_x$][nu][1]=$l_{x,\nu}$,\\
and for nu$\geq$d, with $\nu$=(nu-d)+1:\\
bonds[$i_x$][nu][0]=\&bonds[$i_{x-\ohat{\nu}}$][nu-d][0],\\
bonds[$i_x$][nu][1]=\&bonds[$i_{x-\ohat{\nu}}$][nu-d][1],\\
\&xnbr is a reference to the nearest-neighbor lookup table "xnbr", where for nu<d, with $\nu$=nu+1, we have xnbr[$i_x$][nu]=$i_{x+\ohat{\nu}}$, and for nu>=d, with $\nu$=(nu-d)+1: xnbr[$i_x$][nu]=$i_{x-\ohat{\nu}}$.\\
\&wh is a reference to the array "wh" of weight ratios:\\
wh[n]=$W_{\lambda}\of{n+2}/W_{\lambda}\of{n}$, where $W_{\lambda}\of{n}$ is defined in eq. \eqref{eq:weightf},\\
h=$\kappa$ is the hopping parameter,\\
mu=$\mu$ is the chemical potential,\\
sr=$s_{r}$ is the radial source,\\
Ns: number of sites per spatial dimension,\\
Nt: number of sites in the temporal direction,\\
d: number of dimensions,\\
sweeplen: number of attempted worms or shifts of x0 per sweep, can be tuned to obtain sweeps with fixed average number of local updates.\\
\&ni: reference to the variable "ni", containing the total charge $\sum\limits_{x}\,k_{x,4}$,\\
\&corrs and \&corrt are references to the variables "corrs" and "corrt", containing the spatial and temporal (zero-momentum) propagators, e.g. corrt[t]$\sim\frac{1}{V}\sum_{x_0}$\usebox{\smlmat}.\\
The notation x\%N for x$\in\mathbb{Z}$, N$\in\mathbb{N}$, refers to $\of{x \bmod N}\in\{0,\ldots,N-1\}$ and the function randi(N) returns a random integer in the range $\{0,\ldots,N-1\}$ while rand() is understood to return a random floating point number in the semi-open interval $[0,1)$.}
\label{alg:wormupdate}

\begin{algorithmic}\small
\Function{sweep}{\&x0, \&sites, \&bonds, \&xnbr, h, mu, sr, Ns, Nt, d, sweeplen, \&ni, \&corrs, \&corrt, \&cond}
  \State V=Nt*pow(Ns,d-1);
  \State nc=1;
  \If{sr>0}
    \State nc=2;
  \EndIf
  \State imax=sweeplen;\Comment{\parbox[t][0em][t]{.45\linewidth}{Call on average imax/(2*nc) times the routine WORM\_UPDATE(). If source is none zero, then nc=2 and also COND\_WORM\_UPDATE() will on average be called imax/(2*nc) times. The remaining 50\% of the cases, x0 is changed.}}
  \For{i=0; i<imax; ++i}
    \State tm=randi(2*nc);
    \If{tm>=nc}
       \State x0=(x0+1+randi(V-2))\%V;\\
\algstore{myalg0}
\end{algorithmic}
\end{algorithm}

\begin{algorithm}
\begin{algorithmic}\small
\algrestore{myalg0}
    \ElsIf{tm==0}
      \State WORM\_UPDATE(x0, sites, bonds, xnbr, wh, h, mu, sr, Ns ,Nt ,ni ,corrs ,corrt);
    \Else
      \State COND\_WORM\_UPDATE(x0, sites, bonds, xnbr, wh, h, mu, sr, Ns ,Nt ,ni, cond);
    \EndIf
  \EndFor
\EndFunction\\

\Function{\textbf{worm\_update}}{\&x0, \&sites, \&bonds, \&xnbr, \&wh, h, mu, sr, Ns, Nt, d, \&ni, \&corrs, \&corrt}
  \State c0=randi(2);\Comment{\parbox[t][0em][t]{.5\linewidth}{c0=0: tail of worm is a source; c0=1: tail of worm is sink.}}
  \State x=x0;
  \State c=c0;
  \State rho=2*nc*0.25*wh[sites[x][0]];
  
  \If{rho>=1 || rand()<rho}\Comment{\parbox[t][0em][t]{.5\linewidth}{acceptance test for insertion of source and sink at x.}}
     \State sites[x][0]+=2;
     \State sites[x][1]+=2;
  \Else
    \State \Return;
  \EndIf\\
  \If{c==0}\Comment{\parbox[t][0em][t]{.5\linewidth}{Source or sink at x?}}
    \State del=1;
  \Else
    \State del=-1;
  \EndIf
  \If{sr>0}
    \State mnu=2*d+2;\Comment{\parbox[t][0em][t]{.5\linewidth}{if the radial source is non-zero, two additional moves are possible at each step of the worm.}}
    \State nc=2;
  \Else
    \State mnu=2*d;\Comment{\parbox[t][0em][t]{.5\linewidth}{if the source is zero, the worm can evolve only by moving its head from one site to one of its 2*d neighboring sites.}}
    \State nc=1;
  \EndIf\\
  
  \State cx=array(d,0);\Comment{\parbox[t][2em][t]{.5\linewidth}{cx[nu] will contain the relative distance in $\nu$-direction (with $\nu$=(nu+1)) between head an tail of the worm.}}
  \While{true}
    \State tc=(c+1)\%2;
    \For{i=0; i<d-1; ++i}
      \State corrs[cx[i]][c0][tc]+=1;\Comment{\parbox[t][2em][t]{.5\linewidth}{increase the histogram for the current head-tail separation in each spatial direction.}}
    \EndFor
    \State corrt[cx[d-1]][c0][tc]+=1;\Comment{\parbox[t][2em][t]{.5\linewidth}{increase the histogram for the current head-tail separation in temporal direction.}}
      \State nu=randi(2*mnu);
      \If{nu>=mnu}\Comment{\parbox[t][2.5em][t]{.35\linewidth}{update l or q-variables or terminate worm if possible:}}
\algstore{myalg1}
\end{algorithmic}
\end{algorithm}

\begin{algorithm}
\begin{algorithmic}\small
\algrestore{myalg1}
        \If{c==c0 \&\& x==x0}\Comment{\parbox[t][0em][t]{.35\linewidth}{if head and tail are on the same site and consist of a source-sink pair: propose to terminate the worm by removing them,}}
          \State rho=1./(2*nc*0.25*wh[sites[x][0]-2]);
          \If{rho>=1 || rand()<rho}
            \State sites[x][0]-=2;
            \State sites[x][1]-=2;
            \State break;
          \EndIf
        \Else\Comment{\parbox[t][0em][t]{.35\linewidth}{if head and tail are not on the same site or if they do not consist of a corresponding source-sink pair: attempt to update either the l or the q-variables.}}
          \If{nu-mnu<2*d}
            \State L\_UPDATE(sites, bonds, xnbr, wh, V, d);
          \Else
            \State Q\_UPDATE(sites, wh, V);
          \EndIf
        \EndIf

      \ElsIf{nu<2*d}\Comment{\parbox[t][0em][t]{.35\linewidth}{attempt to move the head to a neighboring site.}}
        \State tnu=nu\%d;
        \State xn=xnbr[x][nu];
        \State tdk=1-2*floor(nu/d);\Comment{\parbox[t][0em][t]{.35\linewidth}{determines if nu corresponds to a positive or negative direction.}}
        \State k=bonds[x][nu][0];
		\State l=bonds[x][nu][1];
		\State kn=k+tdk*del;\Comment{\parbox[t][0em][t]{.4\linewidth}{the factor of tdk is there because k-variables in negative direction enter the delta function constraint at x with a negative sign and must therefore be changed in the opposite direction.}}
		\State ch=abs(k)-l;
		\State chn=abs(kn)-l;
		\State dak=chn-ch;
		\If{dak>0}
		  \State rho=0.5*h*wh[sites[xn][0]]/chn;
		\Else
		  \State rho=ch/(0.5*h*wh[sites[x][0]-2]);
		\EndIf
		\If{tnu==(d-1)}
		  \If{tdk*del>0}
		    \State rho*=exp(2*mu);
		  \Else
		    \State rho*=exp(-2*mu);
		  \EndIf
		\EndIf
		\If{rho>=1 || rand()<rho}\Comment{\parbox[t][0em][t]{.4\linewidth}{acceptance test for moving the head of the worm from site x to site xn while changing bonds[x][nu][0] from k to kn.}}
		  \State bonds[x][nu][0]=kn;
		  \If{dak>0}
		    \State sites[xn][0]+=2*dak;
		    \State sites[xn][1]+=2*dak;
		  \Else
		    \State sites[x][0]+=2*dak;
		    \State sites[x][1]+=2*dak;
		  \EndIf
		  \State cx[tnu]+=tdk;
\algstore{myalg2}
\end{algorithmic}
\end{algorithm}

\begin{algorithm}
\begin{algorithmic}\small
\algrestore{myalg2}
		  \If{tnu==(d-1)}\Comment{\parbox[t][0em][t]{.4\linewidth}{if the head of the worm is moved in time direction, update ni.}}
		    \State ni+=tdk*del;
		    \State cx[tnu]=cx[tnu]\%Nt;
		  \Else
		    \State cx[tnu]=cx[tnu]\%Ns;
		  \EndIf
		  \State x=xn;
		\EndIf
      \Else\Comment{\parbox[t][0em][t]{.4\linewidth}{attempt to move the head to a distant location by inserting monomers.}}
		\State cn=nu-2*d;
		\State xn=randi(V-1);\Comment{\parbox[t][0em][t]{.4\linewidth}{choose random new location xn (distinct from x).}}
		\If{xn>=x}
		  \State ++xn;
		\EndIf
		\State p1=sites[x][2];
		\State q1=sites[x][3];
		\State p1n=p1+del;
		\State dap1=abs(p1n)-abs(p1);
		\If{dap1>0}
		  \State rho=sr/(abs(p1n)+q1);
		\Else
		  \State rho=(abs(p1)+q1)/(0.25*sr*wh[sites[x][0]-2]);
		\EndIf
		\If{cn==0}
		  \State deln=1;
		\Else
		  \State deln=-1;
		\EndIf
		\State p2=sites[xn][2];
		\State q2=sites[xn][3];
		\State p2n=p2-deln;
		\State dap2=abs(p2n)-abs(p2);
		\If{dap2>0}
		  \State rho*=0.25*sr*wh[sites[xn][0]]/(abs(p2n)+q2);
		\Else
		  \State rho*=(abs(p2)+q2)/sr;
		\EndIf
		\If{rho>=1 || rand()<rho}\Comment{\parbox[t][0em][t]{.4\linewidth}{acceptance test for removing the head of the worm at x and inserting it again at xn.}}
		  \State sites[x][2]=p1n;
		  \If{dap1<0}
		    \State sites[x][0]-=2;
		    \State sites[x][1]-=2;
		  \EndIf
		  \State sites[xn][2]=p2n;
		  \If{dap2>0}
		    \State sites[xn][0]+=2;
		    \State sites[xn][1]+=2;
		  \EndIf
\algstore{myalg4}
\end{algorithmic}
\end{algorithm}

\begin{algorithm}
\begin{algorithmic}\small
\algrestore{myalg4}
		  \State xi=x;
		  \State xin=xn;
		  \For{tnu=0; tnu<d-1; ++tnu}\Comment{\parbox[t][0em][t]{.4\linewidth}{compute new distance between the worm's head and tail in each direction.}}
		    \State cx[tnu]+=((xin\%Ns)-(xi\%Ns));
		    \State cx[tnu]=cx[tnu]\%Ns;
		    \State xin=floor(xin/Ns);
		    \State xi=floor(xi/Ns);
		  \EndFor
	      \State cx[d-1]+=((xin\%Nt)-(xi\%Nt));
	      \State cx[d-1]=cx[d-1]\%Nt;
		  \State x=xn;
		  \State c=cn;
		  \State del=deln;
		\EndIf
      \EndIf
  \EndWhile
  \State \Return;
\EndFunction\\

\Function{cond\_worm\_update}{\&x0, \&sites, \&bonds, \&xnbr, \&wh, h, mu, sr, Ns, Nt, d, \&ni, \&cond}
  \State c=randi(2);
  \State x=x0;
  \If{c==0}
    \State del=1;\Comment{\parbox[t][0em][t]{.4\linewidth}{charge at worm's tail positive or negative?}}
  \Else
    \State del=-1;
  \EndIf
  \State p1=sites[x][2];
  \State q1=sites[x][3];
  \State p1n=p1-del;
  \State dap1=abs(p1n)-abs(p1);
  \State rho=2.;
  \If{dap1>0}
    \State rho*=0.25*sr*wh[sites[x][0]]/(abs(p1n)+q1);
  \Else
    \State rho*=(abs(p1)+q1)/sr;
  \EndIf
  \If{rho>=1 || rand()<rho}\Comment{\parbox[t][0em][t]{.4\linewidth}{acceptance test for inserting an external source and a monomer at x. If not accepted: terminate routine.}}
    \State sites[x][2]=p2n;
    \If{dap1>0}
      \State sites[x][0]+=2;
      \State sites[x][1]+=2;
    \EndIf
  \Else
    \State \Return;
  \EndIf
  \State tc=(c+1)\%2;
  \State mnu=2*d;
\algstore{myalg5}
\end{algorithmic}
\end{algorithm}

\begin{algorithm}
\begin{algorithmic}\small
\algrestore{myalg5}
  \While{true}
    \State cond[tc]+=1;
    \State nu=randi(4*mnu);\Comment{\parbox[t][0em][t]{.4\linewidth}{four possible moves: move head of worm, update a l-variable, terminate worm or update a q-variable.}}\vspace{25pt}
    \If{nu<mnu}\Comment{\parbox[t][0em][t]{.4\linewidth}{attempt to move the worms head.}}
      \State tnu=nu\%d;
      \State xn=xnbr[x][nu];
      \State tdk=1-2*floor(nu/d);
      \State k=bonds[x][nu][0];
      \State l=bonds[x][nu][1];
      \State kn=k+tdk*del;
      \State ch=abs(k)+l;
      \State chn=abs(kn)+l;
      \State dak=chn-ch;
      \If{dak>0}
        \State rho=0.5*h*wh[sites[xn][0]]/chn;
      \Else
        \State rho=ch/(0.5*h*wh[sites[x][0]-2]);
      \EndIf

      \If{tnu==(d-1)}\Comment{\parbox[t][0em][t]{.4\linewidth}{if proposed move is in time-direction, take chemical potential into account.}}
        \If{tdk*del>0}
          \State rho*=exp(2*mu);
        \Else
          \State rho*=exp(-2*mu);
        \EndIf
      \EndIf
      \If{rho>=1 || rand()<rho} 
        \If{tnu==(d-1)} 
          \State ni+=tdk*del;
        \EndIf
        \If{dak>0}
          \State sites[xn][0]+=2*dak;
          \State sites[xn][1]+=2*dak;
        \Else
          \State sites[x][0]+=2*dak;
          \State sites[x][1]+=2*dak;
        \EndIf
        \State bonds[x][nu][0]=kn;
        \State x=xn;
      \EndIf
    \ElsIf{nu<2*mnu}\Comment{\parbox[t][0em][t]{.4\linewidth}{attempt to update a l-variable.}}
      \State L\_UPDATE(sites,bonds,xnbr,wh,V,d);
    \ElsIf{nu<3*mnu}\Comment{\parbox[t][0em][t]{.4\linewidth}{attempt to terminate the worm by replacing the external charge at the head by a monomer.}}
      \State p1=sites[x][2];
      \State q1=sites[x][3];
      \State p1n=p1+del;
      \State dap1=abs(p1n)-abs(p1);
      \State rho=1/(2*nc);
\algstore{myalg6}
\end{algorithmic}
\end{algorithm}

\begin{algorithm}
\begin{algorithmic}\small
\algrestore{myalg6}
      \If{dap1>0}
        \State rho*=sr/(abs(p1n)+q1);
      \Else
        \State rho*=(abs(p1)+q1)/(0.25*sr*wh[sites[x][0]-2]);
      \EndIf
      \If{rho>=1 || rand()<rho}
        \State sites[x][2]=p1n;
        \If{dap1<0}
          \State sites[x][0]-=2;
          \State sites[x][1]-=2;
        \EndIf
        \State x0=x;
        \State break;
      \EndIf
    \Else\Comment{\parbox[t][0em][t]{.4\linewidth}{attempt to update a q-variable.}}
      \State Q\_UPDATE(sites, wh, V);
    \EndIf 
  \EndWhile
  \State \Return; 
\EndFunction\\

\Function{l\_update}{\&sites, \&bonds, \&xnbr, \&wh, V, d}
  \State x=randi(V);
  \State nu=randi(d);
  \State del=1-2*randi(2);
  \State l=bonds[x][nu][1];
  \State ln=l+del;
  \If{ln>=0}
    \State xn=xnbr[x][nu];
    \State ak=abs(sites[x][nu][0]);
    \If{del>0}
      \State rho=0.25*wh[sites[x][0]]*wh[sites[xn][0]]*h*h/(ln*(ak+ln));
    \Else
      \State rho=l*(ak+l)/(h*h*0.25*wh[sites[x][0]-2]*wh[sites[xn][0]-2]);
    \EndIf
    \If{rho>=1 || rand()<rho}
      \State bonds[x][nu][1]=ln;
      \State sites[x][0]+=2*del;
	  \State sites[x][1]+=2*del;
	  \State sites[xn][0]+=2*del;
	  \State sites[xn][1]+=2*del;
    \EndIf
  \EndIf
  \State \Return;
\EndFunction\\
\algstore{myalg8}
\end{algorithmic}
\end{algorithm}

\begin{algorithm}
\begin{algorithmic}\small
\algrestore{myalg8}
\Function{q\_update}{\&sites, \&wh, V}
  \State x=randi(V);
  \State del=1-2*randi(2);
  \State q=sites[x][3];
  \State qn=q+del;
  \If{qn>=0}
    \State ap=abs(sites[x][2]);
    \If{del>0}
      \State rho=0.25*wh[sites[x][0]]*sr*sr/(qn*(ap+qn));
    \Else
      \State rho=q*(ap+q)/(sr*sr*0.25*wh[sites[x][0]-2]);
    \EndIf
    \If{rho>=1 || rand()<rho}
      \State sites[x][0]+=2*del;
	  \State sites[x][1]+=2*del;
	  \State sites[x][3]=qn;
    \EndIf
  \EndIf
  \State \Return;
\EndFunction
\end{algorithmic}
\end{algorithm}
\end{appendices}

\end{document}